\documentclass[a4paper,fleqn]{cas-dc}

\usepackage{color}
\usepackage{bm}
\usepackage{graphicx}
\usepackage{natbib}
\usepackage{amsmath,amssymb}
\usepackage{accents}
\usepackage{float}
\usepackage{wrapfig}
\usepackage{layout}
\usepackage{braket}
\usepackage{setspace}
\usepackage{url}
\let\%\relax 
\DeclareFontFamily{OT1}{pxr}{}
\DeclareFontShape{OT1}{pxr}{m}{n}{<->pxr}{}
\usepackage{scalerel,stackengine}
\stackMath
\newcommand\reallywidecheck[1]{%
\savestack{\tmpbox}{\stretchto{%
  \scaleto{%
    \scalerel*[\widthof{\ensuremath{#1}}]{\kern-.6pt\bigwedge\kern-.6pt}%
    {\rule[-\textheight/2]{1ex}{\textheight}}
  }{\textheight}%
}{0.5ex}}%
\stackon[1pt]{#1}{\scalebox{-1}{\tmpbox}}%
}
\newcommand{\beq}{\begin{eqnarray}}
\newcommand{\eeq}{\end{eqnarray}}
\newcommand{\nn}{\nonumber}
\newcommand{\varv}{v}
\newcommand{\varw}{w}

\newcommand{\tPt}{${}^{3}P_{2}$ }

\newcommand{\ii}{i}
\newcommand{\F}{\mathrm{F}}
\newcommand{\dd}{\mathrm{d}}

\newcommand{\n}{\mathrm{n}}

\newcommand{\red}{\textcolor{black}}
\newcommand{\blue}{\textcolor{black}}
\newcommand{\green}{\textcolor{black}}

\newcommand{\sub}[1]{{\bf {#1}.}}
\usepackage[normalem]{ulem}

\begin{document}
\emergencystretch 3em

\def\floatpagepagefraction{1}
\def\textpagefraction{.001}
\shorttitle{Non-Abelian Anyons and Non-Abelian Vortices  
in Topological Superconductors}
\shortauthors{Y. Masaki, T. Mizushima and M. Nitta}

\title[mode = title]{Non-Abelian Anyons and Non-Abelian Vortices 
in Topological Superconductors}

\author[1,2]{Yusuke Masaki}[orcid=0000-0001-6891-7008]


\address[1]{Department of Physics, Tohoku University, Sendai, Miyagi 980-8578, Japan}
\address[2]{Research and Education Center for Natural Sciences, Keio University, Hiyoshi 4-1-1,
Yokohama, Kanagawa 223-8521, Japan}

\author[3]{Takeshi Mizushima}[orcid=0000-0002-7313-6094]

\address[3]{Department of Materials Engineering Science, Osaka University, Toyonaka, Osaka 560-8531, Japan}
\ead{mizushima@mp.es.osaka-u.ac.jp}
\cormark[1]

\author[2,4,5]{Muneto Nitta}[orcid=0000-0002-3851-9305]

\address[4]{Department of Physics, Keio University, Hiyoshi 4-1-1, Japan}
\address[5]{International Institute for Sustainability with Knotted Chiral Meta Matter (SKCM$^2$), Hiroshima University, Kagamiyama 1-3-2, Higashi-Hiroshima, Hiroshima 739-8511, Japan}
\ead{nitta@phys-h.keio.ac.jp}
\cormark[1]
\cortext[1]{Corresponding author}

\begin{abstract}
Anyons are particles obeying statistics of neither bosons nor fermions. Non-Abelian anyons, whose exchanges are described by a non-Abelian group acting on a set of wave functions, are attracting a great attention because of possible applications to topological quantum computations. Braiding of non-Abelian anyons corresponds to quantum computations. The simplest non-Abelian anyons are Ising anyons which can be realized by Majorana fermions hosted by vortices or edges of topological superconductors, $\nu =5/2$ quantum Hall states, spin liquids, and dense quark matter. While Ising anyons are insufficient for 
universal quantum computations, Fibonacci anyons present in $\nu =12/5$ quantum Hall states can be used for universal quantum computations. Yang-Lee anyons are non-unitary counterparts of Fibonacci anyons. Another possibility of non-Abelian anyons (of bosonic origin) is given by vortex anyons, which are constructed from non-Abelian vortices supported by a non-Abelian first homotopy group, relevant for certain nematic liquid crystals, superfluid $^3$He, spinor Bose-Einstein condensates, 
and high density quark matter. Finally, there is a unique system admitting two types of non-Abelian anyons, Majorana fermions (Ising anyons) and non-Abelian vortex anyons. That is $^3P_2$ superfluids (spin-triplet, $p$-wave paring of neutrons), expected to exist in neutron star interiors as the largest topological quantum matter in our universe.  
\end{abstract}


\maketitle

\section*{Key points/objectives}

\begin{itemize}
\item{Unlike fermions and bosons, the exchange of non-Abelian anyons is described by unitary transformations that operate on topologically protected degenerate ground states. The adiabatic exchanges of two or more non-Abelian anyons manipulate quantum information stored in a set of degenerate ground states, leading to topological quantum computations.}
\item{The simplest non-Abelian anyons are Ising anyons which are predicted to exist as Majorana zero modes in topological superconductors, fractional quantum Hall states, Kitaev materials, and dense quark matter.}
\item{Another non-Abelian anyons are vortex anyons, which are constructed from non-Abelian vortices supported by a non-Abelian first homotopy group. Such vortex anyons exist in liquid crystals, superfluid $^3$He, spinor Bose-Einstein condensates, and high density quark matter.}
\item{$^3P_2$ superfluids, which are expected to be realized in neutron stars, provide unique systems simultaneously admitting two kinds of non-Abelian anyons, i.e., Ising anyons and vortex anyons.}
\end{itemize}

\section{Introduction}
In three spatial dimensions, 
all particles are either bosons or fermions 
in quantum physics, that is, 
a wave function of multi-particle states is  
symmetric (antisymmetric) under 
the exchanges of two bosons (fermions). 
On contrary, in two spatial dimensions, 
there exist exotic particles classified to
neither bosons nor fermions,  
{\it anyons}. 
A wave function of two anyons receives 
a nontrivial phase factor under their exchanges~\cite{Leinaas:1977fm,Wilczek:1982wy}.
Such exotic particles play essential roles in 
fractional quantum Hall states~\cite{Halperin:1984fn,Arovas:1984qr}, 
and have been experimentally observed for 
$\nu=1/3$ fractional quantum Hall states~\blue{\cite{bar20,Nakamura:2020}}.

Recently, yet exotic particles attracted great attention, that is, {\it non-Abelian anyons}. 
Non-Abelian anyons are described by 
a set of multiple wave functions, 
and 
the exchanges of two non-Abelian anyons  
lead to unitary matrix operations 
on a set of wave functions. 
They have been theoretically predicted to exist in 
 $\nu=5/2$ fractional quantum Hall states~\cite{Moore:1991ks,nay96}, 
 topological superconductors (SCs) and superfluids (SFs)~\cite{Read:1999fn,Ivanov:2000mjr,Kitaev:2001kla},
 and spin liquids~\cite{Kitaev2006,Motome2020}, 
and experimental observation is pursued.  
Non-Abelian anyons are attracting significant interests owing to 
the possibility to offer a platform of 
topologically protected quantum computations  realized by
braiding of non-Abelian anyons~\blue{\cite{Kitaev:1997wr,Kitaev:2008,
Nayak:2008zza,
pachos_2012,sarma_majorana_2015,Field:2018}}. 
Since the Hilbert space and braiding operations are topologically protected, 
they are robust against noises  
in contrast to the conventional quantum computation methods. 
Recently, it has been reported that 
non-Abelian braiding and fusions has been experimentally realised in
a superconducting quantum processor, where the fusion
and braiding protocols are implemented using a quantum
circuit on a superconducting quantum processor
\cite{Andersen:2022xmz}, thereby opening 
a significant step to realize topological quantum computations.

\blue{One of the main routes to realize non-Abelian anyons 
is based on Majorana fermions in topological SCs~\cite{Ivanov:2000mjr,Kitaev:2001kla,alicea2012,Leijnse2012,Beenakker2013,Silaev2014,Elliott2015,Sato2016,mizushimaJPSJ16,sato_topological_2017-1,Beenakker2020,mar22}.}
Majorana fermions were originally proposed 
in high energy physics to explain neutrinos; 
they are particles that coincide with 
their own anti-particles~\cite{majorana_teoria_1937}.
In condensed matter physics, 
Majorana fermions are localized at vortices 
or edge of materials for which several protocols of non-Abelian braiding were proposed. Non-Abelian anyons constructed from Majorana fermions are 
so-called Ising anyons. 
They are not enough for 
universal quantum computations, 
and thus some non-topological process should be included 
\cite{Nayak:2008zza}. 
In contrast, another type of anyons called Fibonacci anyons \cite{tre08} 
can offer a promising platform for universal topological quantum computation that all quantum gates are implemented by braiding manipulation in a topologically protected way~\cite{preskill,bon05,hor07}. 
Such Fibonacci anyons 
are proposed to exist in $\nu=12/5$ quantum Hall states, 
 a junction made of a conventional SC and 
 a $\nu=2/3$ fractional quantum Hall state ~\cite{mon14}, 
 interacting Majorana fermions realized in a septuple-layer structure of topological SCs~\cite{hu18}, 
 and Rydberg atoms in a lattice~\cite{les12}. 
Yang-Lee anyons are also proposed 
as non-unitary counterparts of 
Fibonacci anyons, obeying nonunitary non-Abelian statistics~\cite{ard11,fre12,san22}.

The aforementioned anyons are all quasiparticle excitations composed of fermions.
On the other hand, 
a different type of non-Abelian anyons 
can be composed of bosons 
in certain ordered states accompanied 
with symmetry breakings $G \to H$.
They are called 
non-Abelian vortex anyons, 
whose exchange statistics are non-Abelian due to 
non-Abelian vortices, that is quantum vortices supported by 
a non-Abelian first homotopy (fundamental) group 
of order parameter manifolds, $\pi_1(G/H)$~\cite{Bais:1980vd,
Wilczek:1989kn,Bucher:1990gs,Brekke:1992ft,Lo:1993hp,
Lee:1993gk,Brekke:1997jj}.\footnote{
The term ``non-Abelian'' \red{on vortices} depends on the context. 
In the other contexts (in particular in high energy physics), 
vortices in a symmetry breaking 
$G \to H$ 
with 
non-Abelian magnetic fluxes are often called non-Abelian even 
though $\pi_1(G/H)$ 
the first homotopy group 
is Abelian 
\cite{Hanany:2003hp,Auzzi:2003fs,Eto:2006pg,Shifman:2007ce,Eto:2013hoa}. 
\red{In condensed matter physics, vortices with Majorana 
fermions in their cores are also sometimes 
called non-Abelian vortices 
(because they are non-Abelian anyons). 
In this article, the term ``non-Abelian'' on 
vortices is used only for vortices with 
non-Abelian first homotopy group $\pi_1$.
}
}
Such non-Abelian vortices exist in  
 liquid crystals~\blue{\cite{Poenaru1977,vol77,Mermin:1979zz,Lavrentovich2001}},
\blue{$^3$He SFs \cite{Balachandran:1983pf,salomaaRMP,Volovik:2003fe}},
spinor Bose-Einstein condensates 
(BECs)~\cite{Semenoff:2006vv,Kobayashi:2008pk,Kobayashi:2011xb,Borgh:2016cco}, 
and high density quark (QCD) matter \cite{Fujimoto:2020dsa,Fujimoto:2021wsr,Fujimoto:2021bes,
Eto:2021nle}.
Non-Abelian braiding of vortex anyons in spinor BECs 
and its application to quantum computations were proposed \cite{PhysRevLett.123.140404}. 

In addition to these systems admitting one type of 
non-Abelian anyons, 
there is the unique system simultaneously 
admitting two kinds of non-Abelian anyons, 
Ising anyons based on Majorana fermions 
and non-Abelian vortex anyons.
It is a $^3P_2$ SF, 
spin-triplet and $p$-wave paring with 
the total angular momentum two 
\cite{Hoffberg:1970vqj,Tamagaki1970,Takatsuka1971,Takatsuka1972,Richardson:1972xn}. 
Such $^3P_2$ SFs are expected 
to be realized by neutrons, 
relevant for neutron star interiors
\cite{Sedrakian:2018ydt}. 
 $^3P_2$ SFs are the largest topological SFs 
 in our universe 
 \cite{Mizushima:2016fbn} 
 and admit non-Abelian vortices \cite{Masuda:2016vak}. 
 Non-Abelian vortices host Majorana fermions in their cores 
 \cite{Masaki2022}, thus behaving as non-Abelian anyons. 

The purpose of this article is to summarise 
these non-Abelian anyons of various types.
After introducing basics of non-Abelian anyons 
in Sec.~\ref{sec:NA}, 
we describe non-Abelian anyons in fermionic and bosonic systems, 
based on Majorana fermions 
and non-Abelian first homotopy group  
in Secs.\ref{sec:NA-topological-SCs} and \ref{sec:vortex},
respectively.
In Sec.~\ref{sec:3P2}, we introduce $^3P_2$ SFs as the unique system simultaneously 
admitting two kinds of non-Abelian anyons. 
We summarize this article in Sec.~\ref{sec:summary}



\section{Non-Abelian anyons}
\label{sec:NA}

\subsection{Braid group and quantum statistics}

\begin{figure}
\includegraphics[width=85mm]{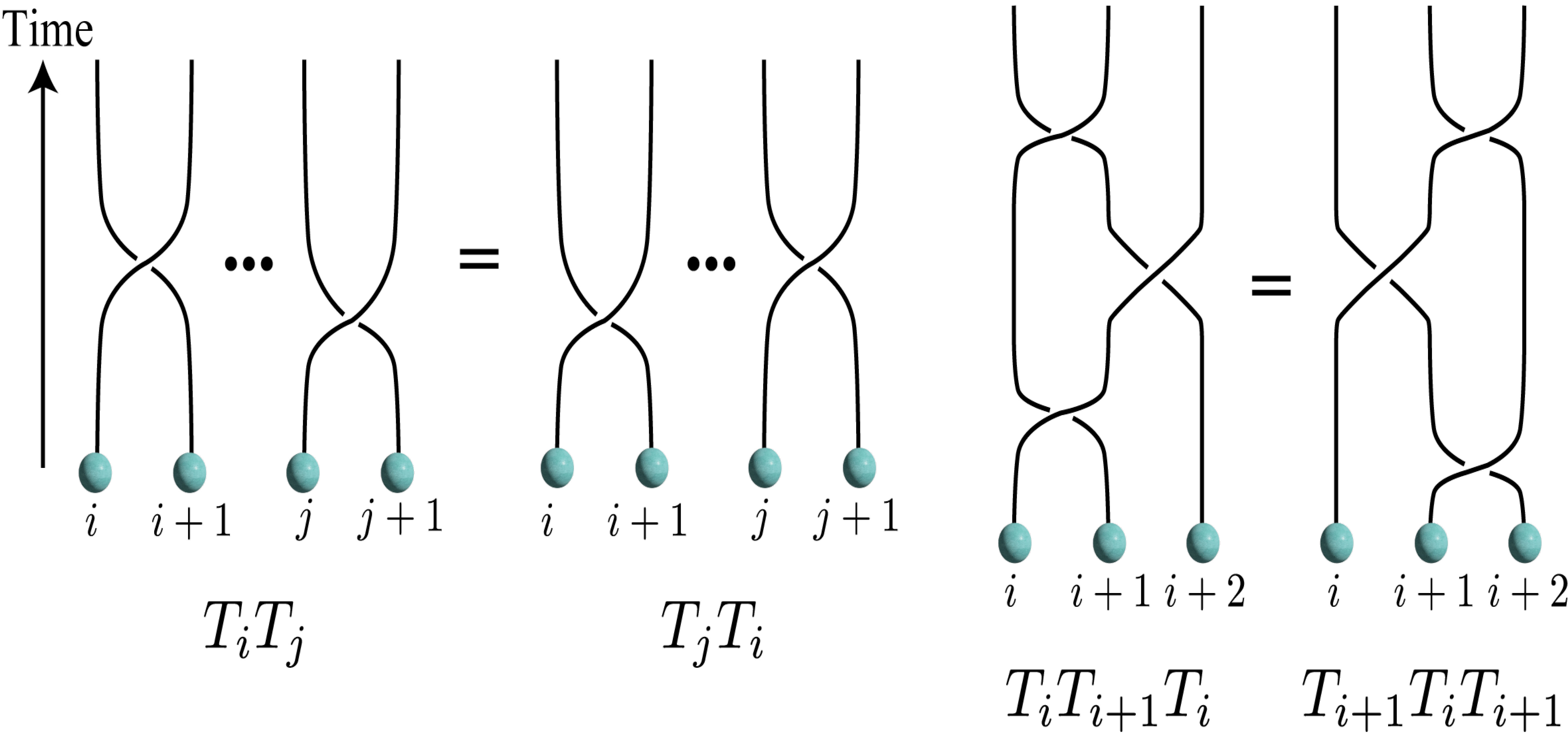}
\caption{Schematic of the braid relations in Eqs.~\eqref{eq:YB} and \eqref{eq:YB2}: $T_iT_j = T_j T_i$ for $|i-j|\ge 2$ (left) and $T_i T_{j} T_i = T_{j}T_iT_{j}$ for $|i-j|=1$ (right).} 
\label{fig:braid}
\end{figure}

Here we consider pointlike topological defects (e.g., vortices in two-dimensional {\it spinless} SCs) which behave as identical particles in a two-dimensional plane. The exchange of $n$ particles in three or higher dimension is
described by the symmetric group $S_n$~\cite{Leinaas:1977fm}. There are two one-dimensional representations of $S_n$, $\pm 1$, due to even/odd permutation and $+1$ ($-1$) corresponds the Bose (Fermi) statistics. Two dimension is special and the exchange of particles is given by the braid group $B_n$~\cite{wu84}. The braid of particles is expressed as a set of operators $T_k$ $(1\le k \le n-1)$ that exchange the neighboring $k$th and $(k + 1)$th particles in an anticlockwise direction. The operators obey the relations (see Fig.~\ref{fig:braid})
\begin{gather}
T_i T_{j} T_i = T_{j}T_iT_{j},  \quad \mbox{for $|i-j|=1$}
\label{eq:YB}, \\
T_iT_j = T_j T_i, \quad \mbox{for $|i-j|\ge 2$}.
\label{eq:YB2}
\end{gather}

The exotic statistics of particles represented by the braid group stems from the relation $T^{-1}_i\neq T_i$. In the one dimensional representation, the generator of $T_i$ is given by a phase factor that a wave function under the exchange of particles acquires, $\tau_j\equiv \tau(T_j)=e^{i\theta_j}$ ($0\le \theta_j < 2\pi$). The relation in Eq.~\eqref{eq:YB} implies that the exchange operation of any two particles induces the same phase factor $\tau_1=\tau_2=\cdots = \tau_{n-1}=e^{i\theta}$. The phase factor characterizes the quantum statistics of particles~\cite{wu84}, and the absence of the relation $T^2_i=1$ allows for the fractional (anyon) statistics with neither $\theta=0$ (bosons) nor $\theta = \pi$ (fermions). In addition to the one-dimensional representation, the braid group has non-Abelian representations. In Sec.~\ref{sec:Majorana}, we will show that the generators of the braid of Majorana zero modes are noncommutative as $[\tau_i,\tau_j] \neq 0$ for $|i-j|=1$ and the pointlike defects hosting Majorana modes behave as non-Abelian anyons.

Although the braid group is trivial in three dimensions, a three-dimensional model with pointlike topological defects which host Majorana modes and obey the non-Abelian statistics has been proposed~\cite{teo10}. The non-Abelian statistics of the defects, which behave as hedgehogs, can be interpreted as the projective ribbon permutation statistics~\cite{fre11a,fre11b}. Such non-Abelian statistics enables to construct three-dimensional networks of topological superconducting wires supporting Majorana modes.

\subsection{Non-Abelian anyons}

In three spatial dimensions, the statistics of particles is determined by their intrinsic spins. According to the spin statistics theorem, all particles with integer (half-integer) spin are bosons (fermions). 
A pair formed by two fermions with the spin $1/2$ behaves as a boson, and the spin of the composite particle obeys the ``fusion rule'' $\frac{1}{2}\otimes \frac{1}{2} = 0 \oplus 1$, where ${0}$ and $1$ denote the spin singlet and triplet states, respectively. When particles are trapped in a two dimensional plane, however, there is another possibility that is neither fermions nor bosons, i.e., anyons. In general, anyons can be characterized by the topological charge, the fusion rule ($N^{c}_{ab}$), the associative law (the $F$-matrix), and the braiding operation (the $R$-matrix)~\cite{preskill,Nayak:2008zza,pachos_2012}. 

\begin{figure}
\includegraphics[width=80mm]{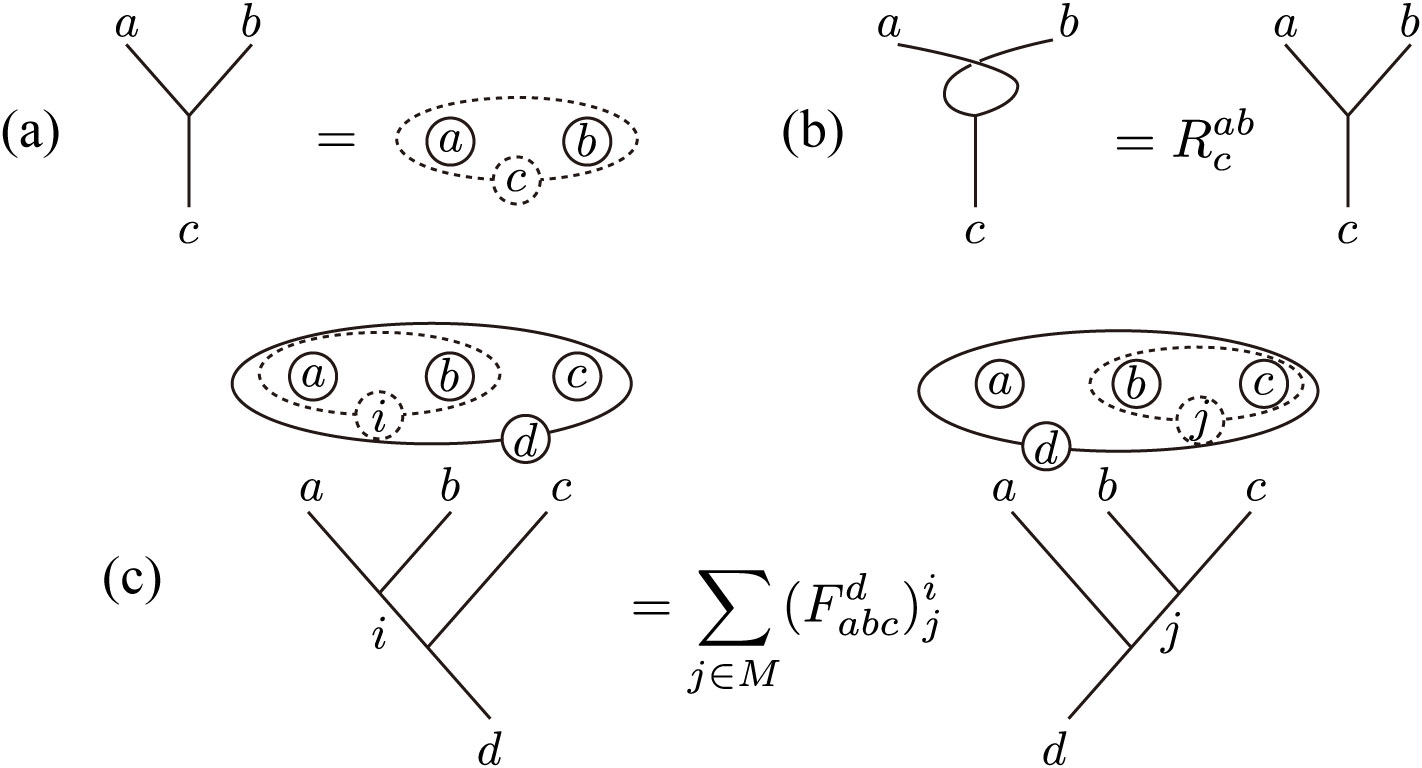}
\caption{(a) Diagram of the fusion of two anyons $a$ and $b$ to an anyon $c$. Diagrammatic expressions of the $R$-matrix (b) and the $F$-matrix (c).} 
\label{fig:anyon}
\end{figure}

Let ${\bf 1}$ and $\{a,b,\cdots \}$ be the vacuum and the different species of particles, respectively. Consider the anyon model spanned by $M=\{{\bm 1},a,b,\cdots \}$ and bring two anyons $a$ and $b$ together. 
The fused particle also belongs to $M$.
The fusion rule is represented by
\beq
a\otimes b = \sum_{c\in M} N_{ab}^c c,\label{eq:fusion-rule}
\eeq
where fusion coefficients $N_{ab}^c$ are non-negative integers. 
Figure~\ref{fig:anyon}(a) shows the diagrammatic expression of the fusion of two anyons with topological charges $a$ and $b$ to an anyon with charge $c$. 
When $N_{ab}^c$ is not zero for only one value of $c$, the fusion of paired $a$ and $b$ anyons is uniquely determined and the anyon is called the Abelian anyon. 
The non-Abelian anyons are characterized by two or more coefficients that satisfy $N_{ab}^c\neq 0$. 
In the context of quantum computation, the fusion rule determines the Hilbert space that encodes quantum information, and quantum computation is implemented by the braiding manipulation of non-Abelian anyons in a topologically protected way. 
Figure~\ref{fig:anyon}(b) shows the diagrammatic expression of the $R$-matrix. 
When one anyon moves around the other, the pairwise anyons acquire a phase. 
As mentioned below, the $R$-matrix describes a phase resulting from the exchange of anyons $a$ and $b$ which fuse to a anyon $c$. 
Let us also consider the fusion of three anyons $a$, $b$, and $c$ into an anyon $d$. 
The outcome of the fusion process is independent of order in which the anyons are to be fused. 
This implies that the fusion process obeys the associative law, $(a\otimes b)\otimes c = a \otimes (b\otimes c)$, which is characterized by the $F$-matrix. 
The $F$ matrix represents the transformation between different fusion bases or the choice of order of fusion, which is expressed as shown in Fig.~\ref{fig:anyon}(c). 
The $R$-matrix and the $F$-matrix are the building-blocks for constructing the braid group in multiple anyon systems~\cite{preskill}.

\sub{Ising anyons}
An example of the non-Abelian anyons is the Ising anyon~\cite{Kitaev2006,Nayak:2008zza,pachos_2012}. 
The Ising anyon model consists of the vacuum ${\bf 1}$, Ising anyons $\sigma$, and  Dirac (complex) fermions $\psi$, which obey the fusion rules
\beq
\sigma \otimes \sigma = {\bf 1} \oplus \psi, \; 
\sigma \otimes \psi = \sigma, \; 
\psi \otimes \psi = {\bf 1}, \;
{\bf 1}\otimes x = x ,\label{eq:Ising}
\eeq
where $x\in\{{\bf 1},\sigma,\psi\}$. 
Consider three Ising anyons, where the two left-most anyons fusing into either ${\bf 1}$ or $\psi$ (see Fig.~\ref{fig:anyon}(c) with $(a,b,c)\rightarrow \sigma$ and $i,j=\{{\bf 1},\psi\}$). 
The $R$-matrix and the $F$-matrix are given, respectively, by 
\begin{gather}
R=\begin{pmatrix}
    R^{\sigma\sigma}_{\bf 1} & 0 \\ 0 & R^{\sigma\sigma}_{\psi}
\end{pmatrix}
= e^{-i\pi/8}\begin{pmatrix}
  1 & 0 \\ 0 & i
\end{pmatrix}
\label{eq:R_ising}, \\
F^{\sigma}_{\sigma\sigma\sigma}
=\frac{1}{\sqrt{2}}
 \begin{pmatrix}
  1 & 1 \\ 1 & -1
\end{pmatrix}.
\end{gather}
The diagonal components of the $R$-matrix are the phases resulting from the counterclockwise exchange of two left-most Ising anyons ($\sigma$) fusing to ${\bf 1}$ or $\psi$, while the twice exchange operation of two right-most Ising anyons is represented by the unitary matrix, $F^{-1}R^2F=e^{-i\pi/4}\sigma_x$. 
Hence, braiding two anyons corresponds to the implementation of the quantum gates acting on the quantum states spanned by ${\bf 1}$ and $\psi$.

\blue{In general, the anyons are described by conformal field theory, corresponding to gapless edge states residing in the boundary of two-dimensional gapped topological phases. For Ising anyons, the theory is the conformal field theory with the central charge $c=1/2$, which describes  critical Ising models such as the two-dimensional Ising model at the point of second-order phase transition~\cite{fra}.}

The Moore-Read state in the fractional quantum Hall state at the filling factor $\nu=5/2$ supports this type of non-Abelian anyons~\cite{Nayak:2008zza}. The Ising anyons can also be realized in the Kitaev's honeycomb model, which is an exactly solvable model of quantum spin liquid states~\cite{Kitaev2006}. In this model, the spins are fractionalized to Majorana fermions coupled to $\mathbb{Z}_2$ gauge fields. The Ising anyons appear as Majorana zero modes bound to the $\mathbb{Z}_2$ flux. Materials including $4d$ or $5d$ atoms with a strong spin-orbit coupling have been proposed as candidates of Kitaev magnets, and the half-integer thermal Hall effect was reported in $\alpha$-RuCl$_3$~\cite{kas18,yam20,yok21,bru22}, which is a signature of Majorana fermions in the chiral quantum spin liquid phase~\cite{ros18,ye18}. 
Apart from materials, Google team reported the demonstration of fusion and braiding rules of non-Abelian Ising anyons on a superconducting quantum processor, where the fusion and braiding protocols are implemented using a quantum circuit on a superconducting quantum processor~\cite{Andersen:2022xmz}. Another platform to realize Ising anyons is a topological SC. In the context of topological SCs, the Ising anyons ($\sigma$) appear as Majorana zero modes bound at their boundaries or topological defects, such as the surface, interface, and vortices~\cite{Read:1999fn,Kitaev:2001kla,Nayak:2008zza,alicea2012,Sato2016,mizushimaJPSJ16}. Pairwise Majorana zero modes form a complex fermion that can define either the unoccupied state (${\bf 1}$) or the occupied state  (${\psi}$) of the zero energy eigenstate, implying $\sigma \otimes \sigma = {\bf 1} \oplus \psi$. The vacuum ${\bf 1}$ corresponds to a condensate of Cooper pairs, while $\psi$ represents a Bogoliubov quasiparticle which can pair into a condensate, i.e., $\psi\otimes \psi = {\bf 1}$. The detailed properties and realization of Majorana zero modes in topological SCs are described in Secs.~\ref{sec:MF} and \ref{sec:plat}.

\sub{Fibonacci anyons}
Another example is the Fibonacci anyons~\cite{tre08}.  The Fibonacci anyon model consists of the vacuum ${\bf 1}$ and the non-trivial anyon $\tau$, which obey the fusion rules
\beq
\tau \otimes \tau = {\bf 1} \oplus \tau, \quad
{\bf 1}\otimes x = x, 
\label{eq:fusion_fibonacci}
\eeq
where $x={\bf 1}, \tau$. The first rule implies that the fusion of two anyons may result in either annihilation or creation of a new anyon, and thus the Fibonacci anyon may be its own anti-particle. Repeated fusions of the $n+1$ $\tau$-anyons  result in either the vacuum or the $\tau$ anyon as $\tau\otimes \tau \otimes \cdots \otimes \tau = a_n\cdot{\bf 1} \oplus b_n \tau$, where $a_n=1$ for $n=2$ and $a_n=n-2$ for $n\ge 3$. The coefficient $b_n$ grows as the Fibonacci series, and the first few values in the sequence are $b_2=1$, $b_3=2$, $b_4=3$, $b_5=5, \cdots$. The $R$-matrix and the $F$-matrix are given, respectively, by 
\begin{gather}
R=\begin{pmatrix}
R^{\tau\tau}_{\bf 1} & 0 \\ 0 & R^{\tau\tau}_{\tau}
\end{pmatrix}
= \begin{pmatrix}
    e^{i4\pi/5} & 0 \\ 0 & -e^{i2\pi/5}
\end{pmatrix}, \\
    F^{\tau}_{\tau\tau\tau} = \begin{pmatrix}
        \phi^{-1} & \phi^{-1/2} \\ \phi^{-1/2} & -\phi^{-1}
    \end{pmatrix},
    \label{eq:FF}
\end{gather}
where $\phi = (1+\sqrt{5})/2$ is the golden ratio. The Fibonacci anyons are described by the level-1 $G_2$ Wess-Zumino-Witten theory with the central charge $c=14/5$~\cite{mon14}, where $G_2$ is the simplest exceptional Lie group. While Ising anyons are not sufficient for universal quantum computation, the Fibonacci anyon systems can offer a promissing platform for universal topological quantum computation that all quantum gates are implemented by braiding manipulation in a topologically protected way~\cite{preskill,bon05,hor07}. 

The existence of the Fibonacci anyons is predicted in the $\nu=12/5$ fractional quantum Hall state that is described by the Read-Rezayi state~\cite{rea99}. It is also proposed that a junction made of a conventional SC and the $\nu=2/3$ fractional quantum Hall state supports the Fibonacci anyons~\cite{mon14}. Fibonacci anyons can also be made from interacting Majorana fermions realized in a septuple-layer structure of topological SCs~\cite{hu18} and from Rydberg atoms in a lattice~\cite{les12}. 

\sub{Yang-Lee anyons}
There are nonunitary counterparts of Fibonacci anyons, which are referred to as Yang-Lee anyons. 
The conformal field theory corresponding to Yang-Lee anyons is nonunitary and Galois conjugate to the Fibonacci conformal field theory~\cite{ard11,fre12}. Because of nonunitarity, the central charge and the scaling dimension for the one nontrivial primary field are negative, $c=-22/5$ and $\Delta=-2/5$, respectively. As shown in Eq.~\eqref{eq:FF}, the $F$-matrix for Fibonacci anyons is given by the golden ratio $\phi$, i.e., one solution of the equation $x^2=1+x$, which is an algebraic analogue of the fusion rule. The $F$-matrix for Yang-Lee anyons is obtained from Eq.~\eqref{eq:FF} by replacing $\phi \rightarrow -1/\phi$ as
\begin{gather}
    F^{\tau}_{\tau\tau\tau}=\begin{pmatrix}
        -\phi & i\sqrt{\phi} \\ i\sqrt{\phi} & \phi
    \end{pmatrix},
    \label{eq:FF2}
\end{gather}
as $-1/\phi$ is the other solution of the equation $x^2=1+x$. In Eq.~\eqref{eq:FF2}, the bases of the $F$-matrix are spanned by the vacuum state ${\bf 1}$ and a Yang-Lee anyon $\tau$.
The $R$-matrix are given by
\begin{gather}
R=\begin{pmatrix}
R^{\tau\tau}_{\bf 1} & 0 \\ 0 & R^{\tau\tau}_{\tau}
\end{pmatrix}
= \begin{pmatrix}
    e^{i2\pi/5} & 0 \\ 0 & e^{i\pi/5}
\end{pmatrix}. 
\end{gather}
Unlike the Fibonicci anyons, braiding two Yang-Lee anyons is represented by a combination of the $R$-matrix and the nonunitary matrix, $F^{-1}RF$. While Yang-Lee anyons obey the same fusion rule as that of Fibonacci anyons given by Eq.~\eqref{eq:fusion_fibonacci}, the $F$-matrix in Eq.~\eqref{eq:FF2} is the nonunitary and the Yang-Lee anyons obey the nonunitary non-Abelian statistics.

The nonunitary conformal field theory with $c=-22/5$ describes the nonunitary critical phenomenon known as the Yang-Lee edge singularity~\cite{cardy}. Let us consider the Ising model with an imaginary magnetic field $ih$ ($h\in \mathbb{R}$). For temperatures above the critical temperature, the zeros of the partition function in the thermodynamic limit, which are referred to as the Lee-Yang zeros, accumulate on the line $h>h_{\rm c}$ and the edge of the Lee-Yang zeros corresponds to the critical point $h_c$~\cite{lee52}. As $h$ ($>h_{\rm c}$) approaches the edge, the density of zeros has a power-law behavior as $|h-h_{\rm c}|^{\sigma}$, which characterizes the critical phenomenon~\cite{kor71,fis78}. For instance, the magnetization exhibits singular behavior with the same critical exponent $\sigma$. The Yang-Lee edge singularity is also realized by the quantum Ising model with a real transverse field and a pure-imaginary longitudinal field~\cite{geh91}.

The quantum Ising model with an imaginary longitudinal field, which supports the Yang-Lee anyons, can be constructed from Majorana zero modes in a network of topological superconducting wires coupled with dissipative electron baths~\cite{san22}. The Majorana modes bound at the end points of one-dimensional topological SCs constitute spin $1/2$ operators. A coupling of Majorana zero modes with electrons in a metallic substrate plays an role of the pure-imaginarly longitudinal field, while the tunneling of Majorana zero modes between neighboring superconducting wires induces a transverse magnetic field. Schemes for the fusion, measurement, and braiding of Yang-Lee anyons are also proposed in Ref.~\cite{san22}. As mentioned above, the Yang-Lee anyons obey the nonunitary non-Abelian statistics. The nonunitary evolution of quantum states has been discussed in connection with measurement-based quantum computation~\cite{ter05,ush17,pir20,zhe21}. 
Although the Yang-Lee anyons with the nonunitary $F$-matrix are not suitable for application to unitary quantum computation, they can be the building-blocks for the construction of measurement-based quantum computation. 
In addition, as the nonunitary quantum gates can be implemented by braiding manipulations, the Yang-Lee anyon systems may offer a quantum simulator for nonunitary time evolution of open quantum systems in a controllable way.

\sub{Vortex anyons}
In this article, we also discuss non-Abelian anyons made of bosonic (topological) excitations in ordered states. The nontrivial structure of the order parameter manifold appears in the liquid crystals, spin-2 BECs, 
the A phase of SF $^3$He, 
dense QCD matter, and $^3P_2$ SFs. The line defects in such ordered systems, such as vortices, are represented by non-Abelian first homotopy group and their topological charges are noncommutative. Such topological defects 
 with noncommutative topological charges behave as non-Abelian anyons, called the non-Abelian vortex anyons. In Sec.~\ref{sec:vortex}, we demonstrate that order parameter manifolds in nematic liquid crystals and spin-2 BECs admit the existence of non-Abelian vortices and show the fusion rules of such non-Abelian vortex anyons.

\section{Non-Abelian anyons in topological SCs}\label{sec:NA-topological-SCs}
\label{sec:Majorana}

\subsection{Majorana zero modes as Ising anyons}
\label{sec:MF}

\sub{Majorana zero modes}
An elementary excitation from superconducting ground states is a Bogoliubov quasiparticle that is a superposition of the electron and hole. The quasiparticle excitations are described by the Bogoliubov-de Gennes (BdG) Hamiltonian 
\begin{gather}
{H} = \sum _{ij} \begin{pmatrix}{\bm \psi}^{\dag}_i,~{\bm \psi}_i\end{pmatrix}
\mathcal{H}_{ij}
\begin{pmatrix}{\bm \psi}_j\\
{\bm \psi}^{\dag}_j\end{pmatrix}, \\
\mathcal{H}_{ij} = \begin{pmatrix}
h _{ij} & \Delta _{ij} \\
\Delta^{\dag}_{ij} & - h^{\ast}_{ij}
\end{pmatrix}.
\label{eq:Hbdg}
\end{gather}
Here, ${\bm \psi}_j$ is the $N$-component vector of the electron field operator and $\mathcal{H}$ is the $2N\times 2N$ hermitian matrix, where $N$ is the sum of the spin degrees of freedom and the number of the lattice sites and so on. 
The $N\times N$ hermitian matrix $h$ describes the normal state Hamiltonian and the superconducting pair potential $\Delta$ obeys $\Delta^{\rm t}=-\Delta$ because of the Fermi statistics, where $a^{\rm t}$ denotes the transpose of a matrix $a$. The BdG Hamiltonian naturally holds the particle-hole symmetry 
\beq
\mathcal{C}\mathcal{H}\mathcal{C}^{-1}=-\mathcal{H},
\label{eq:PHS}
\eeq
where the particle-hole operator $\mathcal{C}=\Theta K $ is an antiunitary operator composed of the unitary operator $\Theta$ and the complex conjugation operator $K$.
The self-conjugate Dirac fermions are called Majorana fermions, where the quantized field ${\bm \Psi}\equiv ({\bm \psi},{\bm \psi}^{\dag})^{\rm t}$ obeys 
\beq
{\bm \Psi} = \mathcal{C}{\bm \Psi}, \quad \mathcal{C}^2 = +1.
\label{eq:selfcharge}
\eeq
We expand the quantized field ${\bm \Psi}$ in terms of the energy eigenstates. The energy eigenstates are obtained from the BdG equation,
\beq
\sum _j \mathcal{H}_{ij}({\bm \varphi}_{E})_j= E({\bm \varphi}_E)_i,
\label{eq:Hbdgr}
\eeq
which describes the quasiparticle with the energy $E$ and the wave function ${\bm \varphi}_E$.
Equation~(\ref{eq:PHS}) guarantees that the quasiparticle state with $E>0$ and ${\bm \varphi}_E$ is accompanied by the negative energy state with $-E$ and ${\bm \varphi}_{-E}=\mathcal{C}{\bm \varphi}_E$. 
Thus, the negative energy states are redundant as long as the particle-hole symmetry is maintained.
Let $\eta_{E}$ be the quasiparticle operator which satisfies the anticommutation relations, $\{ \eta _E, \eta^{\dag}_{E^{\prime}}\} = \delta _{E,E^{\prime}}$ and $\{\eta _E, \eta _{E^{\prime}} \}= \{\eta^{\dag}_E, \eta ^{\dag}_{E^{\prime}} \} = 0$ ($E,E^{\prime}>0$). The self-charge conjugation relation \eqref{eq:selfcharge} then implies that the quasiparticle annihilation operator with a positive energy is equivalent to the creation with a negative energy as
$\eta _{E} = \eta ^{\dag}_{-E}$, and ${\bm \Psi}$ is expanded only in terms of positive energy states as 
${\bm \Psi}({\bm r}) = \sum_{E>0}[ {\bm \varphi}_E\eta_E + \mathcal{C}{\bm \varphi}_{E}\eta^{\dag}_{E}]$. 
The condition \eqref{eq:selfcharge} can be fulfilled by odd-parity SCs. In the absence of spin-orbit coupling, the spin-singlet pair potential is always invariant under the spin rotation, and the particle-hole exchange operator is given by $\mathcal{C}^2= -1$ in each spin sector. Hence, spin-singlet SCs cannot satisfy Eq.~\eqref{eq:selfcharge}. Spin-orbit coupling, however, enables even spin singlet SCs to host Majorana fermions~\cite{sato_topological_2017-1,alicea2012,Sato2016}.

\begin{figure}
\includegraphics[width=85mm]{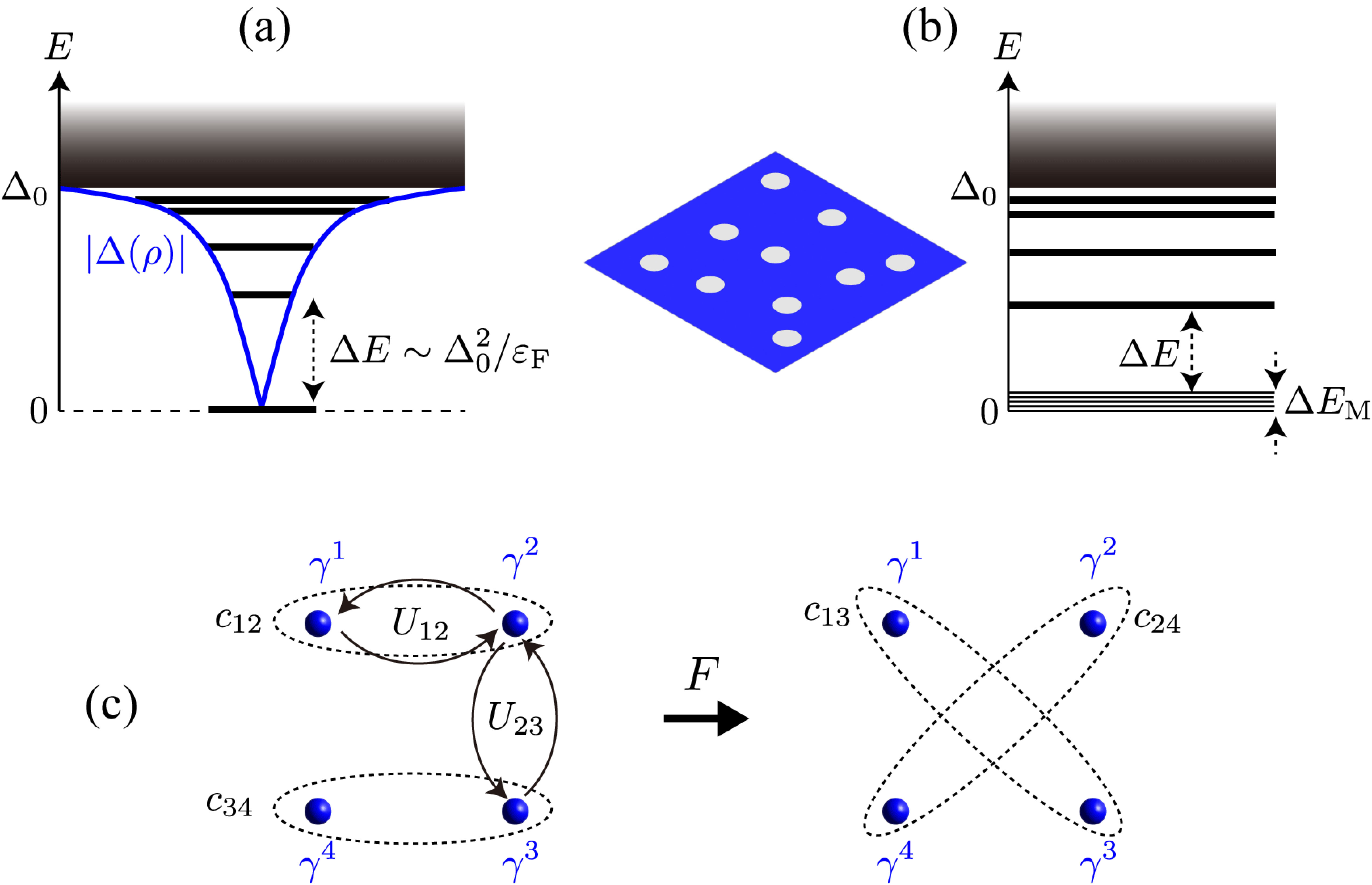}
\caption{Schematic of the quasiparticle states bound at a single vortex (a) and energy spectrum of many vortices (b) in a {\it spinless} SC, where the level spacing of vortex bound states is $\Delta E\sim \Delta^2_0/\varepsilon_{\rm F}$ and $\Delta E_{\rm M}$ denotes the band width of Majorana states bound at each core. (c) Operations of the braiding matrices $U_{12}$ and $U_{23}$ and the $F$-matrix in four Majorana modes.} 
\label{fig:vortex}
\end{figure}

Now, let us suppose that a single zero-energy state exists, and ${\bm \varphi}_0$ is its wave function.
Then, we can rewrite the quantized field to
\beq
{\bm \Psi} = {\bm \varphi}_0\gamma
+ \sum_{E>0}\left[ {\bm \varphi}_E\eta_E + \mathcal{C}{\bm \varphi}_{E}\eta^{\dag}_{E}\right].
\eeq
We have introduced $\gamma$, instead of $\eta_{E=0}$, to distinguish the zero mode from other energy eigenstates. Owing to Eq.~\eqref{eq:PHS}, the zero-energy quasiparticle is composed of equal contributions from the particle-like and hole-like components of quasiparticles, i.e., $\mathcal{C}{\bm \varphi}_0 = {\bm \varphi}_0$. The self-conjugate constraint in Eq.~\eqref{eq:selfcharge} imposes the following relations: 
\beq
\gamma^{\dag} = \gamma,
\eeq
and $(\blue{\gamma})^2=1$ and $\{\blue{\gamma},\eta_{E>0}\}=\{\blue{\gamma},\eta^{\dag}_{E>0} \}=0$.
The quasiparticle obeying this relation is called the Majorana zero mode. The zero energy states appear in a topological defect of topological SCs, such as chiral SCs. In Fig.~\ref{fig:vortex}(a), we show the spectrum of Andreev bound states bound at a vortex in a chiral SC, where the level spacing between the zero mode and the lowest excitation state is $\Delta E\sim \Delta^2_0/\varepsilon_{\rm F}$~\cite{kop91,vol99,Read:1999fn,MatsumotoHeeb2001}. In many vortices, the hybridization between neighboring Majorana modes gives rise to the formation of the band structure with the width $\Delta E_{\rm M}\sim e^{-D/\xi}$~\cite{che09,miz10}, where $D$ and $\xi$ is the mean distance of neighboring vortices and superconducting coherence length, respectively (see Fig.~\ref{fig:vortex}(b)).

The Majorana zero modes exhibit the non-Abelian anyonic behaviors~\cite{Ivanov:2000mjr}. To clarify this, we start with two Majorana zero modes residing in a SC. Using two Majorana operators, $\gamma^{1}$ and $\gamma^2$, we define the new fermion operators $c$ and $c^{\dag}$ as
\beq
c = \frac{1}{2} (\gamma^{1}+i\gamma^{2}), \hspace{3mm}
c^{\dag} = \frac{1}{2} (\gamma^{1}-i\gamma^{2}),
\eeq 
which obey the anticommutation relations, $\{c,c^{\dag}\}=1$ and $\{c,c\}=\{c^{\dag},c^{\dag}\}=0$. The two degenerate ground states are defined as the vacuum $\ket{0}$ and the occupied state of the zero energy state $\ket{1}=c^{\dag}\ket{0}$, respectively, where the former (latter) state is the even (odd) fermion parity. We note that as the BdG Hamiltonian for superconducting states is generally commutable with the parity operator, the fermion parity remains as a good quantum number. For the even (odd) parity sector, the Hilbert space is spanned by using $\ket{0}$ ($\ket{1}$) and excited states that are constructed as $\eta^{\dag}_{E}\eta ^{\dag}_{E^{\prime}}\eta ^{\dag}_{E^{\prime\prime}} \cdots \ket{0} $ ($\eta^{\dag}_{E}\eta ^{\dag}_{E^{\prime}}\eta ^{\dag}_{E^{\prime\prime}} \cdots \ket{1} $). 
The Majorana operators, $\gamma^{1}$, $\gamma^2$, and $i\gamma^1\gamma^2$, act on the Hilbert space as the Pauli matrices $\sigma_x$, $\sigma_y$, and $\sigma_z$, respectively. The eigenstates of the Majorana operators $\gamma^{1}$ and $\gamma^2$ are
given by the superposition \blue{of the degenerate states with different fermion parity, $\ket{0}$ and $\ket{1}$. Hence, the eigenstate of a single Majorana zero mode cannot be a physical state.}

Consider $2N$ Majorana zero modes denoted by $\gamma^j$ ($j=1,\cdots,2N$), where $N$ complex fermions are constructed by the fusion of $i$th and $j$th Majorana zero modes as $c_{ij} = (\gamma^i+i\gamma^{j})/2$. We define the occupation number operator of the complex fermion, 
\beq
n_{ij} \equiv c^{\dag}_{ij}c_{ij} = \frac{1}{2}(1+i\gamma^i\gamma^j).
\label{eq:nij}
\eeq
In a basis that diagonalizes paired Majorana modes $i\gamma^i\gamma^j$, two eigenvalues of the complex fermion, $n_{ij}=0$ and $1$, correspond to the fusion channels ${\bf 1}$ and $\psi$, respectively. 
Hence, the Majorana zero mode is referred to as the Ising anyon. $2^{N}$ degenerate ground states are expressed in terms of the occupation numbers as $\ket{n_{12},n_{34},\cdots}$, which are separated to the sectors of the even/odd fermion parity. 
Here we assume that the temperature of the system is much lower than the level spacing ($\Delta E$) between the Majorana zero mode and the lowest excitation (non-Majorana) state. 
Then, the $2^{N-1}$ degenerate ground states in each fermion parity sector can be utilized as topological qubits, where quantum information is stored in a topologically protected way. 
 
\sub{Braiding Majorana zero modes}
Here we discuss the braiding statistics of Majorana zero modes $\gamma^i$ and $\gamma^j$ and show the non-Abelian statistics of Majorana zero modes~\cite{Ivanov:2000mjr,ali11,cla11}. \blue{While we consider the exchange of Majorana zero modes residing in vortices, the theory is also applicable to Majorana zero modes bound at the end points of one-dimensional topological SCs.}

Let $T_{ij}$ be the braid operators that satisfy Eqs.~\eqref{eq:YB} and \eqref{eq:YB2} and transform $\gamma^i$ and $\gamma^j$ to $e^{\theta_i}\gamma^j$ and $e^{i\theta_j}\gamma^i$, respectively. 
The unitary time-evolution of Majorana zero modes is governed by the Heisenberg equation, $i\frac{d}{dt}\gamma^j(t)=[\gamma^j(t),\mathcal{H}(t)]$. 
The positions of two Majorana modes are adiabatically exchanged in the time interval $[0,T]$. 
The adiabatic condition defines the lower bound for the time scale of the braiding operation, $T$, so that $T$ is much longer than the inverse of the level spacing between the Majorana zero mode and the lowest excitation (non-Majorana) state, $\Delta E$. 
In addition, the upper bound is associated with the band width of Majorana modes $\Delta E_{\rm M}\sim e^{-D/\xi}$ (see Fig.~\ref{fig:vortex}(b)),  i.e., $\Delta E^{-1}\ll T \ll \Delta E^{-1}_{\rm M}$. Within the adiabatic condition, the braiding dynamics of Majorana zero modes can be regarded as the unitary time evolution, $\gamma^j(t) = U^{\dag}_{ij}(t)\gamma^j(0)U_{ij}(t)$. After the braiding operation, $\gamma^i$ ($\gamma^j$) changes to $\gamma^j$ ($\gamma^i$) with an additional phase shift. Then, the braiding operation is represented by 
\beq
U^{\dag}_{ij}\gamma^i U_{ij} = e^{i\theta_i}\gamma^j, \quad 
U^{\dag}_{ij}\gamma^j U_{ij} = e^{i\theta_j}\gamma^i,
\eeq
where $U_{ij}\equiv U_{ij}(T)$ is the unitary operator which describes the exchange operation of two Majorana zero modes $\gamma^i$ and $\gamma^j$, i.e., the representation of $T_{ij}$. 
According to the conditions, $(e^{i\theta_i}\gamma^j)^2 = (U^{\dag}_{ij}\gamma^i U_{ij})^2=1$ and $(\gamma^j)^2=1$, the phase shifts obey $\theta_i=n\pi$ and $\theta_j=m\pi$, where $n,m\in\mathbb{Z}$.
The braiding operations must not change the parity of the occupation number defined in Eq.~\eqref{eq:nij} and thus satisfy $U^{\dag}_{ij}n_{ij}U_{ij}=n_{ij}$, which imposes the condition,
$\theta_1+\theta_2=(2n+1)\pi$, on the phase shift.
As a result, the exchange operation of two Majorana zero modes obtains the following braiding rules: 
\beq
U^{\dag}_{ij}\gamma^{i}U_{ij} = \gamma^{j}, \quad U^{\dag}_{ij}\gamma^{j}U_{ij} = -\gamma^{i}.
\label{eq:braiding}
\eeq

Consider four Majorana zero modes denoted by $\gamma^1$, $\gamma^2$, $\gamma^3$, and $\gamma^4$, which form two complex fermions as $c_{12} \equiv (\gamma^{1}+i\gamma^{2})/\sqrt{2}$ and $c_{34} \equiv (\gamma^{3}+i\gamma^{4})/\sqrt{2}$ (Fig.~\ref{fig:vortex}(c)). When the Majorana mode ``3'' adiabatically encircles the Majorana mode ``2'', both Majorana modes operators acquire the $\pi$ phase shift, $\gamma^{2}\mapsto -\gamma^{2}$ and $\gamma^{3}\mapsto -\gamma^{3}$, corresponding to the twice operation of Eq.~\eqref{eq:braiding}. Therefore, the braiding operation changes the occupation numbers of the complex fermion $n_{12}\equiv c^{\dag}_{12 }c_{12}$ and $n_{34}\equiv c^{\dag}_{34}c_{34}$. 
\blue{For example, the above braiding generates a pair of the complex fermions $\ket{11}$ from their vacuum $\ket{00}$. Here these two states are orthogonal, $\braket{11| 00} = 0$.} 
The braiding rule can be generalized to $2N$ Majorana modes. The $2N$ Majorana modes are fused to $N$ complex fermions, leading to the $2^{N-1}$-fold degeneracy of ground states while preserving fermion parity. 
\blue{As discussed above, when the $i$th and $j$th Majorana modes are exchanged with each other, their operators} behave as 
$\gamma _i \mapsto \gamma _{j}$ and  $\gamma _{j} \mapsto -\gamma _i$. The representation of the braid operator $T_{ij}$ that satisfies Eq.~\eqref{eq:braiding} is given in terms of the zero mode operators as~\cite{Ivanov:2000mjr}
\beq
U_{ij} = e^{i\theta}\exp\left(\frac{\pi}{4}\gamma _{j}\gamma _i\right) = e^{i\theta}\frac{1}{\sqrt{2}}\left(1+\gamma _{j}\gamma _i\right).
\label{eq:Uij}
\eeq
From now on, we omit the overall Abelian phase factor $e^{i\theta}$ as it is not important for quantum computation. Equation~\eqref{eq:Uij} also holds in the case of the Moore-Read state~\cite{nay96}. For $N=1$, there is only a single ground state in each sector with definite fermion parity, and the exchange of two vortices results in the global phase of the ground state by $e^{i\pi /4}$. One can easily find that for $N\ge 2$ the exchange operators $U_{ij}$ and $U_{jk}$ do not commute to each other, $[U_{ij},U_{jk}]\neq 0$, implying the non-Abelian anyon statistics of the Majorana zero modes.

For four Majorana zero modes ($N=2$), twofold degenerate ground states exist in each ferimon-parity sector: $\ket{00} \equiv \ket{{\rm vac}}$ and $\ket{11}= c^{\dag}_{12}c^{\dag}_{34}\ket{\rm vac}$ in the sector of even fermion parity, and $\ket{10}= c^{\dag}_{12}\ket{\rm vac}$ and $\ket{01}= c^{\dag}_{34}\ket{\rm vac}$ in the sector of odd fermion parity. For the even-parity sector, the representation matrix for the exchange of $1\leftrightarrow 2$ and $3\leftrightarrow 4$ \blue{[Fig.~\ref{fig:vortex}(c)]} is given by 
\beq
U _{12} = U _{34} = e^{-i\frac{\pi}{4}} \ket{00}\bra{ 00}
+ e^{i\frac{\pi}{4}} \ket{11}\bra{11}.
\label{eq:U12}
\eeq
This merely rotates the phase of the ground state as in the $N=1$ case. In contrast, the representation matrix for the intervortex exchange [$2\leftrightarrow 3$ in \blue{Fig.~\ref{fig:vortex}(c)}] has the mixing terms of the two degenerate ground states $\ket{00}$ and $\ket{11}$,
\begin{align}
U _{23} =& \frac{1}{\sqrt{2}} \left[ 
\ket{ 00}\bra{00} - i \ket{00}\bra{11} \right. \nn \\
& \hspace{8mm} + \left. \ket{11}\bra{11} - i \ket{11}\bra{00} 
\right].
\end{align}
We note that the choice of the pairing to form the complex fermion is arbitrary. The change of the fused Majorana modes corresponds to the change of the basis from one which diagonalizes $i\gamma^1\gamma^2$ and $i\gamma^3\gamma^4$ to another which diagonalizes $i\gamma^1\gamma^3$ and $i\gamma^2\gamma^4$. The basis 
transformation is represented by the $F$-matrix as 
\begin{align}
    F=\frac{1}{\sqrt{2}}\begin{pmatrix}
    1 & 1 \\ 1 & -1
    \end{pmatrix}.
\end{align}
The braiding matrix in Eq.~\eqref{eq:U12} implies that the exchange of two Majorana zero modes fusing to $\psi$ ($\ket{11}$) acquires an additional $\pi/2$ phase compared to the fusion channel to ${\bf 1}$ ($\ket{00}$). Hence, Eq.~\eqref{eq:U12} satisfies the property of the $R$-matrix in Eq.~\eqref{eq:R_ising}, i.e., $R^{\sigma\sigma}_{\bf 1}=-iR^{\sigma\sigma}_{\psi}$.

\subsection{Platforms for Majorana zero modes}
\label{sec:plat}

The realization of non-Abelian anyons requires to freeze out the internal degrees of freedom of Majorana modes. The simplest example is {\it spinless} $p$-wave SCs/SFs, which emerge from the low-energy part of spinful chiral $p$-wave SCs. To clarify this, we start with spin-triplet SCs, whose pair potential is given by a $2\times 2$ spin matrix~\cite{leg75} 
\begin{align}
\hat{\Delta}({\bm k}) =& \begin{pmatrix}
\Delta_{\uparrow\uparrow}({\bm k}) & \Delta_{\uparrow\downarrow}({\bm k}) \\
\Delta_{\downarrow\uparrow}({\bm k}) & \Delta_{\downarrow\downarrow}({\bm k})
\end{pmatrix}\nn \\
=& i\sigma_{\mu}\sigma_yd_{\mu}({\bm k})
= i\sigma_{\mu}\sigma_yA_{\mu i}\hat{k}_i,
\end{align}
where $\hat{k}_i\equiv k_i/k_{\rm F}$ is scaled with the Fermi momentum $k_{\rm F}$, and the repeated Greek/Roman indices imply the sum over $x,y,z$. Here we omit the spin-singlet component. Owing to the Fermi statistics, the spin-triplet order parameter, ${\bm d}({\bm k})$, obeys ${\bm d}({\bm k}) = - {\bm d}(-{\bm k})$. For  spin-triplet $p$-wave pairing, the most general form of the order parameter is given by a $3\times 3$ complex matrix, $A_{\mu i}\in \mathbb{C}$, where the components are labelled by $\mu, i\in \{x,y,z\}$.

\sub{HQVs with Majorana zero modes in SF $^3$He} We consider the Anderson-Brinkman-Morel (ABM) state~\cite{abm1,abm2}, as a prototypical example of chiral $p$-wave states hosting Majorana zero modes. 
The ABM state is realized in the A-phase of the SF $^3$He, which appears in high pressures and \blue{high temperatures~\cite{vollhardt2013superfluid,Volovik:2003fe}.} At temperatures above the SF transition temperature, $T>T_{\rm c}\approx 1$--$2~{\rm mK}$, the normal Fermi liquid $^3$He maintains a high degree of symmetry
\begin{align}
    G={ SO}(3)_{\bm L} \times { SO}(3)_{\bm S}\times { U}(1),
\end{align}
where ${ SO}(3)_{\bm L}$, ${ SO}(3)_{\bm S}$, and ${ U}(1)$ are the rotation symmetry in space, the rotational symmetry of the nuclear spin degrees of freedom, and the global gauge symmetry, respectively. The tensor $A_{\mu i}$ transforms as a vector with respect to index $\mu$ under spin rotations, and, separately, as a vector with respect to index $i$ under orbital rotations. The order parameter of the ABM state is then given by the complex form
\beq
A_{\mu j} = \Delta e^{i\varphi}\hat{d}_{\mu}(\hat{m}_j+i\hat{n}_j).
\label{eq:ABM}
\eeq
The ABM state is the condensation of Cooper pairs with the ``ferromagnetic'' orbital, and spontaneously breaks the time-reversal symmetry. The orbital part of the order parameter is characterized by a set of three unit vectors forming the triad $(\hat{\bm m},\hat{\bm n},\hat{\bm l})$, where $\hat{\bm l}=\hat{\bm m}\times \hat{\bm n}$ denotes the orientation of the orbital angular momentum of Cooper pairs. The remaining symmetry in the ABM state is 
$H_{\rm A} = { SO}(2)_{S_z}\times { SO}(2)_{L_z-\varphi}\times \mathbb{Z}_2$, where ${ SO}(2)_{S_z}$ is the two-dimensional rotation symmetry in the spin space. The ABM state is also invariant under ${ SO}(2)_{L_z-\varphi}$ which is the combined gauge-orbital symmetry, where the ${ U}(1)$ phase rotation, $\varphi \rightarrow \varphi + \delta\varphi$, is compensated by the continuous rotation of the orbital part about $\hat{\bm l}$, $\hat{m}_j+i\hat{n}_j \rightarrow e^{-i\delta\varphi}(\hat{m}^{\prime}_j+i\hat{n}^{\prime}_j)$. In addition, $\mathbb{Z}_2$ is the mod-2 discrete symmetry $(\hat{\bm d},\hat{\bm m},\hat{\bm n})\rightarrow (-\hat{\bm d},-\hat{\bm m},-\hat{\bm n})$. The manifold of the order parameter degeneracy is then given by
\beq
R_{\rm A} \simeq G/H_{\rm A} \simeq S^2_{\bm S} \times {SO}(3)_{L_z,\varphi}/\mathbb{Z}_2.
\eeq
The two-sphere, $S^2_{\bm S}$, is associated with the variation of  $\hat{\bm d}$. The degeneracy space has an extra $\mathbb{Z}_2$ symmetry that the change from $\hat{\bm d}$ to $-\hat{\bm d}$ can be compensated by the phase rotation $\varphi \rightarrow \varphi + \pi$.

\begin{figure}
\includegraphics[width=70mm]{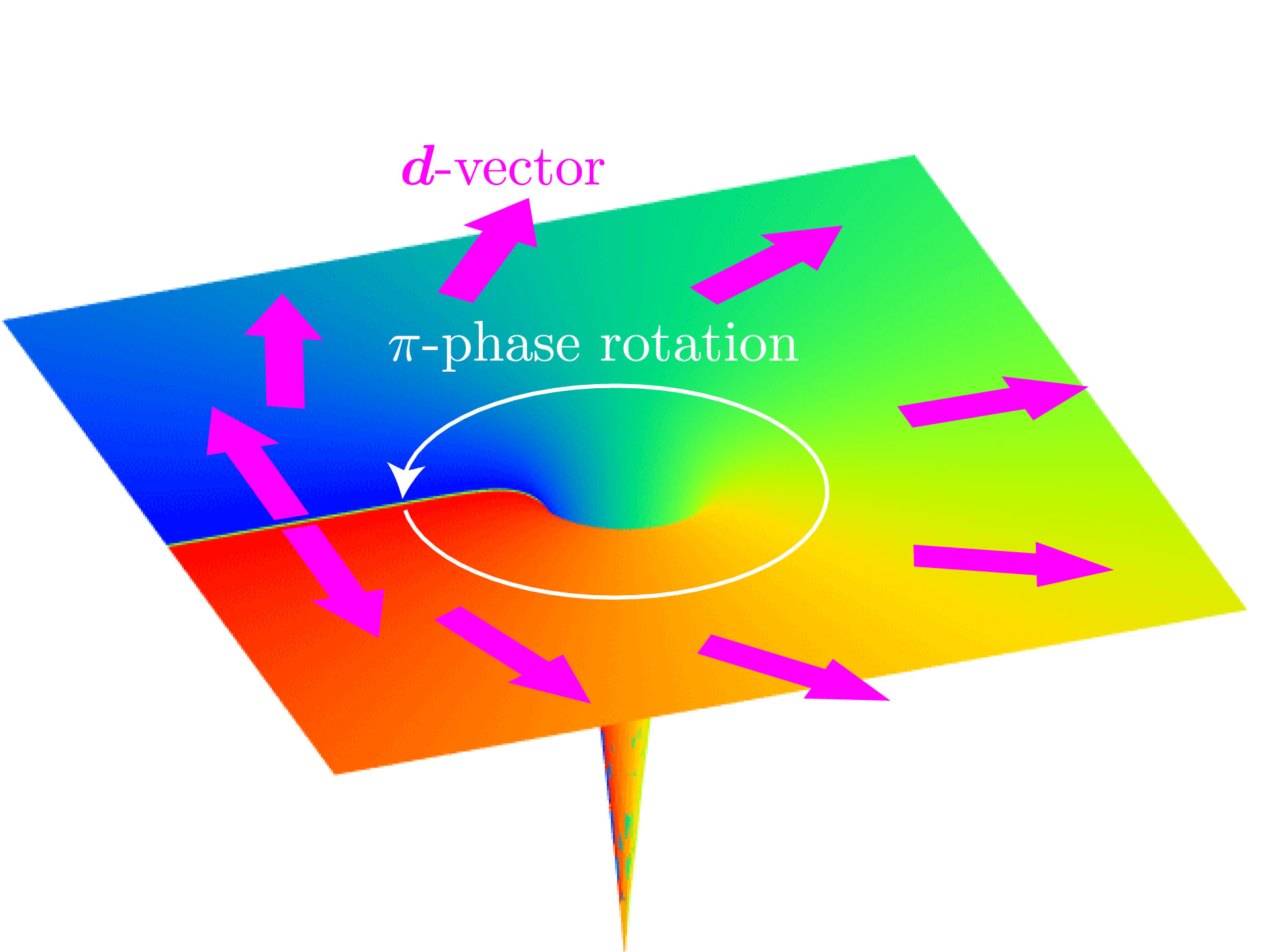}
\caption{Schematic of the HQV realized in the ABM state. The color map shows the ${\rm U}(1)$ phase winding from $\varphi=0$ (red) at $\theta$ to $\pi$ (blue) at $\theta = 2\pi$, and the arrows represent the texture of the $\hat{\bm d}$-vectors shown in Eq.~\eqref{eq:d} with $\kappa^{\rm sp}=1/2$.} 
\label{fig:hqv}
\end{figure}

The topologically stable linear defects in the ABM state are characterized by the group of the \blue{integers modulo 4~\cite{vollhardt2013superfluid,volovik,salomaaRMP,Volovik:2003fe},}
\beq
\pi_1(R_{\rm A}) \simeq \pi_1({SO}(3)/\mathbb{Z}_2) \simeq \mathbb{Z}_4.
\eeq
There exist four different classes of topologically protected linear defects in the dipole-free case.
The four linear defects can be categorized by the fractional topological charge, 
\begin{align}
N_{\rm A} = 0,\; \frac{1}{2}, \; 1, \; \frac{3}{2},
\label{eq:z4-topo-charges}
\end{align}
where $N_{\rm A}=3/2$ is topologically identical to $N_{\rm A}=-1/2$. 
The representatives of $N_{\rm A}=0$ and $N_{\rm A}=1/2$ classes include continuous vortex such as the Anderson-Toulouse vortex and half quantum vortex (HQV), respectively. 
Owing to $N_{\rm A}=2=0$, a pure phase vortex with winding number $2$ is continuously deformed into a nonsingular vortex without a core, that is, the Anderson-Toulouse vortex~\cite{and77}. The $N_{\rm A}=1/2$ vortex is a combination of the half-wound ${\bm d}$-disgyration with a half-integer value of the $U(1)$ phase winding (Fig.~\ref{fig:hqv}). The extra $\mathbb{Z}_2$ symmetry allows us to take the half-integer value of the topological charge, because the $\pi$-phase jump arising from the half-winding of the $U(1)$ phase ($\varphi=\theta/2$) can be canceled out by the change in the orientation of $\hat{\bm d}$ ($\hat{\bm d}\rightarrow-\hat{\bm d}$). The $N_{\rm A}=1$ class includes a pure phase vortex with odd winding number and the radial/tangential disgyrations without phase winding. The latter was originally introduced by de Gennes as ${\bm l}$-textures with a singularity line~\cite{deg73,PhysRevA.9.2676}. 

The vortex with the fractional charge $N_{\rm A}=1/2$ is a harbor for spinless Majorana zero modes. We introduce the center-of-mass coordinate of Cooper pairs, ${\bm R}=(\rho,\theta,z)$, as $\hat{\Delta}({\bm k})\rightarrow \hat{\Delta}({\bm k}, {\bm R})$, where $\rho =\sqrt{x^2+y^2}$. 
The vortex core is located at $\rho = 0$. 
The vortex state is subject to the boundary condition at $\rho \rightarrow \infty$ where the orbital angular momentum of the Cooper pair is aligned to the $z$-axis ($\hat{\bm l}\parallel\hat{\bm z}$) and the $U(1)$ phase $\varphi$ continuously changes from 0 to $2\pi  \kappa$ along the azimuthal ($\theta$) direction, 
\beq
A_{\mu i}(\rho=\infty,\theta)=\Delta e^{i\kappa\theta}\hat{d}_{\mu}(\theta)(\hat{x}_j+i\hat{y}_j).
\label{eq:vortex_ABM}
\eeq
Owing to the spontaneous breaking of the gauge-orbital symmetry, there are two classes for the vorticity $\kappa$: Integer quantum vortices with $\kappa\in\mathbb{Z}$ and HQVs with $\kappa \in\mathbb{Z}/2$. In the HQVs, both the $U(1)$ phase and $\hat{\bm d}$ rotate by $\pi$ about the vortex center (see Fig.~\ref{fig:hqv}). In genral, the $\hat{\bm d}$-texture for HQVs is obtained as 
\beq
\hat{\bm d}(\theta) = \cos(\kappa^{\rm sp}\theta)\hat{\bm x} + \sin(\kappa^{\rm sp}\theta)\hat{\bm y},
\label{eq:d}
\eeq
where $\kappa^{\rm sp}$ denotes the winding of $\hat{\bm d}$. The HQV is characterized by $(\kappa,\kappa^{\rm sp})=(1/2,\pm 1/2)$, while the integer quantum vortex has $(\kappa,\kappa^{\rm sp})=(1,0)$. It is remarkable to notice that since the ABM state is the equal spin pairing state, the order parameter for the HQV is recast into the representation in the spin basis as
\beq
\hat{\Delta} = \Delta \left[ e^{i(\kappa-\kappa^{\rm sp})\theta}\ket{\uparrow\uparrow} + e^{i(\kappa+\kappa^{\rm sp})\theta}\ket{\downarrow\downarrow}\right].
\eeq
For the HQV with $(\kappa,\kappa^{\rm sp})=(1/2,1/2)$ the $\ket{\uparrow\uparrow}$ Cooper pair possesses the spatially uniform phase, while the $\ket{\downarrow\downarrow}$ pair has the phase winding of $2\pi$ around the vortex as in a conventional singly quantized vortex. Thus, the vortex-free state in the $\uparrow$ spin sector exhibits fully gapped quasiparticle excitations, while the low-lying structures in half quantum vortex are effectively describable with a singly quantized vortex in the $\downarrow$ spin sector, i.e., the spin-polarized chiral $p$-wave SC. An odd-vorticity vortex in the spin-polarized chiral system hosts a single {\it spinless} Majorana zero mode that obeys non-Abelian statistics. The existence of non-Abelian anyonic zero modes in half quantum vortices was first revealed by Ivanov, who developed the non-Abelian braiding statistics of vortices with spinless Majorana zero modes~\cite{Ivanov:2000mjr}.

In the bulk A-phase of the SF $^3$He, an obstacle to realizing the HQVs is the formation of continuous vortices with the $\hat{\bm l}$-texture, which are characterized by the topological charge $N=0$. 
As the orientation of the $\hat{\bm l}$-vector is associated with the orbital motion of the Cooper pair, the $\hat{\bm l}$-texture can be uniformly aligned in a parallel plate geometry with thickness $D$. 
The motion of Cooper pairs is confined in the two-dimensional plane and $\hat{\bm l}$ is locked perpendicular to the plates. 
Applying a magnetic field further restricts the orientation of $\hat{\bm d}$ to the two-dimensional plane perpendicular to the applied field, which is a favorable situation to stabilize the HQVs. 
In the parallel plate geometry with $D=12.5~{\mu}{\rm m}$, the measurements of NMR frequency shift observed that the $\hat{\bm d}$-vectors are confined to the two-dimensional plane perpendicular to $\hat{\bm l}$~\cite{yam08}. 
The experiment was performed in a rotating cryostat at ISSP, University of Tokyo. 
The parallel plates are rotated at the angular velocity $\lesssim 12{\rm rad}/{\rm s}$ but no conclusive evidence of HQVs was observed~\cite{yam08}. 
Although the SF $^3$He-A thin film provides an ideal platform for HQVs with Majorana zero modes, the realization of HQVs remains as a challenging task. \blue{We also note that point-like solitons in a superfluid $^3$He film, such as spin disclinations, were shown to behave as anyons with fractional charges~\cite{vol89}.}

\begin{figure}
\includegraphics[width=80mm]{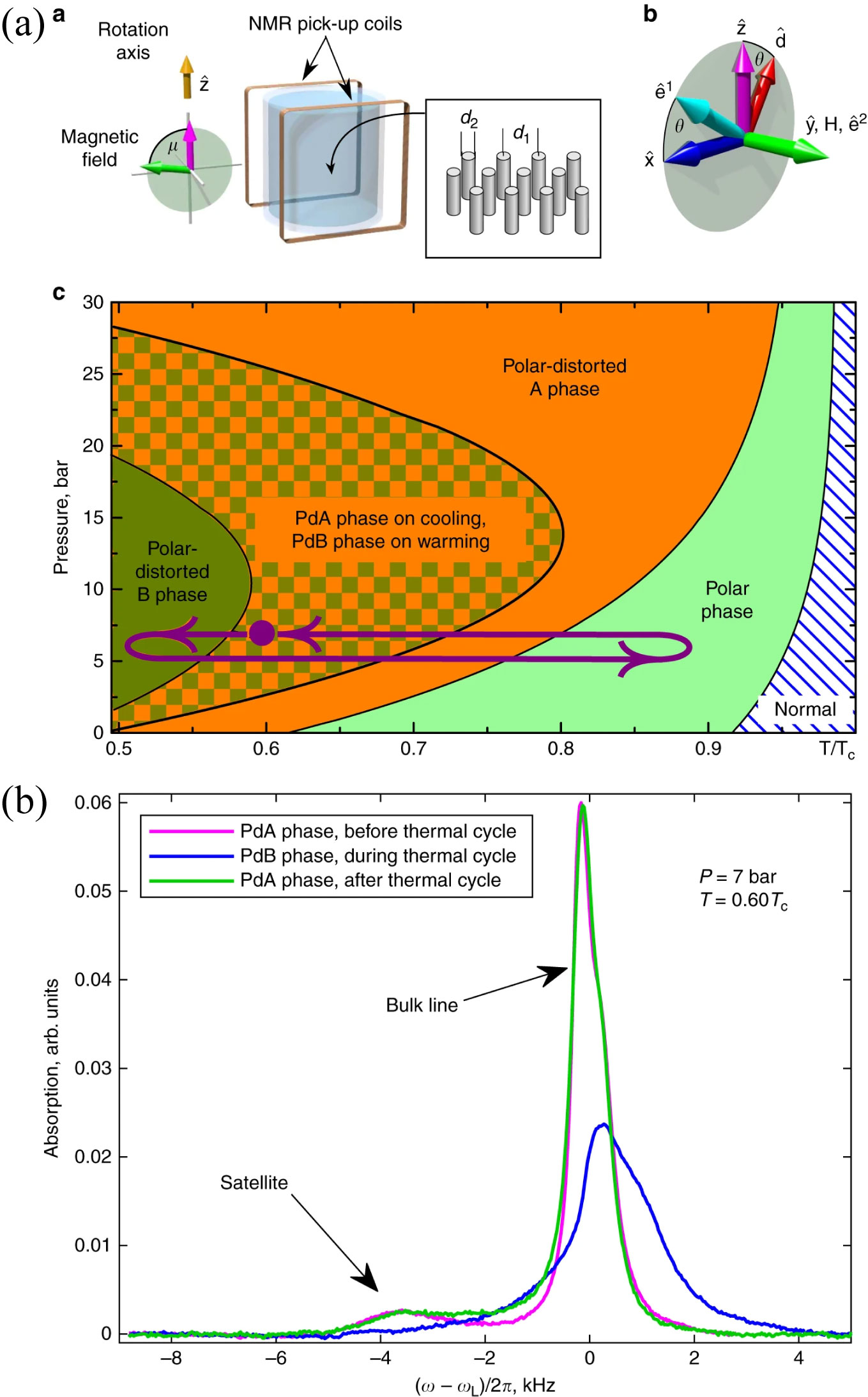}
\caption{(a) The experimental setup and the phase diagram in the liquid $^3$He with nematic disorders and (b) NMR spectra at pressure $P=7~{\rm bar}$ and $T=0.60T_{\rm c}$ in the presence of a magnetic field perpendicular to the anisotropy direction of nematic disorders~\cite{Makinen:2018ltj}, where $T_{\rm c}$ is the critical temperature of the bulk SF $^3$He without disorders. The disorders consist of nearly parallel Al$_2$O$_3$ strands, where the diameter and mean distance are $d_2\approx 8~{\rm nm}$ and $d_1\approx 50~{\rm nm}$, respectively. 
In (b), the main peaks with green and magenta colors correspond to the signal of the bulk PdA phase, while the satellite peaks originate from the spin excitation bound to the $\hat{\bm d}$-solitons connecting pairs of HQVs. The satellite peak remains unchanged after the thermal cycling illustrated by purple arrows in (a). Both figures are taken from Ref.~\cite{Makinen:2018ltj}.} 
\label{fig:hqv2}
\end{figure}

Another promising route to realize HQVs with Majorana zero modes is to artificially introduce well-controlled disorders with high-porosity aerogel. 
In particular, the polar phase was observed in anisotropic aerogels consisting of uniaxially ordered alumina strands, called the nematic aerogels~\cite{dmi15,hal19} [See Fig.~\ref{fig:hqv2}(a)]. 
The order parameter of the polar phase is given by $A_{\mu i}=\Delta e^{i\varphi}\hat{d}_{\mu}\hat{z}_{i}$, where the orbital state of the Cooper pair ($\hat{z}_{i}$) is confined by the uniaxially anisotropic disorders. 
As in the ABM state in the bulk $^3$He, the texture of the $\hat{\bm d}$-vector concomitant with the half-integer vorticity can realize HQVs in the polar phase. 
In the polar phase, HQVs are energetically preferable to integer quantum vortices at zero magnetic fields and magnetic fields applied along the uniaxial anisotropy~\cite{Nagamura2018,min14,Regan2021}. 
Indeed, the HQVs were experimentally observed in nematic aerogels under rotation~\cite{Autti:2015bta}. 
The HQVs were also created by temperature quench via the Kibble-Zurek mechanism~\cite{rys21}. 
As shown in Fig.~\ref{fig:hqv2}(a), the superfluid phase diagram in nematic aerogels is drastically changed from that of the bulk $^3$He without disorders, where the polar-distorted A and B (PdA and PdB) phases are stabilized in addition to the polar phase. The NMR measurements performed in Ref.~\cite{Makinen:2018ltj} observed that the HQVs survive across the phase transition to the PdA phase. As shown in Fig.~\ref{fig:hqv}, the HQV is accompanied by the $\hat{\bm d}$-soliton in which the $\hat{\bm d}$ orientation rotates. Figure~\ref{fig:hqv2}(b) shows the observed NMR spectra which have satellite peaks in addition to the main peak around the Larmor frequency. The main peak is a signal of the bulk PdA phase. The satellite peak originates from the spin excitation localized at the $\hat{\bm d}$-solitons connecting pairs of HQVs. The order parameter of the PdA phase is given by  $A_{\mu i}=\Delta e^{i\varphi}\hat{d}_{\mu}(\hat{\bm m}+i\varepsilon \hat{\bm n})_i$, where $\hat{\bm m}$ is aligned along the axis of nematic aerogels and $\varepsilon \in (0,1)$ is the temperature- and pressure-dependent parameter on the distortion of Eq.~\eqref{eq:ABM} by nematic aerogels. Although the HQVs in the polar phase host no Majorana zero modes, the low energy structure of the HQVs in the PdA phase is essentially same as that of HQVs in the ABM state and each core may support the existence of {\it spinless} Majorana zero modes~\cite{Makinen:2018ltj,mak22}. 
Hence, the survival of the HQVs in the PdA phase is important as a potential platform for topological quantum computation. 
The elucidation of the nature of the fermionic excitations bound to the HQVs remains as a future problem.

\sub{HQVs in SCs} In contrast to the SF $^3$He, it is difficult to stabilize HQVs in SCs. 
The difficulty is attributed to the fact that the mass (charge) current is screened exponentially in the length scale of the penetration depth $\lambda$, while such a screening effect is absent in the spin current~\cite{Chung:2007zzc}. 
The HQV is accompanied by the spin current in addition to the mass current, while in the integer quantum vortex, the integer vorticity ($\kappa\in\mathbb{Z}$) and uniform $\hat{\bm d}$-texture in Eq.~\eqref{eq:vortex_ABM} induce no spin current.
Therefore, the absence of the screening effect on the spin current is unfavorable for the stability of the HQVs relative to the integer quantum vortices, where the later may host spin-degenerate Majorana zero modes. 
It has also been proposed that configurational entropy at finite temperatures drives the fractionalization of the integer quantum vortices into pairwise HQVs~\cite{chu10}. 
However, no firm experimental evidence for HQVs has been reported in SCs.

\sub{Superconducting nanowires} 
Here we briefly mention recent progress on the search of Majorana zero modes bound at the end points of superconducting nanowires, although we mainly focus on non-Abelian anyons emergent in vortices in this article.

The first idea was proposed by Kitaev that a spinless one-dimensional SC is a platform to host a Majorana zero mode~\cite{Kitaev:2001kla}. 
Although it was initially thought to be a toy model, it was recognized that a system with the same topological properties can be realized by a semiconductor/SC heterostructure~\cite{alicea2012,sato_topological_2017-1}. 
The key ingredients are a semiconductor nanowire and a large Zeeman magnetic field in addition to the proximitized superconducting gap $\Delta$~\cite{sat09,sat10}. 
Consider the Hamiltonian of the semiconductor nanowire under a magnetic field $B$, $h(p) = p^2/2m-\mu + \lambda p \sigma_y -B\sigma_z$, which appears in the diagonal part of Eq.~\eqref{eq:Hbdg}, where $m$ is the mass of an electron and $\lambda$ is the strength of the spin-orbit coupling. 
The spin-orbit coupling is not only a source of nontrivial topology, but is also necessary to realize spinless electronic systems together with the large Zeeman field ($B>\mu$). 
In fact, if the proximitized superconducting gap is taken into account, \blue{a combination of spin-orbit coupling with the superconducting gap induces topological odd-parity superconductivity. The topological phase with the spinless Majorana zero mode appears in the higher magnetic field satisfying $B>B_{\rm c}=\sqrt{\Delta^2+\mu^2}$.}

A direct signature of Majorana zero mode is the quantization of a zero-bias peak in differential conductance, where the height of the zero-bias peak stemming from the Majorana zero mode is predicted to be quantized to the universal conductance value of $2e^2/h$ at zero temperature~\cite{akh11,ful11}. Consider a charge-transfer process through a normal metal-SC junction. When an electron with the incident energy $E$ enters from the normal metal to the SC, it forms the Cooper pair with an electron in the Fermi sea at the interface and a hole is created in the normal side as a consequence of momentum conservation. Such a process in which electrons are retroreflected as holes is called Andreev reflection. Let $\phi_E^{\rm in,out}=[u_E,v_E]^{\rm t}$ be the two-component wavefunctions of the incident and reflected particles, respectively, where $u_E$ ($v_E$) accounts for the electron (hole) component.
We now consider the scattering problem ${\phi}^{\rm out}_E({\bm x}) = S(E){\phi}_E^{\rm in}({\bm x}) $. The $S$-matrix is defined in the particle-hole space as
\beq
S(E) = \begin{pmatrix}
r^{\rm ee}(E) & r^{\rm eh}(E) \\ r^{\rm he}(E) & r^{\rm hh}(E)
\end{pmatrix},
\eeq
which is a unitary matrix $S\in {U}(2)$.
The coefficient $r^{\rm ee}$ ($r^{\rm eh}(E)$) is the amplitude of the normal (Andreev) reflection, and the others are the reflection coefficients of incident holes. 

The conductance $G(E)$ at the normal/SC interface is obtained with the conductance quantum $G_0=e^2/h$ per spin as 
$G(E) = 2G_0 |r^{\rm eh}(E)|^2$. 
The particle-hole symmetry imposes on the $S$ matrix the constraint, $\mathcal{C}S(E)\mathcal{C} ^{-1}=S(-E)$, leading to $\det S(0)=|r^{\rm ee}(0)| ^2-|r^{\rm eh}(0)|^2$ at $E=0$. 
Let $V$ be the unitary matrix defined by $V\phi_{0} = ({\rm Re}u,{\rm Im}u)$. 
In this basis, the scattering matrix $S(0)$ is transformed as $S^{\prime}(0)=VS(0)V^{\dag}$, where  $S^{\prime}(0)\in {O}(2)$ is an orthogonal matrix satisfying $S^{\prime\dag}=S^{\prime -1}=S^{\prime{\rm tr}}$ and $\det S(0) = \det S^{\prime}(0)= \pm 1$. 
Hence, there are two different processes, a perfect normal reflection process ($\det S(0)=+1$) and a perfect Andreev reflection process ($\det S(0)=-1$). 
This discrete value, $\det S(0) = \pm 1$, is a topological invariant representing the parity of the number of Majorana zero modes residing at the interface. When $\det S(0)=-1$, there exists at least one Majorana zero mode, which involves the perfect Andreev reflection process and the quantized conductance, 
\beq
G(0)=\frac{2e^2}{h}.
\label{eq:quantizedMajo}
\eeq
\blue{The analytical expression of the conductance is obtained by solving the BdG eqution and quasiclassical Usadel equation at a interface of the normal metal and unconventional SCs (e.g., $p$- and $d$-wave SCs)~\cite{tan95,kashiwaya00,tan04,tan05}, where the the conductance per spin reaches $2e^2/h$. The quantized conductance is also related to the Atiyah-Singer index when the Hamiltonian holds the chiral symmetry~\cite{ike16}.}

The conductance has recently been measured in a junction system of InSb-Al hybrid semiconductor-SC nanowire devices. The differential conductance yields the plateau behavior where the peak height reaches values $2e^3/h$~\cite{zha21}. 
In the device used in the experiment, the tunnel barrier is controlled by applying a gate voltage to the narrow region between the electrode and the region where the  proximity effect occurs in the semiconductor wire. 
In the vicinity of the barrier, the unintentionally formed quantum dot or nonuniform potential formed in the vicinity of the interface gives rise to the formation of nearly zero-energy Andreev bound states. 
Such Andreev bound states mimic the plateau of $G(0)= 2e^2/h$ even in the topologically trivial region of the magnetic field ($B<B_{\rm c}$)~\cite{liu17,pra20,cay21,yu21,val21}. 
From the plateau of $G(0)$, 
it is not possible to identify whether the observed $G(0) \sim 2e^2/h$ is due to Majorana zero mode or the effect of accidentally formed s bound states.
Alternative experimental schemes to distinguish the trivial and topological bound states have been proposed~\cite{liu18,ric19,yav19,awo19,zhang20,sch20,pan21,liu21,ric21,tha21,che22,sug22}.

\sub{Vortices in Fe(Se,Te)} The iron-based SC Fe(Se,Te) is a candidate of  topologically nontrivial superconducting state supporting Majorana zero modes. 
In the parent material FeSe, the $3d$ electrons of Fe near the Fermi level mainly contribute to superconductivity, and the $p_z$ orbitals of Se appear on the higher energy. 
Substitution of Se with Te causes a band inversion between the $p_z$ and $3d$ orbitals. 
As a result of the band inversion, the normal state of FeTe$_{1-x}$Se$_x$ is topologically nontrivial, and accompanied by the surface Dirac fermions~\cite{zha18}. 
Below the superconducting critical temperature, the superconducting gap is proximitized to the surface Dirac states. 
It was theoretically pointed out that the topological phase with a spinless Majorana zero mode can be realized when the superconducting proximity effect occurs in the Dirac fermion system~\cite{sat03,fu08}. Therefore, the surface state of Fe(Se,Te) is a topological superconucting state. When a magnetic field is applied to Fe(Se,Te), the vortex lines penetrate to the surface state and the zero energy state appears. The wave function of the zero mode is tightly bound at the intersection of the vortex line and the surface~\cite{hos11,kaw15}. 

When a magnetic field is applied to a SC, a quantized vortex penetrates. 
The magnetic flux is accompanied by the $2\pi m$ winding of the $U(1)$ phase and the superconducting gap vanishes. In other words, a quantized vortex can be regarded as a quantum well with a radius of $\xi$ and a height of $\Delta$. 
Hence, the Andreev bound states are formed and the level spacing between them is on the order of $\Delta^2/\varepsilon_{\rm F}$, where the level spacing is about $100$--$200~\mu{\rm eV}$ in Fe(Se,Te). In the topological phase, the lowest level 
has the exact zero energy, and the quasiparticle behaves as Majorana fermion. Ultra-low temperature STM/STS with high energy resolution observed a pronounced zero-bias conductance peak~\cite{mac19}. If a Majorana zero mode is bound to the vortex, the conductance should be quantized to the universal value $2e^2/h$ independent of tunnel barriers. In the tunneling spectroscopy performed while changing the distance between the sample surface and the STM tip, the height of the zero-bias conductance peak is approximately $0.6$ times as large as the universal value stemming from the Majorana zero mode~\cite{zhu20}. 
Although $2e^2/h$ has not been reached, this plateau structure strongly suggests the existence of Majorana zero modes because it cannot be explained by non-Majorana vortex bound states.

\subsection{Topological quantum computation}
Quantum computation based on the braiding of Majorana zero modes is basically along the two directions: implementation of quantum gates and measurement-based quantum computation.
The unitary evolution of the qubits is controlled by a set of discrete unitary operations, i.e., quantum gates. 
The simple example of the quantum gates acting on a single qubit is the set of the Pauli gates, which are given by the three Pauli matrices $X\equiv \sigma_x$, $Y\equiv \sigma_y$, and $Z\equiv \sigma_z$. The multi-qubit gate operation can be implemented by a combination of a set of single-qubit gates and the controlled-NOT (CNOT) gate. According to the Solovay-Kitaev theorem~\cite{nielsen}, an arbitrary single-qubit gate can be approximated by a sequence of the discrete gate operations $(H,S,T)$, where $H=(X+Z)/\sqrt{2}$, $S=\sqrt{Z}$, and $T=\sqrt{S}$ are the Hadamard gate, the single-qubit $4\pi$ rotation, and the $T$-gate ($\pi/8$-gate), respectively. Universal quantum computation can be implemented by a sequence of the single-qubit gates $(H,S,T)$ and the CNOT gate.  

In Majorana qubits, the Pauli gates are expressed in terms of the braiding operations as $X=iU^2_{23}$, $Y=iU^2_{31}$, and $Z=iU^2_{12}$, and the Hadamard and $S$ gates are implemented by a combination of such operations as $H=iU_{12}U_{23}U_{12}$ and $S=e^{i\pi/4}U_{12}$. In addition, the CNOT gate can be implemented by a combination of the measurement and braiding operations in two Majorana qubits (i.e., 8 Majorana zero modes) with a single ancilla Majorana qubit~\cite{bra06}. 
All the Clifford gates ($H$, $S$, CNOT) are realized as a composition of braiding operators in a topologically protected way. According to the Gottesman-Knill theorem, however, quantum circuits only using the elements of Clifford group can be efficiently simulated in polynomial time on a classical computer~\cite{nielsen}. Indeed, degenerate quantum states composed of multiple Ising anyons offer tolerant storage of quantum information and their braiding operations provide all necessary quantum gates in a topologically protected way, except for the non-Clifford $T$-gate. The $T$-gate can be implemented only in a topologically unprotected way. For instance, Karzig {\it et al.} proposed a scheme to implement the $T$-gate in a Y-shaped junction accompanied by 4 Majorana zero modes~\cite{kar16}.

As mentioned in Sec.~\ref{sec:NA}, Fibonacci anyon systems can offer a platform for universal topological quantum computation, where all the single-qubit gates $(H,S,T)$ and the CNOT gate are implemented by the braiding manipulations of the anyons in a topologically protected way~\cite{preskill,bon05,hor07}.

Another way to realize universal quantum computation in Majorana qubits is measurement-based quantum computation. It was realized that all braiding manipulations can be implemented in a topologically protected manner by the measurements of topological charges of pairwise anyons~\cite{bon08,bon09}. The process of the quantum information encoded to Majorana qubits can take place by a series of simple projective measurements without braiding Majorana modes.
The building block is a Coulomb blockaded Majorana box which is made from even numbers of Majorana modes in a floating topological SC~\cite{plu17,kar17,ore20}.

\subsection{Symmetry-protected non-Abelian anyons}
\label{sec:SPnAayon}

Symmetry-protected non-Abelian anyons in spinful SCs and SFs were discussed for class D~\cite{ueno,sato14,fan14} and for class DIII~\cite{liu14,gao16}. In the former (latter) case, mirror reflection symmetry (time reversal symmetry) plays an essential role on the topological protection of non-Abelian nature. In addition, unitary symmetry-protected non-Abelian statistics has been theoretically proposed~\cite{hon22}.  The non-Abelian statistics of vortices supporting multiple Majorana zero modes has also been discussed in Refs.~\cite{Yasui:2010yh} and \cite{Hirono:2012ad} in the context of high-energy physics.

\sub{Mirror and unitary symmetries} As the simple example of symmetry-protected non-Abelian anyons, we consider integer quantum vortices in chiral $p$-wave SCs and $^3$He-A~\cite{sato14}, where each vortex core supports spinful Majorana zero modes. Contrary to the standard wisdom, a pair of Majorana zero modes in integer quantum vortices can be protected by the mirror symmetry and obey the non-Abelian statistics~\cite{ueno,sato14}.  

Let us consider two-dimensional superconducting film and assume that the normal state holds the mirror symmetry with respect to the $xy$-plane, $M_{xy} h ({\bm k}) M^{\dag}_{xy} = h ({\bm k})$, where ${\bm k}=(k_x,k_y)$ is the two-dimensional momentum and the mirror operator is defined as $M_{xy}=i\sigma_z$. 
The nontrivial topological properties are characterized by the first Chern number in each mirror subsector. When $\Delta ({\bm k})$ obeys $M_{xy}\Delta ({\bm k})M^{\rm t}_{xy} = \eta \Delta ({\bm k})$ ($\eta = \pm$),
the BdG Hamiltonian is commutable with the mirror reflection operator  in the particle-hole space, ${\mathcal{M}}^{\eta}$, as
\beq
\left[ {\mathcal{M}}^{\eta}, \mathcal{H}({\bm k}) \right] = 0,\quad
{\mathcal{M}}^{\eta} = 
\begin{pmatrix}
M_{xy} & 0 \\ 0 & \eta M^{\ast}_{xy}
\end{pmatrix}.
\label{eq:commut}
\eeq
Thus, the eigenstates of $\mathcal{H}({\bm k})$ are the simultaneous eigenstates of ${\mathcal{M}}^{\eta}$.
Let $\ket{u^{({\lambda})}_n({\bm k})}$ and $E^{(\lambda)}_n$ be the wavefunction and eigenenergy of the Bogoliubov quasiparticles in each mirror subsector, where the $\lambda=\pm i$ are the eigenvalues of $\mathcal{M}^{\eta}$. Then, the first Chern number is defined in each mirror subsector, 
\beq
{\rm Ch}^{(\lambda)}_1 = \frac{i}{2\pi}\int
\mathcal{F}^{(\lambda)} \in \mathbb{Z},
\label{eq:mirrorCh}
\eeq
where $\mathcal{F}^{(\lambda)}=d\mathcal{A}^{(\lambda)}$ is the Berry curvature in each mirror subsector and  
$\mathcal{A}^{(\lambda)}_{\mu} ({\bm k}) = \sum _{E^{(\lambda)}_n<0}\braket{ u^{({\lambda})}_n({\bm k})| \partial _{k_{\mu}}u^{({\lambda})}_n({\bm k})}$.
The nonzero value ensures the existence of the zero energy state in the $\lambda$ subsector. For the $^3$He-A thin film and chiral $p$-wave SCs, $|{\rm Ch}_1^{(\lambda)}|=1$ in each mirror subsector, implying that integer quantum vortices support two Majorana zero modes per each core. 

Multiple Majorana zero modes behave as non-Abelian anyons only when the mirror subsector holds the particle-hole symmetry, which are referred to as mirror Majorana zero modes. The condition for each mirror subsector to maintain the particle-hole symmetry is given by~\cite{ueno,sato14}
\beq
\left\{ \mathcal{C}, \mathcal{M}^{\eta}\right\} = 0.
\label{eq:mirrorMF}
\eeq
For integer quantum vortices, the condition depends on the orientation of the ${\bm d}$-vector. For $\hat{\bm d}\parallel \hat{\bm z}$, the mirror subsector supports its own particle-hole symmetry, but otherwise the subsector does not maintain the particle-hole symmetry and belongs to class A. 

\begin{figure}
\includegraphics[width=80mm]{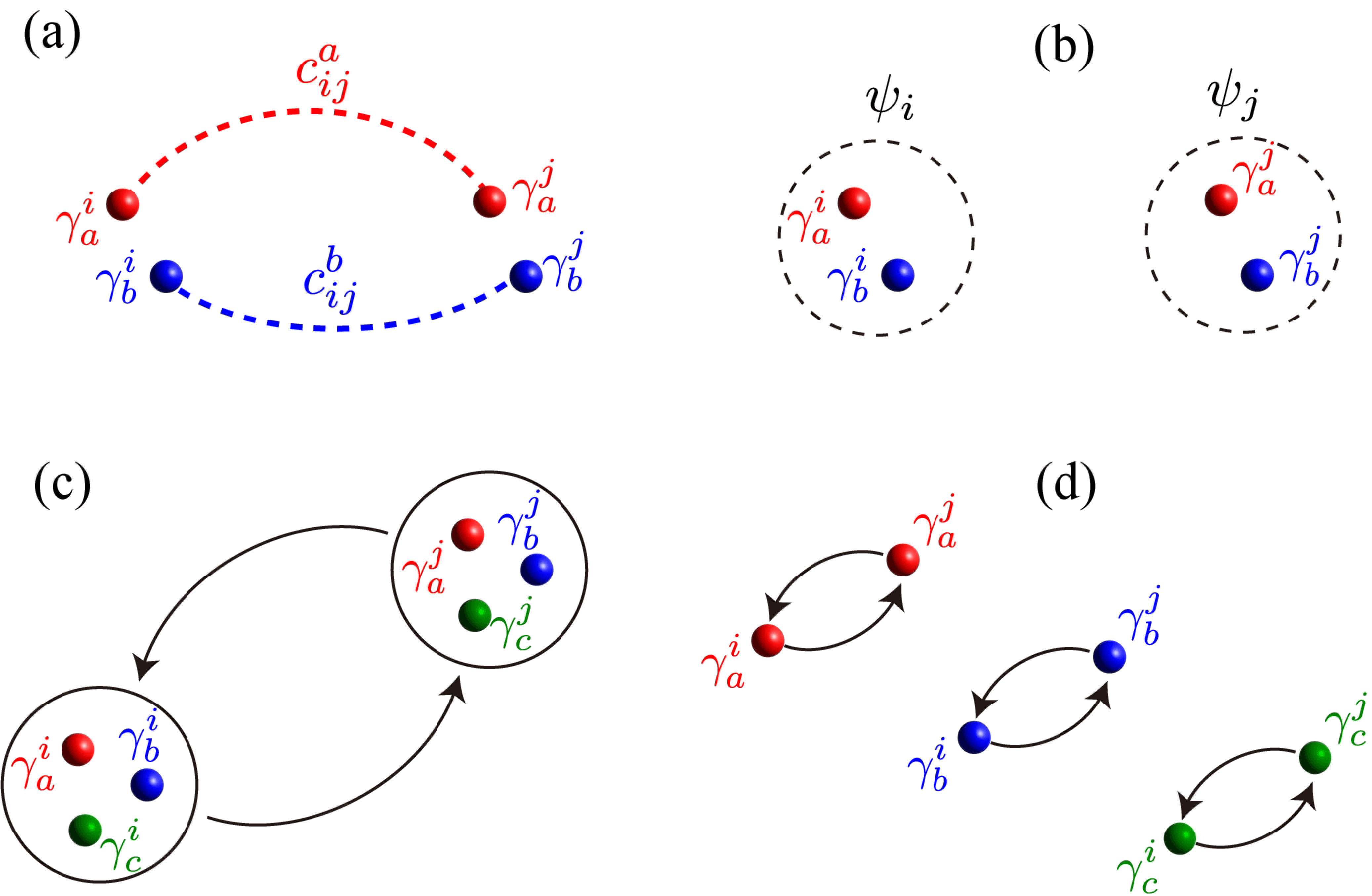}
\caption{Complex fermions formed by intervortex pairs (a) and intravoretx pairs (b), where $\gamma^i_a$ denotes a Majorana zero mode with internal degrees of freedom $a$ (e.g., spin) bound at the $i$th vortex. (c) Schematic of the exchange of two vortices with multiple Majorana zero modes. (d) When the system holds a unitary symmetry such as the mirror symmetry in Eq.~\eqref{eq:commut}, the operation (c) is equivalent to the exchange of two Majorana zero modes in each sector.} 
\label{fig:braiding}
\end{figure}

Now let us clarify the non-Abelian statistics of spin-degenerate Majorana modes. Consider $2N$ integer quantum vortices. Two Majorana zero modes bound at the $i$th vortex are denoted by $\gamma^i_\lambda$ with mirror eigenvalues $\lambda=\pm i$. 
The Majorana zero modes satisfy the self-conjugate
condition $\gamma_{\lambda}^i=(\gamma_{\lambda}^i)^{\dag}$ and the anticommutation relation, $\{\gamma_{\lambda}^i,\gamma_{\lambda^{\prime}}^j\}=2\delta_{i,j}\delta_{\lambda,\lambda^{\prime}}$. 
In contrast to spinless Majorana zero modes, there are two possibilities to form a complex fermion: (i) intervortex pairs, $c^{\lambda}_{ij}\equiv(\gamma_{\lambda}^i+i\gamma_{\lambda}^j)/2$ and (ii) intravortex pairs, $\psi_{i}\equiv(\gamma_{\lambda=i}^i+i\gamma_{\lambda=-i}^i)/2$ [See Figs.~\ref{fig:braiding}(a,b)]. 
In the former case (i), the exchange operators of the $i$th and $j$th vortices are defined in a manner similar to that for spinless Majorana zero modes as
\begin{align}
  U^{\lambda\dag}_{ij} \gamma_{\lambda}^iU^{\lambda}_{ij}  =
  \gamma_{\lambda}^j, \quad
  U^{\lambda\dag}_{ij} \gamma_{\lambda}^jU^{\lambda}_{ij}  = -\gamma_{\lambda}^i.
  \label{eq:Uij2}
\end{align}
The above transformation is realized by the unitary operator,
\begin{align}
U^{\lambda}_{ij} = \exp\left(\frac{\pi}{4}\gamma_{\lambda}^{j}\gamma_{\lambda}^i\right) = \frac{1}{\sqrt{2}}\left(1+\gamma_{\lambda}^{j}\gamma_{\lambda}^i\right),
  \label{eq:Uij3}
\end{align}
Similarly to $U_{ij}$ and $U_{jk}$ in spinless Majorana zero modes, the exchange operators $U^{\lambda}_{ij}$ and $U^{\lambda}_{jk}$ defined in each mirror subsector do not commute to each other, which implies the non-Abelian anyon statistics of integer quantum vortices hosting mirror Majorana zero modes.

In the case (ii) [Fig.~\ref{fig:braiding}(b)], a complex fermion is formed as a local pair of two mirror Majorana modes, which also obeys the non-Abelian statistics~\cite{yas12,sato14}. The expression of the exchange operator is obtained as 
$U_{ij}=1+\psi_j\psi_i^{\dag} + \psi_j^{\dag}\psi_i
-\psi^{\dag}_i\psi_i - \psi^{\dag}_j\psi_j 
+ 2 \psi^{\dag}_j\psi_j\psi^{\dag}_i\psi_i$.
In general, the exchange operators are not commutable with each other, $[U_{ij}, U_{jk}]\neq 0$. As the operator preserves the fermion number $N_{\rm f}=\sum_i\psi_i^\dagger\psi_i$, the degenerate ground states are expressed in terms of the occupation number of the complex fermion in each vortex. For example, let us consider the four-vortex state $\ket{1100}$ where the first and second vortices are accompanied by the Dirac zero modes, while the third and fourth are not. Up to a phase factor, this state changes under $U_{12}$ and $U_{34}$ as $\ket{1100} \rightarrow \ket{1100}$ and $\ket{1100} \rightarrow \ket{1010}$, respectively. The complete representations of the exchange operators $U_{ij}$ are presented in Ref.~\cite{yas12}. The extension to non-Abelian statistics of multiple complex fermions was discussed in Ref.~\cite{Yasui:2012zb}. 

When the BdG Hamiltonian maintains the mirror symmetry in Eq.~\eqref{eq:commut}, as mentioned above, the braiding of vortices supporting spinful Majorana modes is
decomposed into two individual exchanges between
Majorana zero modes $\gamma^i_{\lambda}$ and $\gamma^j_{\lambda}$ in each mirror subsector [see also Figs.~\ref{fig:braiding}(c) and \ref{fig:braiding}(d)] and the exchange matrices are represented by Eqs.~\eqref{eq:Uij2} and \eqref{eq:Uij3}. This argument is applicable for the systems hosting multiple Majorana modes protected by unitary symmetries~\cite{hon22}, where the couplings between Majorana zero modes in each core are excluded by the unitary symmetries.
The braiding of two vortices with $n$ unitary-symmetry-protected multiple Majorana modes generically reduces to the $n$ independent braiding operations in each sector as 
\begin{align}
U_{ij} = \prod_a U_{ij}^{a},
\label{eq:Uija}
\end{align}
where $a$ denotes the internal degrees of freedom of Majorana zero modes. 
The unitary time evolution representing the braiding process
does not dynamically break the unitary symmetry, which prohibits the coupling between Majorana modes in different subsectors labeled by $a$. The mirror reflection symmetry in Eq.~\eqref{eq:commut} is an example of such symmetry-protected non-Abelian anyons. Another example is the non-Abelian vortices in dense QCD matter which trap multiple Majorana zero modes in each core~\cite{Yasui:2010yw,Fujiwara:2011za}. As discussed below, such vortices with multiple Majorana zero modes obey the non-Abelian statistics and the braiding matrices are identical with the elements in the Coxeter group~\cite{Yasui:2010yh,Hirono:2012ad}.

Contrary to unitary symmetries, the braiding process, characterized as a unitary evolution, may dynamically break antiunitary symmetry such as the time-reversal symmetry. We note that in contrast to time-reversal -broken topological SCs such as superconducting nanowires, topological superconductors with time-reversal symmetry do not require an external magnetic field for realizing the topological phase and can keep the topological gap close to the proximity-induced superconducting \blue{gap~\cite{hai19,won12,zha13}.} Hence, the Majorana-Kramers pairs leads to longer decoherence time of the stored quantum information~\cite{liu14}. As mentioned above, however, the braiding process of the Majorana-Kramers pairs may dynamically break the antiunitary symmetry. Such dynamical symmetry breaking gives rise to the local mixing of the Majorana-Kramers pairs~\cite{wol14,wol16,kna20}.
Gao {\it et al.} proposed the conditions to protect the non-Abelian statistics of the Majorana-Kramers pairs~\cite{gao16}. 
In addition, Tanaka {\it et al.} examined by numerical simulations the tolerance of non-Abelian braidings of Majorana-Kramers pairs against two types of perturbations which may cause decoherence of Majorana-Kramers quits~\cite{tak22}: (i) Applied magnetic fields, and (ii) the effect of a gate-induced inhomogeneous potential at junctions of superconducting nanowires. The former break time-reversal symmetry and generate the energy gap of Majorana states, while the latter gives rise to non-Majorana low-energy Andreev bound states~\cite{kel12,liu17,moo18a,moo18b,pan20a,pan20b}. 
The numerical simulation revealed that the non-Abelian braiding is successful when the direction of the applied magnetic field preserves the chiral symmetry in the initial and final states of a braiding process, where the chiral symmetry is a combination of time-reversal symmetry and mirror symmetry~\cite{tak22}. 
Remarkably, this tolerance is preserved even when the intermediate states of the braiding process breaks this symmetry. 
As for (ii), non-Majorana bound states emergent in gate-induced inhomogeneous potentials make Majorana-Kramers qubits vulnerable to quasiparticle poisoning and disturbe the braiding protocol. 
However, the influence can be ignored when the width of the gate-induced potential is sufficiently smaller than the superconducting coherence length.

\sub{Multiple Majorana zero modes and Coxeter group} 
It has been demonstrated that the non-Abelian statistics of vortices hosting three Majorana zero modes has a novel structure, when the Majorana modes have additional ${SO}(3)$ symmetry~\cite{Yasui:2010yh,Hirono:2012ad}. 
The representation of braiding operations in four vortices is given by a tensor product of two matrices, where one is identical to the matrix for the braiding of spinless Majorana zero modes, and the other is a generator of the Coxeter group. 

Let $\gamma_a^i$ ($a=1,2,3$) be the operators of three Majorana zero modes in the $i$th vortex, belonging to the triplet of ${SO}(3)$. Consider four vortices hosting three Majorana zero modes in each core. 
The braiding of these vortices is decomposed into the three individual exchanges between Majorana zero modes $\gamma^i_a$ and $\gamma^j_a$ [see Figs.~\ref{fig:braiding}(c,d)]. 
For each component $a$, Majorana zero modes $\gamma_a^i$ and $\gamma_a^j$ under braiding manipulations transforms in the same way as Eq.~\eqref{eq:Uij2} and the exchange matrices, $U^a_{ij}$, are obtained by $U_{ij}^{\lambda}\rightarrow U^a_{ij}$ in Eqs.~\eqref{eq:Uij2} and \eqref{eq:Uij3}. 
Then, the representations of $U_{ij}$ for vortices with triplet Majorana zero modes are obtained by the $64\times 64$ matrix, $U_{ij}=\prod_{a=1,2,3}U^{a}_{ij}$, as in Eq.~\eqref{eq:Uija}.

We now introduce the complex fermion operators as the nonlocal pairs of Majorana zero modes, $c_{ij}^a=(\gamma_a^i+i\gamma^j_a)/2$. For two vortices (i.e., six Majorana modes), this leads to the $2^3$-fold degenerate ground states and the Hilbert space is spanned by the four basis sets: singlet-even $\ket{{\bm 1}_0}\equiv \ket{0}$ (vacuum) and triplet-even $\ket{{\bm 3}_2}\equiv \frac{1}{2!}\epsilon^{abc}c^{b\dag}_{ij}c^{c\dag}_{ij}\ket{0}$ (occupied by two fermions) for even fermion parity and singlet-odd $\ket{{\bm 1}_3}\equiv \frac{1}{3!}\epsilon^{abc}c^{a\dag}_{ij}c^{b\dag}_{ij}c^{c\dag}_{ij}\ket{0}$ (occupied by three fermions) and triplet-odd $\ket{{\bm 3}_1}\equiv c^{a\dag}_{ij}\ket{0}$ (occupied by single fermion). In the case of four vortices, the ground state degeneracy is $2^6=64$ and the bases of the Hilbert space are singlet (${\bm 1}$), triplet (${\bm 3}$), and quintet (${\bm 5}$) states, which are further divided in terms of the fermion parity. The exchange operator is expressed as a product of two $SO(3)$ invariant unitary operators,
\begin{align}
    U_{ij}=\prod_{a=1,2,3}U^{a}_{ij}=\sigma_{ij} h_{ij}
\end{align}
where both matrices are given in terms of $\gamma^a_i$ as 
\begin{align}
\sigma_{ij}=\frac{1}{2}\left(1-\gamma^1_j\gamma^2_j\gamma^1_i\gamma^2_i
-\gamma^2_j\gamma^3_j\gamma^2_i\gamma^3_i
-\gamma^3_j\gamma^1_j\gamma^3_i\gamma^1_i\right),
\end{align}
and 
\begin{align}
h_{ij}=\frac{1}{\sqrt{2}}\left(
1-\gamma^1_j\gamma^2_j\gamma^3_j\gamma^1_i\gamma^2_i\gamma^3_i
\right).
\end{align}
While the matrices $h_{ij}$ are a natural extension of that originally introduced in Ref.~\cite{Ivanov:2000mjr}, the matrices $\sigma_{ij}$ are proper to vortices with three or more Majorana zero modes. It was demonstrated that the matrices $\sigma_{ij}$ are identified with the elements in the Coxeter group which satisfies the relations
\begin{align}
    (\sigma_{ij})^2=1, \;
    (\sigma_{ij}\sigma_{jk})^3=1, \;
    (\sigma_{ij}\sigma_{kl})^2=1.
\end{align}
The Coxeter group is a symmetry group of polytopes in high dimensions such as a triangle and a tetrahedron. The exchange operators acting on the singlet (${\bf 1}$) and triplet (${\bf 3}$) are identified with the elements in the Coxeter group, such as a 2-simplex (triangle) under the reflections for the singlet and a 3-simplex (tetrahedron) under the reflections for the triplet~\cite{Yasui:2010yh}. The decomposition of the braiding operators $U_{ij}$ in the ${SO}(3)$ triplet was generalized to the Majorana zero modes of arbitrary odd $n_{\rm M}\ge 3$ with ${SO}(n_{\rm M})$ symmetry~\cite{Hirono:2012ad}.

Such multiple Majorana zero modes were 
first found in Refs.~\cite{Yasui:2010yw,Fujiwara:2011za}
inside the cores of 
color flux tubes (``non-Abelian'' vortices) 
\cite{Balachandran:2005ev,Nakano:2007dr,
Eto:2009kg,Eto:2013hoa} 
in high density quark (QCD) matter \cite{Alford:2007xm}. 
They were also discussed in 
edges of a topological superconducting wire coupled to a normal lead~\cite{PhysRevLett.114.116801}.

\section{Non-Abelian vortex anyons}
\label{sec:vortex}

Here we introduce  
non-Abelian anyons made of bosonic excitations,
which \green{are} non-Abelian vortices 
obeying a non-Abelian statistics  
\cite{Bais:1980vd,Wilczek:1989kn,Bucher:1990gs,Brekke:1992ft,Lo:1993hp,
Lee:1993gk,Brekke:1997jj,PhysRevLett.123.140404}. 
These non-Abelian vortices are sometime called fluxons.\footnote{Here, the term ``flux'' originally comes
from the fact that 
they were proposed in gauge theory 
where non-Abelian symmetry is gauged 
in the context of high energy physics  
\cite{Bais:1980vd}.
In this article, we consider only global non-Abelian symmetry relevant for condensed matter physics 
and they do not carry non-Abelian fluxes.
}

\subsection{Non-Abelian vortices}

When a symmetry $G$, that is either global or local, 
is spontaneously broken into its subgroup $H$, 
there appears an order parameter space 
\begin{eqnarray}
 M \simeq G/H
\end{eqnarray}
parameterized by Nambu-Goldstone modes. 
The first homotopy group of the order parameter space
\begin{eqnarray}
 \pi_1 (G/H) \simeq \pi_0 (H) 
\end{eqnarray}
is responsible for the existence of quantum vortices, 
where we have assumed that $G$ is semisimple.
We further have
\begin{eqnarray}
 \pi_0 (H) \simeq H/H_0  
 (\simeq H \mbox{ for a discrete group } H)
 \label{eq:pi0H}
\end{eqnarray}
with a normal subgroup $H_0$ of $H$.
When $\pi_1(G/H)$ is Abelian 
(and thus $H/H_0$ (for semisimple $G$) is Abelian), 
vortices are Abelian, 
while when it is non-Abelian, 
vortices are non-Abelian 
\cite{Balachandran:1983pf,Alford:1990mk,Alford:1990ur,Alford:1992yx}.

One of characteristic features of non-Abelian vortices is that $H$ is not globally defined 
(or multi-valued): 
 the unbroken symmetry 
$H_{\theta}$ depends on the azimuthal angle $\theta$ 
around a non-Abelian vortex as 
$H_{\theta} = g(\theta) H_{\theta =0} g^{-1}(\theta)$ 
with $g(\theta)\in G$, 
$g(0) = {\bf 1}$ and $g(2\pi)$ taking a disconnected 
component of $H/H_0$, see Eq.~(\ref{eq:pi0H}).
Then, after an adiabatic transport along a loop encircling the non-Abelian vortex, the unbroken symmetry at 
$\theta = 2\pi$ does not have to come back: 
$H_{\theta = 2\pi} \neq H_{\theta =0}$. 
This is called {\it topological obstruction}, 
and the symmetry $H$ is topologically broken 
down to a subgroup 
$K \equiv \{h| [h,g(2\pi)]=0, h \in H\}$ which is 
single-valued and globally defined 
\cite{Balachandran:1983pf,Alford:1990mk}.
This is also called {\it topological symmetry breaking}.
When $G$ and $H$ are local gauge symmetries, 
it gives a non-Abelian Aharonov-Bohm effect 
\cite{Bais:1980vd,Wilczek:1989kn,Bucher:1990gs,
Alford:1990mk,Alford:1990ur,Alford:1992yx}.
Another related feature is the existence 
of non-local charges, called Cheshire charges, 
existing among separated vortices 
\cite{Preskill:1990bm,Alford:1990mk,Bucher:1991bc}.

\begin{figure}
\includegraphics[width=\hsize]{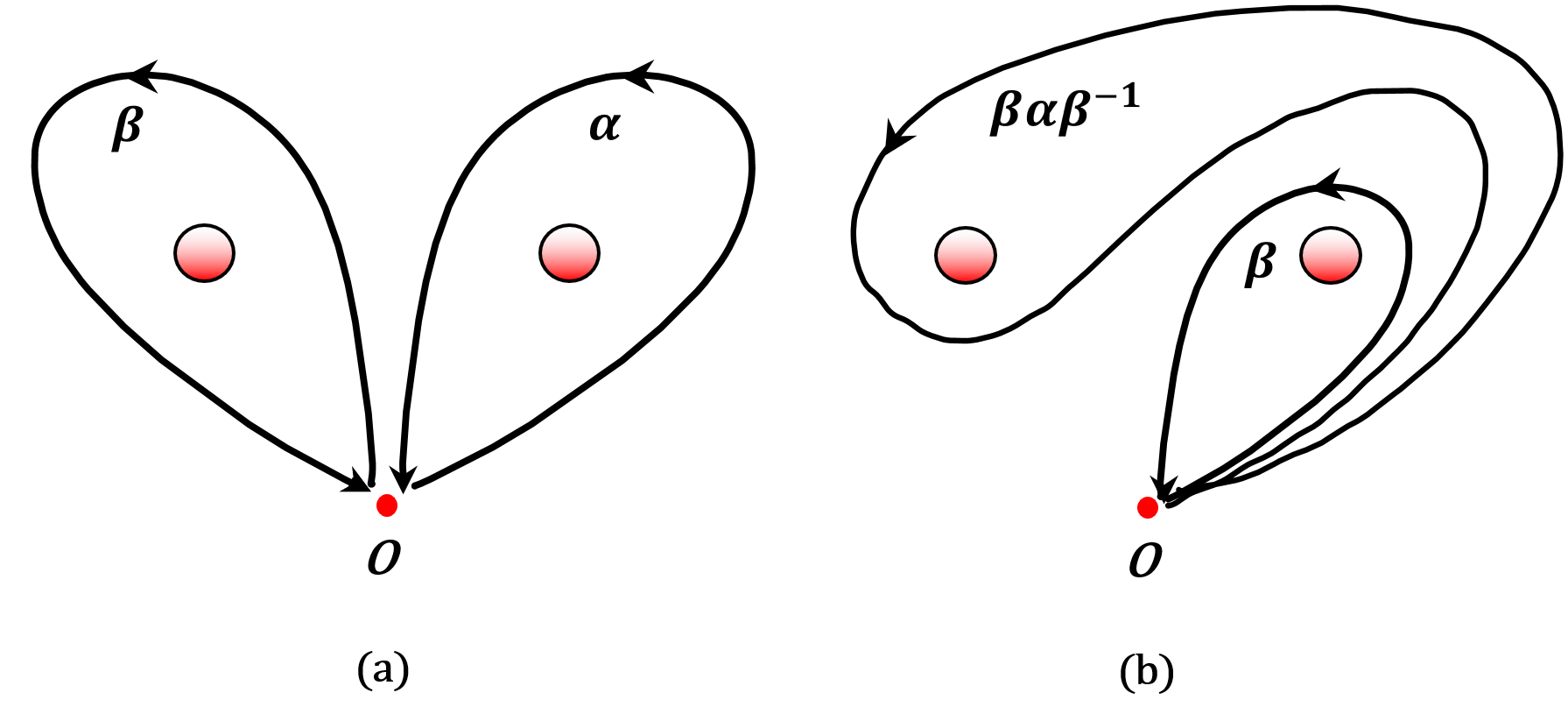}
\caption{Exchange of two vortices. 
(a) Two closed paths $\alpha$ and $\beta$ encircle two vortices. (b) Exchange them counterclockwisely.
} 
\label{fig:exchange}
\end{figure}
Let $\alpha, \beta \in \pi_1(G/H)$ being 
closed passes enclosing two vortices.
Then, if one adiabatically transports 
the vortex $\alpha$ around the vortex 
$\beta$ counterclockwisely, \green{the vortex $\alpha$} transforms 
as (see Fig.~\ref{fig:exchange}\green{(b)})
\begin{eqnarray}
    \alpha \to \beta \alpha \beta^{-1}.
    \label{eq:transform-0}
\end{eqnarray}
Therefore, when one exchanges a pair 
$(\alpha,\beta)$ 
counterclockwisely, 
they transform as 
\begin{eqnarray}
    (\alpha,\beta) \to (\beta,\beta \alpha \beta^{-1}).
    \label{eq:transform}
\end{eqnarray}
The first homotopy group $\pi_1$ is 
a based fundamental group for which every loop 
starts from a fixed point $O$.
Instead of it, the conjugacy classes 
\begin{eqnarray}
   [\alpha] = \beta \alpha \beta^{-1}, \; \forall \beta \in \pi_1(G/H) \label{eq:conj}
\end{eqnarray}
characterize individual vortices as can be seen from Eqs.~(\ref{eq:transform-0}) and (\ref{eq:transform}).
\red{
Two vortices corresponding to two different elements in the same conjugacy class 
are {\it indistinguishable} even though they are not the same.
}

Non-Abelian first homotopy groups significantly control
the dynamics of vortices.
In two spatial dimensions, 
scattering of such non-Abelian vortices was discussed 
in the case that $G$ and $H$ are local gauge symmetries
\cite{Wilczek:1989kn,Bucher:1990gs,Lo:1993hp}.
In three spatial dimensions, vortices are lines.
When two vortex lines collide in three spatial dimensions, 
the fate is determined by 
 corresponding elements of the first homotopy group.
When these elements  commute, the two vortices pass through 
or reconnect each other if these elements are identical.
On the other hand,
when these elements are non-commutative, the third vortex bridging 
them must be created \cite{Poenaru1977, Mermin:1979zz,Kobayashi:2008pk}.

\subsection{Examples of non-Abelian vortices}
\label{sec:vortex-examples}

\sub{Nematic liquid crystals}
Let us start from nematic liquid crystals
focusing on  
uniaxial nematics (UNs), 
and $D_2$- and $D_4$-biaxial nematics~\green{(BNs)}.

{\bf UN liquid crystals}.
The order parameter of \green{UNs} is an unoriented rod. 
The rotational symmetry $G=SO(3)$ is spontaneously broken down to 
$H \simeq O(2)$, 
\green{which implies} the order parameter \green{manifold}
\red{
\begin{eqnarray}
\frac{G}{H} = \frac{SO(3)}{O(2)} 
\simeq \frac{SU(2)}{Pin(2)}
\simeq S^2/{\mathbb Z}_2 
\simeq {\mathbb R}P^2,
\end{eqnarray}
which is a real projective space of two dimensions, 
where $SU(2)$ and $Pin(2)$ 
are double covering groups of 
$SO(3)$ and $O(2)$, respectively.
}
Since the fundamental group $\pi_1 ({\mathbb R}P^2) = {\mathbb Z}_2$ is Abelian, 
this does not admits non-Abelian vortices.
These vortices are called Alice strings 
(when $G$ and $H$ are local gauge symmetries)
\cite{Schwarz:1982ec}. 
\red{
Around the Alice string, $SO(2) \subset H$ 
is not globally defined and $H$ is topologically broken to $K\simeq {\mathbb Z}_2$.
}

{\bf $D_2$-BN liquid crystals}~\blue{\cite{Poenaru1977,vol77,Mermin:1979zz,Balachandran:1983pf}}.

\begin{table*}
\begin{spacing}{1.5}
\begin{tabular}{|c|ccccc|}
\hline
 $g(2\pi)\in \pi_1(G/H)$ & ${\bf 1}_2$ 
 & $-{\bf 1}_2$ & $\pm i \sigma_x$ & $\pm i \sigma_y$ & 
 $\pm i \sigma_z$ \\
 \hline
  $g(\theta)$ & ${\bf 1}_2$  
 &\red{
 $\exp \left( \green{\pm} i \green{\frac{\theta}{2}} {\bf n} \cdot {\mathbb \sigma}\right)$
 }
 & $\exp \left( \pm i \green{\frac{\theta}{4}} \sigma_x\right)$ 
 & $\exp \left( \pm i \green{\frac{\theta}{4}} \sigma_y \right)$
 & $\exp \left( \pm i \green{\frac{\theta}{4}} \sigma_z \right)$ \\
 $O(2\pi)$ & ${\bf 1}_3$ 
 & ${\bf 1}_3$ & $I_x$ & $I_y$ & $I_z$ \\
 $O(\theta)$ & ${\bf 1}_3$ 
 & \red{
  $\exp(i\theta {\bf n} \cdot {\bf L})$
  }
 & \red{$\exp\left(\pm i\frac{\theta}{2}L_x\right)$}
 & \red{$\exp\left(\pm i\frac{\theta}{2}L_y\right)$}
 & \red{$\exp\left(\pm i\frac{\theta}{2}L_z\right)$}     \\
 \hline   
\end{tabular}
\end{spacing}
\caption{Asymptotic configurations of 
spin vortices in $D_2$-BNs: 
\red{
the fundamental group elements 
$g(2\pi) \in \pi_1(G/H)$,
the $SU(2)$ elements $g(\theta) \in SU(2)$,
$O(2\pi)$, and
the $SO(3)$ group elements 
$O(\theta) \in SO(3)$, 
acting on the 
the order parameter $A$ 
as Eq.~(\ref{eq:vortex-config}).
The group actions 
$O(\theta)$ and $g(\theta)$ 
for a spin vortex of $2\pi$ rotation 
($-{\bf 1}_2$) can be a rotation around any axis 
along the unit vector ${\mathbb n}$.
$L_i$ are $3\times 3$ spin-1 matrices 
$[L_i]_{jk} = -i \epsilon_{ijk}$.
}
\label{table:spin-vortices}}
\end{table*}
\red{
Cholesteric liquid crystals (chiral nematic liquid crystals) can 
also have the same order parameter space 
with $D_2$-BN liquid crystals \cite{Lavrentovich2001} 
and all the following discussions hold.
}
The order parameter of \green{BNs} 
is a set of two unoriented rods 
of different lengths orthogonal to each other.
The (unbroken) symmetry is the dihedral group $D_2$ keeping a rectangular invariant.
Thus, the rotational symmetry $G=SO(3)$ is spontaneously broken 
down to 
\begin{align}
    \hspace{-0.7cm} H \simeq D_2 
    &= \left\{ 
    {\bf 1}_3, 
    I_x = \begin{pmatrix}
        \green{+}1 & 0 & 0\\
         0 & \green{-}1 & 0\\
         0 & 0 & \green{-}1
         \end{pmatrix}, \right. \nonumber\\
 & \left. \quad
    I_y = \begin{pmatrix}
         \green{-}1 & 0  & 0\\
         0 & \green{+}1 & 0\\
         0 & 0  & \green{-}1
         \end{pmatrix},
    I_z = \begin{pmatrix}
         \green{-}1 & 0& 0\\
         0 & \green{-}1 & 0\\
         0 & 0 & \green{+}1
         \end{pmatrix}
    \right\}.
\end{align}
\red{
Note that $D_2 \simeq {\mathbb Z}_2 \times {\mathbb Z}_2$ is Abelian.
}
The order parameter manifold is
\begin{eqnarray}
&&  G/H = SO(3)/D_2  \simeq SU(2)/{\mathbb Q}, 
\end{eqnarray}
where  the quaternion group ${\mathbb Q}$
are universal (double) covering group 
of $D_2$: 
\begin{eqnarray}
 {\mathbb Q} = \left\{ \pm {\bf 1}_2,\pm i \sigma_x, \pm i \sigma_y, \pm i \sigma_z \right\}.\label{eq-Q}
\end{eqnarray}
Note that $H^* \simeq {\mathbb Q}$ is a non-Abelian group 
while $H\simeq D_2$ is an Abelian group. 
This breaking can occur for instance by \green{five real-scalar order-parameters}
belonging 
to spin-2 representation of $SO(3)$, 
which is  a traceless symmetric $3 \times 3$ tensor 
$A$ 
with real components, taking a form of  
\begin{eqnarray}
 A = {\rm diag} (1,r,-1-r)  
\end{eqnarray}

The first homotopy group is isomorphic to the quaternion group ${\mathbb Q}$ in Eq.~(\ref{eq-Q})
\begin{eqnarray}
  \pi_1 (SU(2)/{\mathbb Q}) \simeq {\mathbb Q}.
\end{eqnarray}
The elements in Eq.~(\ref{eq-Q}) 
correspond to the ground state (${\bf 1}\green{_2}$), 
spin vortices of $2\pi$ rotation ($-{\bf 1}\green{_2}$), 
and spin vortices of $\green{\pm}\pi$ rotation about 
the $x,y,z$-axes \green{($\pm i\sigma_{x,y,z}$)}. 
\red{
These vortex configurations at the large distance 
can be asymptotically written as
\begin{eqnarray}
   A \sim O(\theta) A_{\theta =0} O^T(\theta), \quad 
   O(\theta) \in SO(3) \label{eq:vortex-config}
\end{eqnarray}
with the azimuthal angle $\theta$.
The concrete forms of $O(\theta)$'s are given in Table~\ref{table:spin-vortices}.
Corresponding $SU(2)$ elements $g(\theta)$ 
can be defined as a double covering of $SO(3)$ 
with $g (2\pi) \in \pi_1(G/H)$.
A spin vortex of $2\pi$ rotation ($-{\bf 1}\green{_2}$)
and a spin vortex of 
$\pi$ rotation ($\pm i \sigma_a$) $(a=x,y,z)$
are given by  
\begin{align}
&
g(\theta) = \exp \left(\pm i \frac{\theta}{2} \sigma \cdot {\bf n} \right)
= \cos \frac{\theta}{2} {\bf 1}_2 \pm i \sigma \cdot {\bf n} \sin \frac{\theta}{2}, \nonumber \\
&
g(\theta) = \exp \left(\pm i \frac{\theta}{4} \sigma_a \right)
= \cos \frac{\theta}{4} {\bf 1}_2 \pm i \sigma_a \sin \frac{\theta}{4}, \quad 
\end{align} 
respectively, 
with a unit three-vector ${\bf n}$, 
giving
the elements of $\pi_1(G/H)$ 
at $\theta = 2\pi$: 
$g(2\pi) =  - {\bf 1}_2$ 
and $g(2\pi) = \pm i\sigma_a$, 
respectively.
While 
the former 
is Abelian \green{because} $[g(2\pi),H^*]=0$ 
(even though $g(2\pi) \neq g(0)$), 
the latter is non-Abelian 
\green{because} $[g(2\pi),H^*]\neq 0$ with 
the universal covering 
$H^* \simeq SU(2)$,\footnote{
Note that $[O(2\pi),H]=0$, 
and thus the criterion of non-Abelian property 
should be considered in 
the universal covering group $H^*$ but not in $H$.
}  
and $H^*$ is topologically broken 
to $K^*_a=\{ \pm {\bf 1}_2, \pm i \sigma_a\}$.
}

The elements in Eq.~(\ref{eq-Q}) 
are grouped to the conjugacy classes, Eq.~(\ref{eq:conj}), consisting of five elements:
\begin{equation}
    \{{\bf 1}_2\}, \{-{\bf 1}_2 \},
    \{\pm i \sigma_x\}, \{\pm i \sigma_y\}, 
    \{\pm i \sigma_z\} .
    \label{eq:conj-Q}
\end{equation}
\red{
This  follows for instance from $(i\sigma_y)^{-1}(i\sigma_x)(i\sigma_y) = - i\sigma_x$,
impling that a vortex corresponding to 
$i \sigma_x$ becomes $-i \sigma_x$ when it travels around 
a vortex corresponding to $i \sigma_y$. 
Thus, the vortices corresponding to $i \sigma_x$ and 
$-i \sigma_x$ are {\it indistinguishable} although they are not the same. 
From $(i\sigma_a)(-i\sigma_a) = {\bf 1}_2$ ($a=x,y,z$),
vortices belonging to the same conjugacy class 
are anti-particles of each other, 
and thus they are similar to Majorana fermions discussed in the last section.
}

In three spatial dimensions, this phenomenon of 
non-commutativity appears as follows:
when two vortex lines corresponding to $i \sigma_x$ and $i \sigma_y$ collide in three spatial dimensions, 
the third vortex 
bridging them is created \cite{Poenaru1977,Mermin:1979zz}.

{\bf $D_4$-BN liquid crystals}.
The order parameter of $D_4$ BNs is a set of two unoriented 
indistinguishable rods of the {\it same} lengths orthogonal to each other.
In this case, the (unbroken) symmetry is a dihedral group $D_4$ keeping a square invariant:
\begin{eqnarray}
 G/H = SO(3)/D_4  \simeq SU(2)/D_4^*,
\end{eqnarray}
where $M^*$ denotes a universal covering group of $M$.
The fundamental group is given by 
\begin{align}
\pi_1 (SU(2)/D_4^*) \simeq D_4^*  \label{eq:D4}
  \end{align}
with the sixteen elements   
\begin{eqnarray}
 \left\{ \pm \bm{1}_2 , \pm i \sigma_x , \pm i \sigma_y, \pm  i \sigma_z , 
 \pm C_4  , \pm \red{C_4^{-1}}, \pm i \sigma_x C_4 ,
     \pm i \sigma_x \red{C_4^{-1}} \right\} 
\end{eqnarray}
with $C_4 \equiv e^{i {\pi \over 4}\sigma_z } = (1/\sqrt 2) ({\bf 1}_2 + i \sigma_z)$. Note $(C_4)^2 = i\sigma_z$ and $(C_4)^4 = -{\bf 1}_2$.
The conjugacy classes consist of the following seven elements 
\begin{eqnarray}
 &&
 \{ \bm{1}_2 \},\{ -\bm{1}_2 \} , 
 \{ \pm i \sigma_x  , \pm i \sigma_y \} , 
 \{ \pm i \sigma_z  \} , 
 \nonumber 
 \\ 
 && 
 \{ C_4,\red{C_4^{-1}} \} , 
\{ -C_4, \red{-C_4^{-1}} \} , \{ \pm i \sigma_x C_4, \pm i \sigma_x \red{C_4^{-1}} \} 
. \quad\quad. \label{eq:D4-conj}
\end{eqnarray}
\red{
Since 
$- C_4 = C_4^3$ 
($- C_4^{-1} = C_4^{-3}$),
vortices of these elements 
have more energy than those of 
$C_4$ and $C_4^{-1}$.
}
This symmetry breaking of 
$D_4$ nematics can be realized by a 
higher rank tensor order parameter \cite{Mietke:2020czn},
in which more generally $D_n$ nematic liquid crystals were also discussed.

\sub{Spinor BECs}
Next we provide examples of 
spinor BECs~\cite{Kawaguchi:2012ii}. 
The order parameter of
spin-2 BECs is 
a traceless symmetric $3 \times 3$ tensor 
 $A$ with 
{\it complex} components transforming under 
the symmetry $G = U(1) \times SO(3)$ as
\begin{eqnarray}
 A \to e^{i\theta} g A g^T, \quad e^{i\theta} \in U(1), \quad g \in SO(3).
\end{eqnarray}
\red{
Phases of condensations with total angular momentum two are classified by Mermin
\cite{Mermin:1974zz}. 
They are nematic, cyclic and ferromagnetic phases.
All phases are theoretically possible in the case of spin-2 BECs. The spin-2 BECs are experimentally realized by $^{87}$Rb atoms 
 for which the phase is
around the boundary between cyclic phase and ferromagnetic 
phase, see e.g. Ref.~\cite{PhysRevA.80.042704}.
Here, we discuss nematic and cyclic phases 
which host non-Abelian vortex anyons.
In the next section, we discuss 
\tPt SF which is in the nematic phase.}

\sub{BN phases in spin-2 BEC}~\cite{Song:2007ca,Uchino:2010pf,Kobayashi:2011xb,Borgh:2016cco}. 
The nematic phase consists of three degenerate phases: the UN, 
$D_2$- and $D_4$-BN phases. 
The order parameters of the UN, 
$D_2$- and $D_4$-BN phases are given by
\begin{align}
{\rm UN}: & A  \sim {\rm diag} (1,-1/2,-1/2)  ,\nonumber\\
D_2\mbox{-BN}:& A  \sim {\rm diag} (1,r,-1-r),\nonumber\\
D_4\mbox{-BN}:& A  \sim {\rm diag} (1,-1,0)  ,
\label{eq:op-nematic}
\end{align}
respectively, where $r \in {\mathbb R}$ and 
$-1<r< -1/2$. 
In the $D_2$-BN phase, the limits $r \to -1/2,-1$ correspond to 
UN and $D_4$-BN phases, respectively. 
The symmetry breaking patterns and 
the order parameter manifolds are
\begin{align}
{\rm UN}: &  {G \over H} = 
U(1) \times {SO(3) \over O(2) } 
\simeq {U(1) \times {\mathbb R}P^2}, \nonumber\\
D_2\mbox{-BN}:&  {G \over H} = U(1) \times {SO(3) \over D_2 } 
\simeq U(1) \times {SU(2) \over {\mathbb Q}}, \nonumber\\
D_4\mbox{-BN}: &  {G \over H} = {U(1) \times SO(3) \over D_4 } 
\simeq {U(1) \times SU(2) \over D_4^*} . \nonumber \\
\label{eq:OPS-UNBN}
\end{align}
These phases are continuously connected by the parameter 
$r$ interpreted as a quasi-Nambu-Goldstone mode. 
The degeneracy is lifted by quantum effects and 
either of these phases remains as the ground state 
\cite{Uchino:2010pf}.

The $D_2$-BN and $D_4$-BN admit non-Abelian vortices.
The order parameters of the UN and $D_2$-BN phases of spin-2 BECs are
merely products of the $U(1)$ phonon and the order parameters of 
the corresponding nematic liquids. 
Thus, we concentrate on the $D_4$-BN phase hereafter.
The fundamental group of the $D_4$-BN phase is given by 
\begin{align}
\pi_1 \left( U(1) \times SU(2) \over D_4^{\ast}\right) \cong 
\mathbb{Z} \times_h D_4^{\ast}  \label{conj-4}
\end{align}
where $\times_h$ is defined in Ref.~\cite{Kobayashi:2011xb}. 
This consists of the following sixteen elements
\begin{align}
  & \Big\{ 
  (N, \pm \bm{1}_2) , 
  (N, \pm i \sigma_x)  , 
  (N, \pm i \sigma_y )  ,
  (N, \pm i \sigma_z )  , \nonumber \\
  & \quad 
  \left(N + {1\over 2} ,\pm C_4\right), 
  \left(N + {1\over 2},\pm i \sigma_x C_4  \right), 
  \nonumber\\  
  & \quad
  \left(N+{1\over 2} ,\pm \red{C_4^{-1}}\right), 
  \left(N+{1\over 2}, \pm i \sigma_x \red{C_4^{-1}} \right)  \Big\}, \label{conj-5}
\end{align}
where 
the first and second elements 
of a pair $(\cdot,\cdot)$ denote 
the circulation $\kappa$ ($U(1)$ winding number)
and an $SU(2)$ element, respectively with 
$N \in {\mathbb Z}$.
Here $C_4 \equiv e^{i {\pi \over 4}\sigma_z } = (1/\sqrt 2) ({\bf 1}_2 + i \sigma_z)$ satisfying $C_4^4 =-{\bf 1}_2$.
The conjugacy classes of Eq.~(\ref{conj-5})
are composed of seven elements for each $N$:
\begin{eqnarray}
{\rm (I)} &&  \{ (N,\bm{1}_2) \},
 \nonumber \\ 
{\rm (II)} &&  \{ (N,-\bm{1}_2) \} , 
 \nonumber \\ 
{\rm (III)} && 
\{ (N,\pm i \sigma_x ) , (N,\pm i \sigma_y ) \} ,
\nonumber \\ 
{\rm (IV)} && \{ (N,\pm i \sigma_z ) \} , 
 \nonumber \\ 
{\rm (V)} && 
\left\{ \left(N+{1\over 2},C_4\right),\left(N+{1\over 2},\red{C_4^{-1}}\right) \right\} , 
 \nonumber \\ 
{\rm (VI)} && 
\left\{ \left(N+{1\over 2},-C_4\right),\left(N+{1\over 2},\red{-C_4^{-1}}\right) \right\} ,
\nonumber \\ 
{\rm (VII)} && 
\left\{ \left(N+{1\over 2},\pm i \sigma_x C_4\right),
   \left(N+{1\over 2},\pm i \sigma_x \red{C_4^{-1}}\right) \right\}. 
\nonumber\\
\label{eq:conj-nematic}
\end{eqnarray}
They describe 
(I) integer vortices ($N=0$ corresponds to the vacuum),
(II) spin vortices of $2\pi$ rotation,
(III),(IV) spin vortices of $\pi$ rotation around 
the \green{$x$, $y$, and $z$} axes, and 
(V) -- (VII) non-Abelian HQVs.

\red{
In Sec.~\ref{sec:3P2}, we see that 
\tPt SFs are 
in the $D_4$-BN phase in a strong magnetic field.
There, a singly quantized vortex 
$(1,{\bf 1})$ splits into two HQVs 
$(1/2,C_4)$ and $(1/2,C_4^{-1})$ 
both in the conjugacy class (V)\footnote{
\red{
Topologically a decay into a pair 
$(1/2,-C_4)$ and $(1/2,-C_4^{-1})$ in 
the conjugacy class (VI) 
is also possible, but energetically disfavored.
}
}. 
When they fuse, they go back to the singly quantized 
vortex $(1,{\bf 1})$. 
However, one of them, e.g.,~$(1/2,C_4^{-1})$ 
can transform to the other $(1/2,C_4)$ 
once some other vortex passes through between them 
because they belong to the same conjugacy class.
If they fuse after that, they become 
$(1,i \sigma_z)$ because $C_4^2=i \sigma_z$ 
which is 
a composite belonging to (IV), 
of a singly quantized vortex and 
a spin vortex of $\pi$ rotation.
}

\sub{Cyclic phase in spin-2 BEC} 
\cite{Semenoff:2006vv,Kobayashi:2008pk,Kobayashi:2011xb}. 
The order parameter of the cyclic phase is
\begin{align}
  A  \sim {\rm diag} (1,e^{2\pi i/3},e^{4\pi i/3}),
  \label{eq:cyclic}
\end{align}
yielding 
the symmetry breaking to the tetrahedral group 
$H \simeq T$ and 
the order parameter manifold, given by
\begin{align}
 {G\over H} = {U(1) \times SO(3)\over T } \simeq {U(1) \times SU(2) \over T^*}  
\end{align}
with the universal covering group $T^*$ of $T$. 
The fundamental group in the cyclic phase is given by 
\begin{align}
\pi_1 \left( {U(1) \times SU(2) \over T^{\ast}}\right) \cong \mathbb{Z} \times_h T^{\ast} 
\label{conj-1}
  \end{align}
 with the 24 elements 
\begin{align}
\Big\{ 
  &(N,\pm \bm{1}_2) , 
   (N, \pm i \sigma_x), (N,\pm i  \sigma_y), 
   (N,\pm i  \sigma_z), \nonumber \\ 
  &
  \left(N+ \frac{ 1}{3} ,\pm C_3\right), 
  \left(N+ \frac{ 1}{3} ,\pm i \sigma_x C_3\right)  , 
  \nonumber \\ 
  &
  \left(N+ \frac{ 1}{3} , \pm i \sigma_y C_3 \right) ,
  \left(N+ \frac{ 1}{3} , \pm i \sigma_z C_3 \right), 
\nonumber \\ 
  &
  \left(N -\frac{ 1}{ 3} , \pm C_3^2 \right) ,
  \left(N -\frac{ 1}{ 3}, \pm i \sigma_x C_3^2 \right) , 
  \nonumber \\ 
  &
  \left(N-\frac{ 1}{ 3}, \pm i \sigma_y C_3^2\right), 
  \left(N-\frac{ 1}{ 3}, \pm i \sigma_z C_3^2\right) \Big\} \label{cyclic-elements}
\end{align}
with the same notation with Eq.~(\ref{conj-5}). 
Here 
$C_3 \equiv 
 (1/2) ({\bf 1}_2 + i \sigma_x + i \sigma_y + i \sigma_z)$ 
 satisfying 
 $(C_3)^3 = - {\bf 1}_2$. 
  The conjugacy classes of Eq.~(\ref{cyclic-elements}) are composed of 
  the following seven elements for each $N$ \cite{Semenoff:2006vv}: 
\begin{eqnarray}
 {\rm (I)}  &&\{ (N,  \bm{1}_2)\}, \nonumber\\
 {\rm (II)} &&\{ (N,-\bm{1}_2)\},  \nonumber\\
{\rm (III)} &&
  \{  (N,  \pm i \sigma_x) , 
      (N,  \pm i  \sigma_y) , 
      (N,  \pm i  \sigma_z)\}, 
 \nonumber\\
{\rm (IV)} && \Big\{ \left(N + \frac{ i}{3} ,  C_3\right)  , 
        \left(N + \frac{ 1}{3} , - i \sigma_x C_3 \right) , 
 \nonumber\\
  &&\quad  
  \left(N + \frac{ 1}{3} , - i \sigma_y C_3 \right), 
  \left(N + \frac{ 1}{3} , - i \sigma_z C_3 \right)  \Big\}, 
 \nonumber\\
{\rm (V)} && \Big\{ \left(N + \frac{ i}{3} , - C_3 \right) , 
   \left(N + \frac{ 1}{3} , i \sigma_x C_3 \right)  ,
    \nonumber\\
  &&\quad \left(N + \frac{ i}{3} ,  i \sigma_y C_3\right), 
  \left(N + \frac{ 1}{3} , i \sigma_z C_3 \right)   \Big\}, \nonumber\\ 
{\rm (VI)} && \Big\{ 
  \left(N - \frac{ 1}{3} , C_3^2\right) , 
  \left(N - \frac{ 1}{3} , i \sigma_x C_3^2 \right)  ,
    \nonumber\\
   &&\quad 
  \left(N - \frac{ 1}{3} , i \sigma_y C_3^2 \right)   ,  
  \left(N - \frac{ 1}{3} , i \sigma_z C_3^2 \right)  \Big\} , 
  \nonumber\\
{\rm (VII)} && \Big\{ 
   \left(N-\frac{ 1}{3} , - C_3^2 \right) , 
   \left(N - \frac{ 1}{3} , - i \sigma_x C_3^2 \right)  ,
    \nonumber\\
  &&\quad 
  \left(N - \frac{ 1}{3} , - i \sigma_y C_3^2 \right) , 
  \left(N -\frac{ 1}{3} ,  - i \sigma_z C_3^2\right) \Big\}.
 \nonumber\\
  \label{eq:conj-cyclic}
\end{eqnarray} 
They describe 
(I) integer vortices ($N=0$ corresponds to the vacuum),
(II) spin vortices of $2\pi$ rotation,
(III) spin vortices of $\pi$ rotation around 
the $x,y,z$ axes, and 
(IV) -- (VII) 1/3-quantum vortices.

\sub{Other examples}
Another example in condensed matter physics can be found in vortex lines in the dipole-free A-phase of SF $^3$He \cite{Balachandran:1983pf,salomaaRMP}.
An example in high energy physics can be found 
in color flux tubes (``non-Abelian'' vortices) in high density QCD matter \cite{Fujimoto:2020dsa,Fujimoto:2021wsr,Fujimoto:2021bes,
Eto:2021nle}.

\subsection{Fusion rules for non-Abelian vortex anyons 
}
\label{sec:vortex-fusion}

Two non-Abelian vortices belonging to the same 
conjugacy class are indistinguishable 
even if they are not identical.
For instance, a vortex $a \in \pi_1(G/H)$ 
and a vortex $b = c a c^{-1}$ with $c \in \pi_1(G/H)$ are not 
the same and thus $a$ and anti-vortex of $b$ 
cannot pair annihilate.
Nevertheless they are indistinguishable with a help of $c$. 
This leads a class of non-Abelian statistics 
\cite{Brekke:1992ft,Lo:1993hp,
Lee:1993gk,Brekke:1997jj}.
Such non-Abelian anyons are called non-Abelian vortex anyons.

In order to discuss 
 non-Abelian vortex anyons, 
first we define anyons 
$\{{\bf 1}, \sigma ,\tau\}$ 
appropriately 
by conjugacy classes of non-Abelian vortices.
Then, 
the fusion rule (\ref{eq:fusion-rule}) 
for non-Abelian vortex anyons can be 
written as 
\cite{PhysRevLett.123.140404}
\begin{align}
   & \tau \otimes \tau 
    = N^{\bf 1}_{\tau\tau}{\bf 1} 
    \oplus N^{\sigma}_{\tau\tau} \sigma
    \oplus N^{\tau}_{\tau\tau} \tau ,\nonumber\\
   & \tau \otimes \sigma = \tau, \quad
    \sigma \otimes \sigma = {\bf 1}, \quad
    x \otimes {\bf 1} = x,\label{eq:fusion}
\end{align}
with $x \in \{{\bf 1}, \sigma ,\tau\}$. 

\begin{table*}
    \centering
    \begin{tabular}{c||c|c|c|c|c|c|c}
 & I$_0$   & II$_0$ & III$_0$ 
 & ${\rm IV}_0$ & ${\rm V}_0$ 
 & ${\rm VI}_{-1}$ & ${\rm VII}_{-1}$ 
 \\
                      \hline
 ${\bf 1} = {\rm I}_0$  & I$_0$   & II$_0$ & III$_0$ 
  & ${\rm IV}_0$ & ${\rm V}_0$ 
  & ${\rm VI}_{-1}$ & ${\rm VII}_{-1}$ 
  \\
 $\sigma = {\rm II}_0$  & II$_0$  & I$_0$ & III$_0$ 
  & ${\rm V}_0$ & ${\rm IV}_0$ 
  & ${\rm VII}_{-1}$ & ${\rm VI}_{-1}$ 
  \\
$\tau = {\rm III}_0$    & III$_0$ & III$_0$
 & $6{\rm I}_0 \oplus 6 {\rm II}_0 \oplus 4{\rm III}_0$
 & $3{\rm IV}_0 \oplus 3 {\rm V}_0$ 
 & $3{\rm IV}_0 \oplus 3 {\rm V}_0$ 
 & $3{\rm VI}_{-1} \oplus 3 {\rm VII}_{-1}$ 
 & $3{\rm VI}_{-1} \oplus 3 {\rm VII}_{-1}$ 
  \\
${\rm IV}_0$ &  ${\rm IV}_0$ & ${\rm V}_0$ 
 & $3{\rm IV}_0 \oplus 3 {\rm V}_0$ 
 & $3{\rm VI}_0 \oplus {\rm VII}_0$ 
 & ${\rm VI}_0 \oplus 3 {\rm VII}_0$ 
 & $4{\rm I}_0 \oplus 3 {\rm III}_0$ 
 & $4{\rm II}_0 \oplus 2 {\rm III}_0$ 
  \\
${\rm V}_0$ & ${\rm V}_0$ &  ${\rm IV}_0$
 & $3{\rm IV}_0 \oplus 3 {\rm V}_0$ 
 & ${\rm VI}_0 \oplus 3{\rm VII}_0$ 
 & $3{\rm VI}_0 \oplus {\rm VII}_0$ 
 & $4{\rm II}_0 \oplus 2 {\rm III}_0$ 
 & $4{\rm I}_0 \oplus 2 {\rm III}_0$ 
  \\ 
${\rm VI}_{-1}$ & ${\rm VI}_{-1}$ & ${\rm VII}_{-1}$
 & $3{\rm VI}_{-1} \oplus 3 {\rm VII}_{-1}$ 
 & $4{\rm I}_0 \oplus 2{\rm III}_0$ 
 & $4{\rm II}_0 \oplus 2{\rm III}_0$ 
 & $3{\rm IV}_{-1} \oplus {\rm V}_{-1}$ 
 & ${\rm IV}_{-1} \oplus 3 {\rm V}_{-1}$ 
  \\
${\rm VII}_{-1}$ & ${\rm VII}_{-1}$ & ${\rm VI}_{-1}$
 & $3{\rm IV}_{-1} \oplus 3 {\rm VII}_{-1}$ 
 & $4{\rm II}_0 \oplus 2{\rm III}_0$ 
 & $4{\rm I}_0 \oplus 2{\rm III}_0$ 
 & ${\rm IV}_{-1} \oplus 3 {\rm V}_{-1}$ 
 & $3{\rm IV}_{-1} \oplus 2 {\rm V}_{-1}$   \\
    \end{tabular}
    \caption{Fusion rule for vortex anyons 
    in a cyclic spin-2 BEC~\cite{PhysRevLett.123.140404}. 
    \red{The subscript on the conjugacy class denotes the $U(1)$ winding number $N$.}
    \label{tab:fusion-cyclic}}
    \centering
    \begin{tabular}{c||c|c|c|c|c|c|c}
  & I$_0$   & II$_0$ & III$_0$ &  IV$_0$ 
  & V$_0$ & VI$_0$ & VII$_0$ \\ \hline
 ${\bf 1} = {\rm I}_0$  & I$_0$   & II$_0$ & III$_0$ & IV$_0$ & V$_0$ & VI$_0$ & VII$_0$\\
 $\sigma = {\rm II}_0$  & II$_0$  & I$_0$ & III$_0$ & IV$_0$ & VI$_0$ & V$_0$ & VII$_0$ \\
$\tau_1 = {\rm III}_0$ & III$_0$ & III$_0$ 
 & $4{\rm I}_0 \oplus 4 {\rm II}_0 \oplus 4{\rm IV}_0$ 
  & 2III$_0$ & 2VII$_0$ & 2VII$_0$ 
  & $4{\rm V}_0 \oplus 4{\rm VI}_0$\\
$\tau_2 = {\rm IV}_0$    & IV$_0$ & IV$_0$& 2III$_0$ 
  & $2{\rm I}_0 \oplus 2 {\rm II}_0$
  & ${\rm V}_0 \oplus {\rm VI}_0$ 
  & ${\rm V}_0 \oplus {\rm VI}_0$ & 2VII$_0$ \\
${\rm V}_0$ & ${\rm V}_0$ & ${\rm VI}_0$
  & $2{\rm VII}_0$
  & ${\rm V}_0 \oplus {\rm VI}_0$
  & $2{\rm I}_1 \oplus {\rm IV}_1$
  & $2{\rm II}_1 \oplus {\rm IV}_1$
  & $2{\rm III}_1$ \\
${\rm VI}_0$ & ${\rm VI}_0$ & ${\rm V}_0$  
  & $2{\rm VII}_0$  
  & ${\rm V}_0 \oplus {\rm VI}_0$
  & $2{\rm II}_1 \oplus {\rm IV}_1$
  & $2{\rm I}_1 \oplus {\rm IV}_1$
  & $2{\rm III}_1$ \\
${\rm VII}_0$ & ${\rm VII}_0$ & ${\rm VII}_0$ 
  & $4{\rm IV}_0 \oplus 4 {\rm V}_0$
  & $2{\rm VII}_0$ & $2{\rm III}_1$ & $2{\rm III}_1$ 
  & $4{\rm I}_1 \oplus 4 {\rm II}_1 \oplus 4{\rm IV}_1$ \\
    \end{tabular}
    \caption{Fusion rule for for vortex anyons 
    in a $D_4$-BN spin-2 BEC 
    \cite{PhysRevLett.123.140404}.
    \red{The subscript on the conjugacy class denotes the $U(1)$ winding number $N$.}
    \label{tab:fusion-nematic}}
\end{table*}

\sub{Non-Abelian vortex anyons in $D_2$-BN}
\red{Let us discuss the simplest case. 
For $D_2$-BN liquid crystals 
or $D_2$-BN phases of  spin-2 BECs, the conjugacy class
is given in Eq.~(\ref{eq:conj-Q}).
We then define 
\beq
 {\bf 1} = \{ + {\bf 1}_2 \}, \quad
 \sigma = \{- {\bf 1}_2 \}, 
 \quad \tau = \{ \pm i \sigma_a \}
\eeq
with $a$ being either $x$, $y$, or $z$. 
The coefficient $N^c_{ab}$ counts how many ways 
anyons $a$ and $b$ fuse to an anyon $c$. 
The relations 
$(-i \sigma_a)(+i \sigma_a) = 
(+i \sigma_a)(-i \sigma_a) = 
+ {\bf 1}_2$ 
and 
$(-i \sigma_a)(-i \sigma_a) 
= (+i \sigma_a)(+i \sigma_a) 
= -{\bf 1}_2$ 
lead 
$N^{\bf 1}_{\tau\tau} = 2$ 
and 
$N^{\sigma}_{\tau\tau} = 2$, respectively.
Furthermore, $\tau$ anyons do not fuse to $\tau$ in this case and then $N^{\tau}_{\tau\tau} =0$.
We thus reach  
\begin{align}
   & \tau \otimes \tau 
    = 2 {\bf 1}_2 
    \oplus 2 \sigma,\nonumber\\
   & \tau \otimes \sigma = \tau, \quad
    \sigma \otimes \sigma = {\bf 1}, \quad
    x \otimes {\bf 1} = x,\label{eq:fusion-Q}
\end{align}
with $x \in \{{\bf 1}, \sigma ,\tau\}$. 
Since two \green{$\tau$-}anyons fuse to two different 
anyons ${\bf 1}$ and $\sigma$, 
$\tau$ anyons are non-Abelian anyons
Comparing this fusion rule 
with that of the Ising anyons in Eq.~(\ref{eq:Ising}), 
we find that 
these non-Abelian vortex anyons are similar to 
the Ising anyons. 
In fact, a $\tau$ aynon coincide with its anti-particle, 
and so similar to a Majorana fermion.
}

\red{
Gathering all $\tau_a = \{ \pm \green{i}\sigma_a\}$ ($a=x,y,z$)
together, we obtain the following fusion formula 
from the relation $(i\sigma_a)(i\sigma_b) = - \delta_{ab} 
+ \epsilon_{abc} (-i \sigma_c)$:
  \begin{align}
   & \tau_a \otimes \tau_b 
    = 2 \delta_{ab} {\bf 1} 
    \oplus 2 \delta_{ab} \sigma 
    \oplus 2 \epsilon_{abc} \tau_c
    \nonumber\\
   & \tau_a \otimes \sigma = \tau_a, \quad
    \sigma \otimes \sigma = {\bf 1}, \quad
    x \otimes {\bf 1} = x,\label{eq:fusion-Q2}
\end{align}  
with $x \in \{{\bf 1}, \sigma ,\green{\tau_x, \tau_y, \tau_z}\}$.
In this case, there are the three non-Abelian anyons $\tau_a$ 
coupled to each other.
}

\sub{Non-Abelian vortex anyons  in spin-2 BECs}

Non-Abelian vortex anyons  in spin-2 BECs 
were studied in \cite{PhysRevLett.123.140404}. 
For the cyclic phase in a spin-2 BEC,  
$\{{\bf 1}, \sigma ,\tau\}$ are defined by 
the conjugacy classes (I), (II) and (III) 
with $N=0$ in Eq.~(\ref{eq:conj-cyclic}).
Then, we have 
(see Table~\ref{tab:fusion-cyclic})
$N^{\bf 1}_{\tau\tau}=6, \;
N^{\sigma}_{\tau\tau}=6 , \;
N^{\tau}_{\tau\tau}=4$:
\begin{align}
   & \tau \otimes \tau 
    = 6 {\bf 1} 
    \oplus 6 \sigma \oplus 4 \tau,\nonumber\\
   & \tau \otimes \sigma = \tau, \quad
    \sigma \otimes \sigma = {\bf 1}, \quad
    x \otimes {\bf 1} = x,\label{eq:fusion-cyclic}
\end{align}
with $x \in \{{\bf 1}, \sigma ,\tau\}$.
\red{
The non-Abelian 1/3-quantum vortices in (IV)--(VII) 
are also non-Abelian anyons 
but one cannot restrict to the $N=0$ sector, 
and needs infinite numbers of anyons 
for a closed algebra.
}

The fusion rule for vortex anyons 
in the $D_4$-BN phase of a spin-2 BEC 
was also obtained by 
the conjugacy classes (I)--(IV) 
in Eq.~(\ref{eq:conj-nematic}) \cite{PhysRevLett.123.140404}.
In this case, $\tau$ in the cyclic phase 
is split into $\tau_1 = {\rm III}_0$ and $\tau_2 = {\rm IV}_0$.
Then, we have 
$N^{\bf 1}_{\tau_1\tau_1}=4, \;
N^{\sigma}_{\tau_1\tau_1}=4 , \; 
N^{\tau_1}_{\tau_1\tau_1}=4, 
N^{\bf 1}_{\tau_2\tau_2}=2, \;
N^{\sigma}_{\tau_2\tau_2}=2$ 
with the other components zero:
\begin{align}
   & \tau_1 \otimes \tau_1 
    = 4 {\bf 1} 
    \oplus 4 \sigma \oplus 4 \tau_1,\nonumber\\
  & \tau_2 \otimes \tau_2 
    = 2 {\bf 1} \oplus 2 \sigma ,\nonumber\\
   & \tau_i \otimes \sigma = \tau_i, \quad
    \sigma \otimes \sigma = {\bf 1}, \quad
    x \otimes {\bf 1} = x,\label{eq:fusion-D4}
\end{align}
with $i=1,2$ and $x \in \{{\bf 1}, \sigma ,\tau_i\}$.
See Table~\ref{tab:fusion-nematic}.
The non-Abelian HQVs in (V)--(VII) 
are also non-Abelian anyons,  
but again
 infinite numbers of anyons are necessary for a closed algebra.
The same fusion rule in Eq.~(\ref{eq:fusion-D4}) should hold for a $D_4$-BN liquid crystal, as seen in Eq.~(\ref{eq:D4-conj}).

Including Bogoliubov modes, 
the algebra of vortex anyons is extended to
a quantum double \cite{Koornwinder:1999bg,PhysRevLett.123.140404}.
An application  
to quantum computation was also proposed in Ref.~\cite{GenetayJohansen:2021udt}.


\section{$^3P_2$ topological SFs}\label{sec:3P2}
\green{Finally we discuss the two-fold non-Abelian object in a novel topological SF called a $^3P_2$ SF: The two-fold non-Abelian nature is attributed to the fermionic part (Sec.~3) and the vortex part (Sec.~4).} The \tPt SF is a condensate of 
spin-triplet ($S=1$) $p$-wave ($L=2$) Cooper pairs with total angular momentum $J=2$. In the following, we briefly review $^3P_2$ SFs in Sec.~\ref{sec:overview}. In Sec.~\ref{sec:hqv}, the stability of a pair of HQVs compared with a singly quantized vortex is discussed. In Sec.~\ref{sec:majorana-in-hqv}, we show that a Majorana fermion, a non-Abelian Ising anyon, exists in the core of each HQV.

\subsection{Overview of $^3P_2$ topological SFs}\label{sec:overview}
Study of $^3P_2$ SFs  was initiated since 1970s~\cite{Hoffberg:1970vqj,Tamagaki1970,Takatsuka1971,Takatsuka1972,Richardson:1972xn,Sedrakian:2018ydt}. 
On the basis of the analysis of the phase-shifts from nucleon-nucleon scattering in a free space, spin-singlet $s$-wave $^1S_0$ symmetry is a dominant attractive channel in the inner crust region ($\rho \lesssim 0.5 \rho_0$, where $\rho_0 = 0.17 \mathrm{fm}^{-3}$ is called the nuclear density)~\cite{Tamagaki1970}. The critical temperature was estimated as $10^{8} - 10^{10}$ K, which is two orders of magnitude smaller than the Fermi energy. 
For further increase in density $0.7\rho_0 \lesssim \rho \lesssim 3 \rho_0$ in the inner core region, the $^1S_0$ superfluidity is suppressed, and instead the attractive channel of the $^3P_2$ grows.  
Because of the attractively strong spin-orbit force, the $^3P_2$  Cooper pair channels with high angular momentum become more attractive than other $^3P$ pairing symmetry in contrast to the atomic physics where the lower total angular momentum state is favored as in the SF $^3$He-B phase. 
The anisotropic form of the Cooper pair is thought to affect the Cooling rate of nuetron stars.
Extreme conditions of neutron stars are not only high density, but also rapid rotation, and a strong magnetic field. Especially, neutron stars accompanied by magnetic field ranging from $10^{15}$ -- $10^{18}$ G are called magnetars.
Rich structures of exotic condensates and vortices explained below may be the origin of the observed {\it pulser glitches}, that is, the sudden increase of the angular momentum of neutron stars. Among them, the existence of non-Abelian HQVs in high magnetic fields is responsible for the scaling law of the glitches without any fitting parameters~\cite{Marmorini:2020zfp}.

\begin{figure}
\includegraphics[width=\hsize]{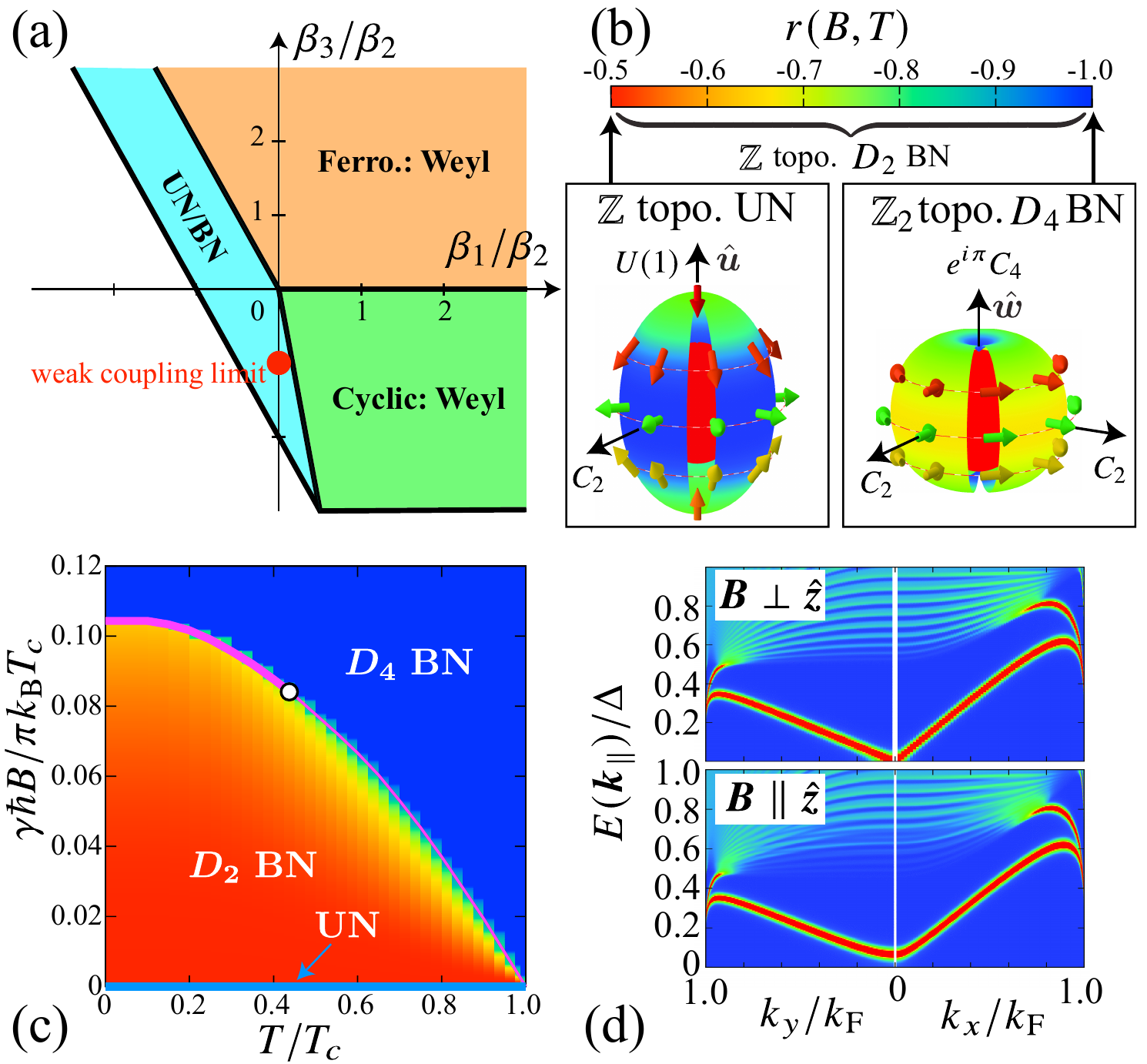}
\caption{(a) Phase diagram without magnetic field based on the GL theory. (b) Gap structure of the nematic state. (c) Phase diagram with magnetic field in the weak coupling limit. (d) The momentum resolved  local density state at the surface perpendicular to the $\hat{z}$ axis. The magnetic field is applied parallel (perpendicular) to the surface [top (bottom) panel]. Figures are adapted from Ref.~\cite{Mizushima:2016fbn} \copyright 2022 American Physical Society.} 
\label{fig:GL-tri}
\end{figure}

In terms of the symmetry point of view,  
\tPt SFs spontaneously break 
the global gauge symmetry $U(1)_\varphi$ and the simultaneous rotational symmetry in the spin-momentum space $SO(3)_J$ which originally exist in the normal state inside neutron stars. 
The order parameter of $^3P_2$ SFs is spanned by 3 by 3 traceless symmetric tensor $A_{\mu i}$, where the 2 by 2 gap function in momentum space can be given by
$\hat{\Delta}(\bm{k}) = A_{\mu i}\bar{k}_i\hat{\sigma}_{\mu}i\hat{\sigma}_y$. 
For later convenience, we introduce a representation based on the total angular momentum along the quantization axis, $M = -2,\cdots, 2$: 
\begin{align}
A_{\mu i} = \sum_{M = -2}^{2}\gamma_M [\Gamma_M]_{\mu i},
\end{align}
where $\gamma_M$ is a complex wave function of the Cooper pair in the angular momentum sector $M$. A basis set $\Gamma_M$ is given for the quantization axis $\hat{w}$ by
\begin{align} 
[\Gamma_{\pm2}]_{\mu i}
&= \frac{
(\hat{u} \pm i \hat{v})_\mu
(\hat{u} \pm i \hat{v})_i}{2},\\
[\Gamma_{\pm1}]_{\mu i}
&= \mp\frac{
(\hat{u} \pm i \hat{v})_\mu
\hat{w}_i
+ \hat{w}_\mu (\hat{u} \pm i \hat{v})_i
}{2},\\
[\Gamma_{0}]_{\mu i} &= \frac{
-\hat{u}_\mu\hat{u}_i
-\hat{v}_\mu\hat{v}_i
+2\hat{w}_\mu\hat{w}_i
}{\sqrt{6}},
\end{align}
where $\hat{u}$, $\hat{\varv}$, and $\hat{\varw}$ constitute the orthonormal triad. 

From a phenomenological aspect, 
the Ginzburg-Landau (GL) energy functional
invariant under the $SO(3)_J$ and $U(1)_\varphi$ symmetry was obtained for $^3P_2$ SFs as \cite{Richardson:1972xn,Fujita1972,Sauls:1978lna,Muzikar:1980as,Sauls:1982ie}
\begin{align}
\mathcal{F}
&= \alpha \mathrm{tr}[A A^*] 
+ \beta_1 |\mathrm{tr}A^2|^2 \nonumber \\
&\quad + \beta_2 (\mathrm{tr}[AA^*])^2
+ \beta_3 \mathrm{tr}[A^2A^{*2}].
\end{align}
It should be remarked that spin-2 spinor BECs ($L=0, S=2$) and $d$-wave superconductivty ($L=2, S=0$) have similar order parameter structures.  
The GL functional for the $^3P_2$ Cooper pairs was minimized by Sauls and Serene by using the correspondence to the $L = 2$ GL functional which was solved by Mermin~\cite{Sauls:1978lna, Mermin:1974zz}. The ground state solutions are classified into three types of phases determined by $\beta_{i = 1,2,3}$, as shown in Fig.~\ref{fig:GL-tri}(a): nematic phases with time reversal symmetry and the cyclic and ferromagnetic phases as non-unitary state. 
The microscopic derivation of the GL parameters clarifies that the ground state is the nematic phases in the weak coupling limit~\cite{Sauls:1978lna,Sauls1980}. The order parameter tensor of a nematic phase is given by 
$A_{\mu i} = \Delta[
\hat{u}_{\mu}\hat{u}_{i} 
+r
\hat{\varv}_{\mu}\hat{\varv}_{i}
-(1 + r)
\hat{\varw}_{\mu}\hat{\varw}_{i}
]$ with a real number $r \in [-1, -1/2]$ [see also Eq.~\eqref{eq:op-nematic}].
The nematic phase has a continuous degeneracy 
which is lifted by either a magnetic field or sixth-order terms
in the GL free energy. For the GL parameters derived from the microscopic model, the UN phase for $r = -1/2$
is favored at zero magnetic field while $D_2$-BN ($-1 < r < -1/2$)
and $D_4$-BN $r = - 1$ phases are favored for moderate and strong magnetic
fields, respectively~\cite{Masuda:2015jka}, relevant for magnetars. The schematic images of the gap structures for the UN and $D_4$-BN states are shown in Fig.~\ref{fig:GL-tri}(b), where the arrows account for the momentum-dependent direction of the $d$-vector, defined by $d_\mu(\bm{k}) = \sum_{i} A_{\mu i} k_{i}$. 
We will discuss other possible states later in the context of the fermionic topology.

The microscopic model of $^3P_2$ SFs in terms of the fermion degrees of freedom was constructed by Richardson~\cite{Richardson:1972xn} and Tamagaki and Takatsuka~\cite{Tamagaki1970,Takatsuka1971,Takatsuka1972}:
The starting microscopic Hamiltonian is composed of the one-body term $H_{1}$ and the interaction term $H_{2}$ given, respectively, in the following forms:
\begin{align}
H_{1} &= \int \dd \bm{r} \vec{\psi}^{\dagger}(\bm{r})[\hat{h}_{N}(-\ii \bm{\nabla})]\vec{\psi}(\bm{r}),\\
H_{2} &= - \int \dd \bm{r} \sum_{\alpha\beta= x,y,z}\dfrac{g}{2}T_{\alpha\beta}^{\dagger}(\bm{r})T_{\alpha\beta}(\bm{r}).
\end{align}
In the first line, $[\vec{\psi}(\bm{r})]_{\sigma} = \psi_{\sigma}(\bm{r})$ and  $[\hat{h}_{N}(-i\bm{\nabla})]_{\sigma,\sigma^{\prime}}$ is the sum of the kinetic energy $h_{0}(-\ii \bm{\nabla})\delta_{\sigma,\sigma^{\prime}} = (-\nabla^{2}/2m - \mu)\delta_{\sigma,\sigma^{\prime}}$ measured from the chemical potential $\mu$, and the Zeeman energy $-\bm{B}\cdot\hat{\bm{\sigma}}$ 
for a magnetic field $\bm{B} = B \bm{n}$. The $^{3}P_{2}$ force with coupling strength $g$ $>$ $0$ is represented by $H_{2}$ with the pair-annihilation operator $T$ defined by
$T_{\alpha\beta}(\bm{r})$ $=$ $\sum_{\sigma\sigma^{\prime}}[t_{\alpha\beta,\sigma\sigma^{\prime}}(-\ii \bar{\bm{\nabla}})\psi_{\sigma^{\prime}}(\bm{r})]\psi_{\sigma}(\bm{r})$, where $\bar{\bm{\nabla}}\equiv k_\F^{-1}\bm{\nabla}$. 
We have introduced a spin-momentum coupling in the pair force via the 2 $\times$ 2 matrix in spin space $\hat{t}_{\alpha\beta}$ defined by
$\hat{t}_{\alpha\beta}(-\ii \bar{\bm{\nabla}})
$ $=$ $\ii \hat{\sigma}_{y}
\left\{\left[
\hat{\sigma}_{\alpha}(-\ii \bar{\nabla}_{\beta})
+ \hat{\sigma}_{\beta}(-\ii\bar{\nabla}_{\alpha})\right]/2\sqrt{2}-\delta_{\alpha\beta}\hat{\bm{\sigma}}\cdot(-\ii \bar{\bm{\nabla}})/3\sqrt{2}
\right\}$.
Within the mean field approximation, the BdG equation is derived as,
\begin{align}
&\check{\mathcal{H}}_{}(-i\bm{\nabla},\bm{R})\vec{u}_{\nu}(\bm{R}) = \epsilon_{\nu} \vec{u}_{\nu}(\bm{R}),\\
&\check{\mathcal{H}}_{}(-i\bm{\nabla},\bm{R}) =
\begin{pmatrix}
\hat{h}_{N}(-i\bm{\nabla}) & \hat{\Delta}(-i\bm{\nabla}, \bm{R}) \\
-[\hat{\Delta}(
-i\bm{\nabla},\bm{R})]^* & -[\hat{h}_{N}(-\ii \bm{\nabla})]^*
\end{pmatrix},
\label{eq:bdg}
\end{align}
where the gap matrix is given by
$\hat{\Delta}(-i\bm{\nabla},\bm{R}) = \sum_{\alpha,\beta}\frac{i \hat{\sigma}_{\alpha}\hat{\sigma}_{y}}{2k_{\F}}\{2A_{\alpha\beta}(\bm{R})(-i\nabla_{\beta})
+ [(-i\nabla_{\beta})A_{\alpha\beta}(\bm{R})]\}$.
This model was also used to determine the aforementioned GL parameters in the weak coupling limit and was directly investigated within a quasiclassical approximation recently in Ref.~\cite{Mizushima:2016fbn}. 
Even in the presence of the magnetic field, the quasiclassical approximation where the size of the Fermi surface is treated as an infinite one, predicts that a nematic state is the most stable with parameter $r$ determined by the strength of the magnetic field as in the GL theory. \green{The $T$-$B$ phase diagram is shown in Fig.~\ref{fig:GL-tri}(c). As explained later, the ferromagnetic state appears near $T_c$ when the finite-size correction of the Fermi surface is taken into account.}
The microscopic analysis reveals the existence of the tricritical point on the phase boundary between the $D_4$-BN and $D_2$-BN states in the $T$-$B$ phase diagram shown in Fig.~\ref{fig:GL-tri}(c): at temperatures below the tricritical point, the phase transition becomes discontinuous~\cite{Mizushima:2016fbn}. 
The existence of the tricritical point was later confirmed in 
Ref.~\cite{Mizushima:2019spl}
also in the GL theory with higher order terms (up to the eighth order \cite{Yasui:2019unp}).
In addition to such drastic change in critical phenomena, the advantage of the microscopic model lies in studies of the fermionic topology of the superfluidity. In the following, we explain the fermionic topology of the possible phases of the $^{3}P_2$ SFs.

\sub{Nematic state} In the weak coupling limit without a magnetic field, the UN state is the ground state. 
The order parameter tensors of the UN state and the BN state under a magnetic field are already explained above. 
On the basis of the symmetry of the BdG Hamiltonian, the nematic state is revealed as a topological SF with time reversal symmetry (a
class DIII in the classification of topological insulators and
SCs). The analysis of the BdG equation in the presence of the boundary clarifies the existence of the gapless surface bound Majorana fermion, the hallmark character of the topological states. 
The surface Majorana fermion has an Ising spin character~\cite{Chung2009,Nagato2009,Volovik2010,Mizushima2011}, that is, the only external field coupled to the Ising spin gives a mass gap to the Majorana fermion as shown in Fig.~\ref{fig:GL-tri}(d). The magnetic field direction giving the gap to the Majorana fermion is the direction perpendicular to the surface, denoted by $\bm{n}_{\perp}$. This is because the surface Majorana fermion is protected not only by the time reversal symmetry but also another  key symmetry, which is the magnetic $\pi$-rotation symmetry about $\bm{n}_{\perp}$. The magnetic $\pi$-rotation symmetry is also called the $\bar{P}_3$ symmetry\footnote{
Here we use the bar to distinguish the magnetic $\pi$-rotation about the $x$-axis introduced later. 
}. The magnetic $\pi$-rotation is the combined operation of the time-reversal and the $\pi$ rotation. The chiral operator, defined by the combination of particle-hole operation $\mathcal{C}$ and the magnetic $\pi$-rotation $\bar{P}_3$ as $\Gamma = \mathcal{C}\bar{P}_3$, commutes with the Hamiltonian $\mathcal{H}(\bm{k}_{\perp})$ with magnetic field $\bm{B}\cdot \bm{n}_{\perp} 
= 0$ and the chiral symmetric momentum $\bm{k}_{\perp} = k_{\perp}\bm{n}_{\perp}$ such that $\Gamma:\bm{k}_\perp\to \bm{k}_\perp$. Using this commutation relation, a one-dimensional winding number can be introduced as $w_{\mathrm{1d}} = -\frac{1}{4\pi i}\int dk_{\perp} \mathrm{Tr}[\Gamma \mathcal{H}(\bm{k}_\perp)\partial_{k_\perp}\mathcal{H}(\bm{k}_\perp)] = 2$, which is unchanged unless the symmetry is broken or the bulk gap is closed. The $P_3$ symmetry and thus the chiral symmetry is broken by the magnetic filed $\bm{B}\cdot \bm{n}_{\perp}\neq 0$ which gives a mass gap to the Majorana fermion (the bottom panel of Fig.~\ref{fig:GL-tri}(d), where $\bm{n}_{\perp}\parallel \hat{z}$).

\sub{Cyclic state} For $-6\beta_1 < \beta_3 < 0$ based on the basis of $SO(3)_J$ invariant GL functional, the cyclic state is the ground state. 
The order parameter tensor has a form of
$A_{\mu i} = \Delta[\hat{u}_{\mu}\hat{u}_i + \omega\hat{\varv}_{\mu}\hat{\varv}_i + \omega^2\hat{\varw}_{\mu}\hat{\varw}_i]$ with $\omega = e^{2\pi i/3}$ [see also Eq.~\eqref{eq:cyclic}], which is unique except for trivial $SO(3)_J$ and $U(1)_\varphi$ rotations.
In the absence of magnetic field, the quasiparticle excitation energy consists of two branches $E_{\pm }(\bm{k}) = [h_0(\bm{k})^2 + |\bm{d}(\bm{k})|^2 \pm |\bm{d}(\bm{k})\times \bm{d}^*(\bm{k})|]^{1/2}$, where $E_+(\bm{k})$ is full gap and $E_-(\bm{k})$ has 8 nodal points $\pm \bm{k}_{\alpha = 1,\cdots,4}$, each of which is a Weyl point. The subscript $\alpha$ denotes the 4 vertices of a tetrahedron, and $\pm$ represents the monopole charge of the Weyl point. 
Because of the topological nature of Weyl SCs/SFs, there exist surface zero energy states which connect two oppositely charged Weyls points on the projected momentum space normal to the surface direction. 

\sub{Ferromagnetic state} For $|\beta_1| - \beta_1 < \beta_3$, the ferromagnetic state described by $A_{\mu i} = \Delta (\hat{u}_{\mu} + i \hat{\varv}_{\mu}) (\hat{u}_i + i \hat{\varv}_i)$ minimizes the GL functional. This non-unitary state is also unique except for trivial $SO(3)_J$ and $U(1)_\varphi$ rotations. The order parameter is equivalent to the $A_1$ state of the SF $^{3}$He. The bulk quasiparticle spectrum  consists of two parts: one is a normal fluid, and the other is a Weyl SF with a pair of oppositely charged Weyl fermions. The ferromagnetic state is also realized in the limit of a strong magnetic field or the vicinity of the critical temperature even in the weak coupling limit owing to the finite-size correction of the Fermi surface.

\sub{Magnetized nematic state}
Recently, even in the weak coupling limit, non-unitary states are predicted in a magnetic field~\cite{Mizushima2021}. As mentioned above, quasiclassical approximation, which treats the size of the Fermi surface as infinity, allows only the nematic states, which is unitary. 
By contrast, in the presence of the finite-size correction of the Fermi surface, a magnetic field along the $\hat{w}$-direction induces the non-unitary component into the nematic state with $r\in(-1, -1/2)$ as
$A_{\mu i} = \Delta[
\hat{u}_\mu\hat{u}_i + 
r \hat{v}_\mu \hat{v}_i
+ i \kappa (
\hat{u}_\mu\hat{v}_i + 
\hat{v}_\mu \hat{u}_i)
- (1 + r)\hat{w}_{\mu}\hat{w}_i
]$ or equivalently
\begin{align}
A_{\mu i} = \Delta
\begin{pmatrix}
1 & i\kappa & 0 \\
i\kappa &  r & 0 \\
0 & 0 & -(1 + r)
\end{pmatrix}. \label{eq:Mag.BN}
\end{align}
For simplicity, Ref.~\cite{Mizushima2021}
studies the case of $r = - 1$, i.e., in a strong magnetic field. In the quasiclassical limit, $\kappa = 0$ and the $D_4$-BN state is realized. The order parameter tensor in the angular momentum representation is reduced to 
$A_{\mu i} = \Delta [\Gamma_{2} + \Gamma_{-2}]_{\mu i}$, which shows the same gap amplitudes for $M = \pm 2$. The finite size correction of the Fermi surface induces a finite $\kappa$. In the angular momentum representation, the order parameter tensor becomes 
$A_{\mu i} = \Delta
[(1 + \kappa)\Gamma_2
+ 
(1 - \kappa)\Gamma_{-2}
]_{\mu i}
$. Thus, $\kappa$ represents the imbalance between the $M = \pm 2$ sectors. In terms of the  gap matrix in the spin basis, 
the Cooper pairs are decomposed into two spin polarized sectors $\ket{\uparrow\uparrow}$ and $\ket{\downarrow\downarrow}$ with different gap amplitudes. 
Each spin sector has a polarized orbital angular momentum state, and thus, a pair of the Weyl fermions appears at the north pole and the south pole in each spin sector. The orbital angular momentum between these two spin states are opposite, which means that two Weyl fermions with opposite helicity at each pole. This is in contrast to the $A_2$ phase proposed for the SF $^3$He.
In the limit of a strong magnetic field \green{or the vicinity of the critical temperature}, $\kappa \to 1$, the ferromagnetic state is realized. However, we note that this imbalance originates from the finite size effect of the Fermi surface. The effect is parametrized by $\Delta/\varepsilon_{\F}$, which is usually small for conventional neutron stars, and thus the imbalance $\kappa$ is also expected to be small. 

\sub{\green{Vortices in $^3P_2$ SFs}}
\green{The vortices admitted in $^3P_2$ SFs are basically the same as those in spin-2 BECs, which are discussed in Sec.~\ref{sec:vortex-examples}.}  
Table~\ref{tab:3p2-phase-symmetry-f.group} summarizes  remaining symmetry, order parameter manifold, and fundamental group of the topological defects of the above-discussed possible phases. 
\green{Non-Abelian vortex anyons are included in the $D_2$- and $D_4$-BN states and the cyclic state as discussed in Sec.~\ref{sec:vortex-fusion}}. 
\green{
Non-Abelian vortex anyons in the $D_2$-BN states are spin vortices of $\pi$ rotation, as shown in Eq.~\eqref{eq:fusion-Q2}. The $D_4$-BN states admit HQVs (V)--(VII) in Eq.~\eqref{eq:conj-nematic} in addition to spin vortices (III) and (IV) in Eq.~\eqref{eq:conj-nematic} as non-Abelian vortex anyons (see also Table~\ref{tab:fusion-nematic}). 
} The cyclic states admit 1/3-quantum vortices (IV)--(VII) in Eq.~\eqref{eq:conj-cyclic} as well as spin vortices (III) in Eq.~\eqref{eq:conj-cyclic} as non-Abelian vortex anyons (see also Table~\ref{tab:fusion-cyclic}).

Especially, the non-Abelian vortex in the $D_4$-BN state is the main target in the remaining part. We explain the vortices in the $D_4$-BN state in the beginning of the next subsection. 
\begin{table*}[t]
 \caption{Summary of order parameters ($r,\kappa$ in Eq.~\eqref{eq:Mag.BN}), remaining symmetry ($H$), ($R \simeq G/H$), and topological vortices $\pi_1(R)$ in possible phases, taken from Ref.~\cite{Mizushima2021}. See also the references therein. }
 \label{tab:3p2-phase-symmetry-f.group}
 \centering
\begin{tabular}{ccccc}
\hline
\hline
phase & $r$ \& $\kappa$ in Eq.~\eqref{eq:Mag.BN} & $H$ & $R \simeq G/H$ & $\pi_1(R)$\\
\hline
UN & $r = -1/2$ \& $\kappa = 0$ & $D_\infty \simeq O(2)$ & $U(1)\times \mathbb{R}P^2$ & $\mathbb{Z}\oplus\mathbb{Z}_2$
\\
\hline
$D_2$-BN & $r\in(-1,-1/2)$ \& $\kappa = 0$ & $D_2$ & $[U(1)\times SO(3)]/D_4$ & $\mathbb{Z}\oplus\mathbb{Q}$
\\
$D_4$-BN &$r = -1$ \& $\kappa = 0$ & $D_4$ & $[U(1)\times SO(3)]/D_4$ & $\mathbb{Z}\times_h D_4^*$
\\
\hline
Cyclic & $r = e^{i2\pi/3}$ \& $\kappa = 0$ & $T$ & $[U(1)\times SO(3)]/T$ & $\mathbb{Z}\times_h T^*$
\\
\hline
Mag. $D_2$-BN & $r \in (-1,-1/2)$ \& $\kappa \in (0,1)$ & 0 & $U(1)\times SO(3)$ & $\mathbb{Z} \oplus\mathbb{Z}_2$ \\
Mag. $D_4$-BN & $r = -1$ \& $\kappa = -1$  & $C_4$ & $[U(1) \times SO(3)]/\mathbb{Z}_4$& $\mathbb{Z} \times_hC_4^*$ \\
\hline
FM  & $r = - 1$ \& $\kappa = 1$ & $U(1)_{J_z + 2\Phi}$ & $SO(3)_{J_z - 2\Phi}/\mathbb{Z}_2$& $\mathbb{Z}_4$ \\
\hline
\end{tabular}
\end{table*}

\begin{figure}
\includegraphics[width=\hsize]{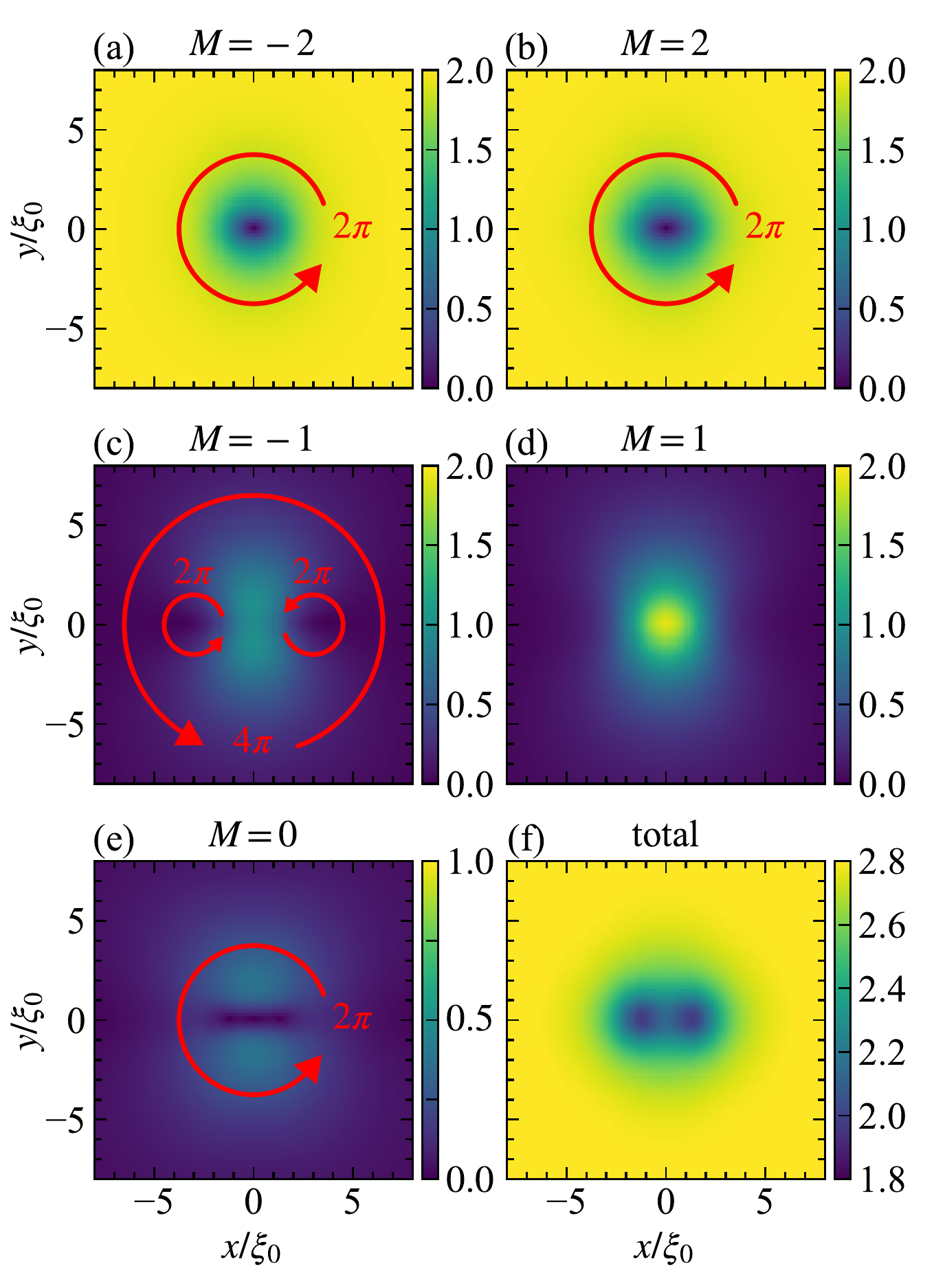}
\caption{
        Gap structure of the $d$ vortex in the $D_4$-BN state ($T=0.4 T_c$ and $B = 0.5 T_c)$. (a)-(e) Amplitude of each angular momentum sector. Red arrows indicate the phase winding structures. (f) Total amplitude of the order parameter defined by $[\frac{3}{2}\mathrm{Tr}\braket{\hat{\Delta}^\dagger\hat{\Delta}}_{\F}]^{1/2}$.
        }
\label{fig:d-vortex}
\end{figure}
\subsection{Half quantum vortex in $D_4$-BN state}
\label{sec:hqv}
\sub{Vortices in $D_4$-BN state}
The $D_4$-BN state, which can be thermodynamically stabilized by a large magnetic field, has a pair of point nodes at the north and south poles along the direction of the magnetic field ~\cite{Masuda:2015jka,Mizushima:2016fbn,Yasui:2018tcr,Mizushima:2019spl}. 
Hereafter, we focus on the $D_4$-BN state with a pair of point node along $\hat{w} = \hat{z}$. The homogeneous order parameter tensor has a diagonal form~\cite{Sauls:1978lna}:
$A = \Delta\mathrm{diag}(1, -1, 0)$ [Eq.~\eqref{eq:op-nematic}],
which is invariant under a $D_4$ group. 
(A similar order parameter form is taken in the planar state of SF $^3$He \cite{Makhlin2014,Silaev2014}, but different topological defects are resulted from different order parameter manifolds.)

\begin{figure*}
\includegraphics[width=\hsize]{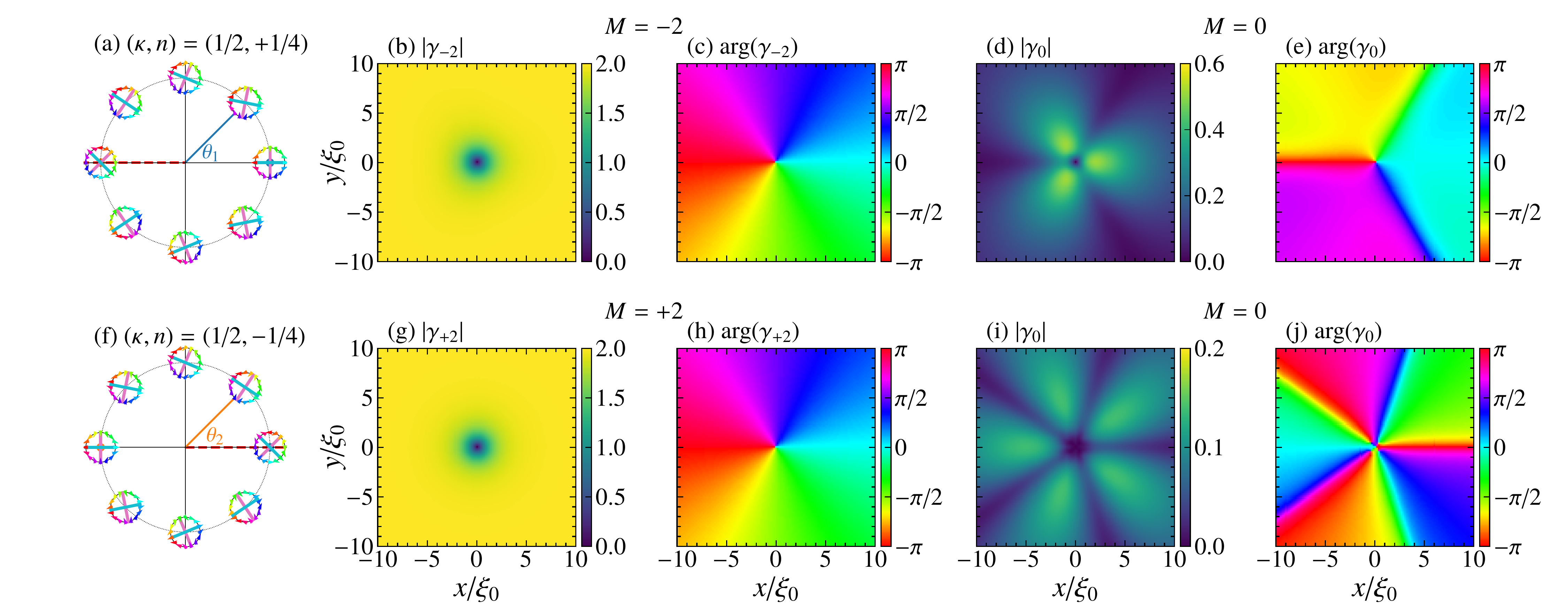}
\caption{HQV with $(\kappa,n) =(1/2, +1/4)$ 
        [(a) - (e)] and $(\kappa,n) =(1/2, -1/4)$ 
        [(f) - (j)]. The boundary conditions are sketched in (a) and (f). The amplitudes of $\gamma_M$ for $M = -2$ and $0$ ($+2$ and $0$) are shown in (b) and (d) [(g) and (i)], respectively. The phases of $\gamma_M$ for $M = -2$ and $0$ ($+2$ and $0$) are shown in (c) and (e) [(h) and (j)], respectively.
        Figures are adapted from Ref.~\cite{Masaki2022} \copyright 2022 American Physical Society.
        }
\label{fig:1/4}
\end{figure*}

\begin{figure*}
\begin{center}
\includegraphics[width=\hsize]{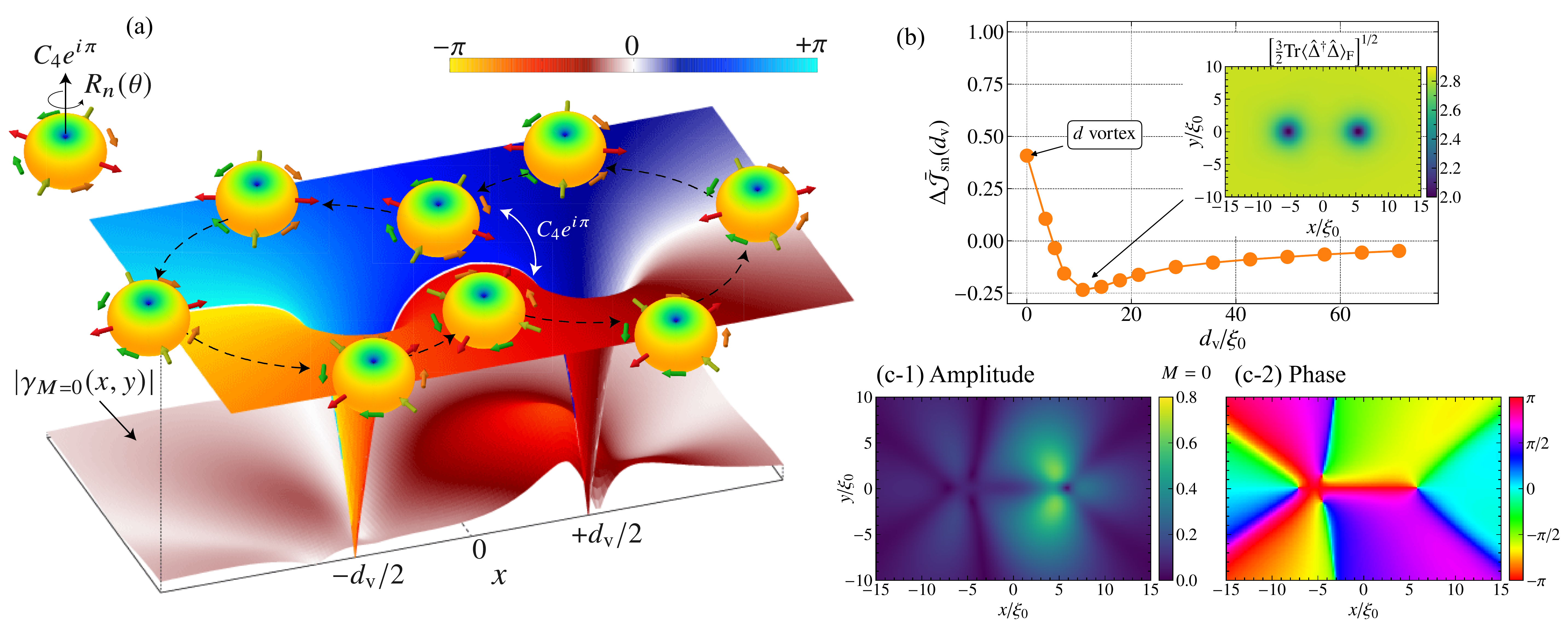}
\caption{
(a) Schematic image of a molecule of non-Abelian HQVs in a $D_{4}$-BN state at a cross section perpendicular to the two parallel vortex lines, characterized by $(\kappa,n)=(1/2,+1/4)$ at $x=d_{\mathrm{v}}/2$ and $(1/2,-1/4)$ at $x=-d_{\mathrm{v}}/2$. 
Its spin-momentum structure is shown by objects with color arrows representing $d$ vectors. The color map on the surface shows the $U(1)$ phase, and the bottom plot shows the induced component also shown in panel (c-1). 
(b) Interaction energy of two HQVs as a function of their separation $d_{\mathrm{v}}$. 
The inset shows the total amplitude of the order parameter 
for the HQV molecule whose intervortex distance is indicated by the arrow. The free energy is scaled as $\bar{\mathcal{J}}_{\mathrm{sn}} = \mathcal{J}_{\mathrm{sn}}/(\nu_{\n} T_{\mathrm{c}}^{2} \xi_{0}^{2} \Omega_{z})$, where $\Omega_{z}$ is the length of the system in the $z$ direction, and $\nu_{\n}$ is the density of states at the Fermi energy in the normal state. The energy at $d_{\mathrm{v}} = 0$ is the free energy of the $d$ vortex shown in Fig.~\ref{fig:d-vortex}. (c-1) and (c-2) The amplitude and the phase of the induced component $\gamma_0$. The intervortex distance $d_{\mathrm{v}}$ is the same as that in the inset of panel (b). Figures are adapted from Ref.~\cite{Masaki2022} \copyright 2022 American Physical Society.
}
\label{fig:hqv_all}
\end{center}
\end{figure*}

As already discussed in Sec.~\ref{sec:vortex}, the order parameter manifold in the $D_{4}$-BN state is then characterized by the broken symmetry, $R \simeq [U(1)\times SO(3)]/D_{4}$, see Eq.~\eqref{eq:OPS-UNBN}. The topological charges of line defects are characterized by the first homotopy group, Eq.~(\ref{conj-1})
\begin{align}
\pi_{1}(R) \simeq  \mathbb{Z}\times_{h} D_{4}^{\ast}
\end{align}
where $\times_{h}$ denotes a product 
defined in Ref.~\cite{Kobayashi:2011xb}.
\green{This ensures two different classes of topological line defects: Vortices with commutative topological charges and vortices with non-commutative topological charges. The former includes integer vortices 
\red{(I) in Eq.~(\ref{eq:conj-nematic})} 
with or without internal structures, while an example of the latter is the non-Abelian HQV, 
\red{(V)--(VII) in Eq.~(\ref{eq:conj-nematic}).}} Integer vortices are characterized by vorticity  quantized to an integer number because of the singlevaluedness of the order parameter. 
Fractionally quantized case such as a HQV usually has a phase jump ($\pi$-phase jump for a HQV) along the contour surrounding the vortex. 
In the $D_{4}$-BN state, however, the phase discontinuity is compensated by the phase originating from the discrete rotation of the $D_4$ symmetry. Thereby HQVs are topologically allowed. Here, we particularly focus on the case where the vorticity is along the $\hat{z}$-direction accompanying the $\pi$-phase jump compensated by the $C_4$ rotation about the $\hat{z}$-axis, as indicated in Figs.~\ref{fig:GL-tri}(b), \ref{fig:1/4}(a), \ref{fig:1/4}(f), and \ref{fig:hqv_all}(a).
Such HQVs belong to the group (V) or (VI) in Eq.~\eqref{eq:conj-nematic}.

Hereafter, we review possible vortices when the vorticity and the point nodes of the $D_4$-BN state are along the $\hat{z}$-direction. 
As a basis set of the order parameter tensor $A_{\mu i}$, the eigenstates of the $z$ component of the total angular momentum, denoted by $J_z\Gamma_M = M \Gamma_M$, are convenient: $A_{\mu i}(\rho,\theta) = \sum_{M=-2}^{2} \gamma_{M}(\rho,\theta) [\Gamma_{M}]_{\mu i}$, where the cylindrical coordinate system $\bm{R} = (\rho, \theta)$ is introduced with assumption of uniformity along the $z$-direction. 
In the region far from the vortex core $\rho\to \infty$, there is no radial dependence, and 
the order parameter modulation is described by the $U(1)$ phase rotation characterized by vorticity $\kappa$ and the simultaneous rotation in the spin-orbit space of Cooper pairs. For the latter rotation by angle $\varphi$ about the $\hat{z}$-axis, the other components of the triad, $\hat{x}$, and $\hat{y}$, are transformed as
\begin{align}
R(\varphi): \hat{x} &\to \cos \varphi \hat{x} + \sin \varphi \hat{y}, \\
R(\varphi): \hat{y} &\to -\sin \varphi \hat{x} + \cos \varphi \hat{y}.
\end{align}
Using $\Gamma_M$, the $U(1)$ phase winding and the spin-orbit rotation for non-zero bulk components \green{(i.e., $\rho\to\infty$)} are  expressed as
\begin{align}
A_{\mu i}(\theta) 
&= \sum_{M=-2}^2 \gamma_{M,\infty} e^{i\kappa\theta - M \varphi}[\Gamma_M]_{\mu i}     \nonumber \\
&\equiv \sum_{M=-2}^2 \gamma_{M,\infty} e^{i (\kappa - nM) \theta}[\Gamma_M]_{\mu i}. \label{eq:bc}
\end{align}
Especially, in the $D_4$-BN state with the point nodes along the $\hat{z}$-axis, $|\gamma_{2,\infty}| = |\gamma_{-2,\infty}|$ and $\gamma_{\pm1,\infty} = \gamma_{0.\infty} = 0$. Here in the second line, we parametrize $\varphi = n \theta$ along the contour surrounding the vortex, where the spin-orbit rotation can be regarded as the $n$-fold $J$-disgyration\footnote{
\green{The term ``$J$-disgyration" accounts for a singularity in the spin-orbit space, i.e., the total angular momentum ($J$) space. Spin disgyrations and orbital disgyrations are introduced after Eq.~\eqref{eq:z4-topo-charges}. }
}. Note that $n$ is not necessarily an integer. 
The boundary condition of a vortex characterized by $(\kappa,n)$ is given by Eq.~\eqref{eq:bc}. 

\sub{Axisymmetric singly quantized vortex}
The integer vortices belong to a class $\kappa \in \mathbb{Z}$. Even in the singly quantized case $\kappa = \pm 1$, however, there are many possibilities because of differences in the internal structure and the disgyration in the spin-orbit space. As a simple case, the axisymmetic vortex can be described as $A_{\mu i}(\rho,\theta) = \sum_{M=-2}^2 \gamma_M(\rho)e^{i(\kappa - M)\theta}[\Gamma_M]_{\mu i}$ in a whole region, which includes complex radial functions $\gamma_{M}(\rho)$ to be determined. 
\green{The disgyration of the axisymmetric vortex is 1-fold ($n=1$ in Eq.~\eqref{eq:bc}), that is, the $2\pi$ rotation of the triad around the vortex. Thus, the axisymmetric vortex belongs to the conjugacy class (II) in Eq.~\eqref{eq:conj-nematic}\footnote{
\green{In Sec.~\ref{sec:vortex-examples}, 
we discuss the spin vortex of $2\pi$ rotation in spin-2 BECs.}}.} 
Although the loss in the kinetic energy becomes large because of the $J$-disgyration, there is a nice analogy with vortices in SF $^3$He-B phase. As in the SF $^3$He-B phase, the symmetry of the vortices can be characterized by three discrete symmetries called $P_{1,2,3}$ symmetries~\cite{sal85,salomaaRMP}, which are discussed in the next subsection\footnote{
The $P_3$ symmetry is already introduced when defining the one-dimensional winding number in the bulk nematic state. 
}. 
The internal structures $\gamma_{-1,0,1}(\rho)$ characterize a classification based on these discrete symmetries, which allows several vortices as summarized in Table~\ref{tab:axisym-vortex}.
Within an axisymmetric condition, only $o$ vortex or $v$ vortex is realized as a self-consistent solution. The $o$ vortex has a singular core and is the most symmetric one so that it meets all the discrete symmetries. By contrast, the core of the axisymmetric $v$ vortex is occupied by the SF component $\gamma_{\pm1}$ for $\kappa =  \pm1$, and it has the $P_2$ (magnetic mirror reflection) symmetry, but does not have the $P_1$ (inversion) or $P_3$ (magnetic $\pi$-rotation about the $x$ axis) symmetries. Energetically, the $v$ vortex is more stable than the $o$ vortex because of a gain in the condensation energy in the core region. However, the induced component of the $v$ vortex which occupies the core is suppressed by a magnetic field .

\begin{table}[h]
    \centering
    \begin{tabular}{c|c|c|c|c}
    vortex & $\gamma_{M=0}(\rho)$ & $\gamma_{M=\pm1}(\rho)$ & core & symmetry\\ \hline
    $o$ vortex & real & zero & singular & $P_1$, $P_2$, $P_3$\\
    $u$ vortex & complex & zero & singular & $P_1$ \\
    $v$ vortex & real & real & coreless & $P_2$\\
    $w$ vortex & real & imaginary & coreless & $P_3$\\
    $uvw$ vortex & complex & complex  & coreless & -
    \end{tabular}
    \label{tab:axisym-vortex}
    \caption{Internal structures of axisymmetric vortices. The second and third columns show order parameters of the core structure. The fourth column is coreless (singular) whether the vortex core is occupied (unoccupied) by some SF components. The last column accounts for the preserved symmetry.
    }
\end{table}

\sub{Non-axisymmetric singly quantized vortex}
Without the axisymmetric condition, the lower energy condition for a singly quantized vortex is given by $(\kappa,n) = (\pm1, 0)$ to reduce the loss of the gradient energy. 
Such a vortex belongs to the conjugacy class (I) in Eq.~\eqref{eq:conj-nematic}.
In this case, so called double-core vortex ($d$ vortex) can be a self-consistent solution~\cite{Masaki2022} as in the SF $^3$He-B phase~\cite{thunebergPRL86,Salomaa1986}. 
The $d$ vortex is also called the non-axisymmetric $v$ vortex because it has the $P_2$ symmetry but does not have the $P_1$ and $P_3$ symmetries~\cite{tsutsumiPRB15}. Note that the axial symmetry is spontaneously broken for the $d$ vortex in the SF $^3$He-B phase, while the symmetry is broken by the boundary condition for the $d$ vortex in a \tPt SF. 
Figure~\ref{fig:d-vortex} shows the order parameter structure of the $d$ vortex in the $D_4$-BN state for $T=0.4 T_c$ and $B = 0.5 T_c$ with critical temperature $T_c$. In the panels, the unit of length is the coherence length defined by $\xi_0 = v_{\F}/2\pi T_c$ with Fermi velocity $v_\F$, where $\braket{\cdots}_{\F}$ accounts for the Fermi surface average. The bulk components $\gamma_{\pm2}$ have a conventional vortex structure with a single phase winding [see panels (a) and (b)]. The double core structure can be clearly observed in the total amplitude defined by $[\frac{3}{2}\mathrm{Tr}\braket{\hat{\Delta}^\dagger\hat{\Delta}}_{\F}]^{1/2}$ in panel (f). Such a structure is due to the occupations of the core by $\gamma_{M = \pm1}$ as shown in panels (c) and (d).
As can be seen below, the $d$ vortex is not the lowest energy state among the vortex states with $(\kappa,n)=(\pm1,0)$ in the $D_4$-BN state and it splits into two HQVs. The $d$ vortex in the SF $^3$He-B phase is the most stable vortex solution in the low pressure region~\cite{kasamatsuPRB19,Regan2019}. 
A $d$ vortex is realized as the lowest energy state in the UN and the $D_2$-BN states.

\sub{HQVs}
The HQVs can be characterized by $\kappa = \pm 1/2$. The possibility of the half-quantized vortex in the $D_4$-BN state was pointed out long back in 1980~\cite{Sauls1980} on the basis of the topological consideration of the form of $A_{\mu i}$. Since $\kappa$ is a half odd integer, the $U(1)$ phase part gives the minus sign under the spatial rotation $\theta \to \theta + 2\pi$.
By rotating the triad by angle $\pm \pi/2$ while going around the vortex, which is denoted by  $n = \pm1/4$ in Eq.~\eqref{eq:bc}, the above minus sign can be compensated by the other minus sign stemming from the disgyration. This topological consideration is represented by the boundary condition Eq.~\eqref{eq:bc} with $(\kappa, n) = (\pm 1/2, \pm 1/4)$, which belongs to the conjugacy class (V) or (VI) in Eq.~\eqref{eq:conj-nematic}.
The gradient energy of an isolated HQV far from the core is a half of that of an singly quantized vortex described by $(\kappa,n) = (\pm1,0)$, namely, the energy of two HQVs is the same as that of the singly quantized vortex except for contributions from the vortex cores.  
In other words, whether the singly quantized vortex can split into two HQVs depends on the energy of their internal structures and the interaction energy between two HQVs.

Although the above topological consideration does not take into account the core structure, recently, the whole structures of isolated HQVs were studied using the phenomenological GL theory~\cite{Masuda:2016vak,Kobayashi:2022moc,Kobayashi:2022dae}, and the microscopic quasiclassical theory~\cite{Masaki2022}. It is proposed that an integer vortex with $\kappa = 1$ can split into two HQVs with $(\kappa,n) = (1/2, 1/4)$ and $(\kappa,n) = (1/2, -1/4)$~\cite{Masuda:2016vak}. Here we assume that $\kappa > 0 $ without loss of generality. 
The configuration of the two HQVs given as one for $n_1 = +1/4$ with its core at $\bm{R}_1$ and the other for $n_2 = -1/4$ with its core at $\bm{R}_2$ has the following boundary condition at $\bm{R}$:
\begin{align}
A(\bm{R}) \sim \sum_{M=\pm2} e^{\ii (\kappa - n_1M)\theta_{1} + \ii (\kappa - n_2 M)\theta_{2}}\gamma_{M}\Gamma_{M}, \label{eq:bc-twohqvs}
\end{align}
where $\theta_{1}$ ($\theta_2$) is the angle of $\bm{R} - \bm{R}_1$ ($\bm{R} - \bm{R}_2$).
Since the difference between $\theta_1$ and $\theta_2$ is negligible for $|\bm{R}| \to \infty$, the phase part behaves as 
$\exp( 2 i \kappa \theta)$ [$\theta \simeq (\theta_1 + \theta_2)/2$], which means the boundary conditions for the two HQVs and the singly quantized vortex are equivalent in the limit $\rho\to \infty$. In the following, we explain the internal structure of each HQV, and the interaction energy of the two HQV vortex through the comparison of the singly quantized vortex~\cite{Masaki2022}. Particularly, the non-axisymmetric internal structure of each HQV is important to the interaction energy.

\sub{Isolated HQV $(\kappa,n) = (1/2, +1/4)$}
The two disgyrations of the triads, given by $n=\pm 1/4$, are inequivalent because one is parallel to the vorticity, while the other is antiparallel. Here we first explain the antiparallel case $(n,\kappa) = (1/2, 1/4)$. In the above notation, by setting $\bm{R}_1 = (0,0)$ and $\bm{R}_2 = (-\infty, 0)$, and thus setting $\theta_2 \to 0$, the boundary condition is obtained from Eq.~\eqref{eq:bc-twohqvs}. In this case, the relative phase of $\gamma_{\pm2}$ is zero, and the schematic image of the disgyration is drawn in Fig.~\ref{fig:1/4}(a). Each circular object shows the directions of $d$-vectors by colored arrows, whose color bar is the same as in panel (c). There are also the cyan and magenta lines, which represent the directions of the eigenvalues $+1$ and $-1$ of the order parameter $A$, respectively. \green{This HQV is $(1/2, C_4)$ in the conjugacy class (V) in Eq.~\eqref{eq:conj-nematic}.} 
The self-consistent solutions of the non-trivial components are shown in panels (b)--(e). Here $\gamma_2$, not shown, is almost uniform without phase winding, as expected from the boundary condition $\kappa - n_1 M = 1/2 - 1/4 * 2 = 0$. The other bulk component $\gamma_{-2}$ is a conventional singular vortex with the single phase winding shown in panel~(c). The amplitude and the phase of the internal structure of $\gamma_0$ are, respectively, shown in panels~(d) and (e). The phase of the internal structure is oppositely winding against $\gamma_{-2}$, which can be understood by analogy with a vortex in a spinless chiral $p$-wave SC. We assume, for simplicity, $\bar{k}_z = 0$, and then write $\bar{k}_\chi = (k_x + i \chi k_y)/k_{\F} = e^{i \chi \alpha}$ with the sign of the angular momentum $\chi=\pm$. With this notation, the gap function is written as 
\begin{align}
\hat{\Delta}(\bm{k}_\F,\bm{R})
&= 
[-\gamma_2(\bm{R}) \bar{k}_+ + \gamma_0(\bm{R})/\sqrt{6} \bar{k}_-]\ket{\uparrow\uparrow}\nonumber \\
&+
[\gamma_{-2}(\bm{R}) \bar{k}_- - \gamma_0(\bm{R})/\sqrt{6} \bar{k}_+]\ket{\downarrow\downarrow}.
\label{eq:hqv-spinsector}
\end{align}
In each spin sector, two opposite orbital-angular-momenta, $\bar{k}_\pm$, are mixed owing to the induced component $\gamma_0$. In a spinless chiral $p$-wave SC, the induced component has the opposite angular momentum relative to the component which has a non-zero order parameter far from the vortex core, i.e., $\chi_i = - \chi_b $~\cite{Heeb1999,MatsumotoHeeb2001}. Here we call the latter component the {\it bulk component}, and $\chi_b$ ($\chi_i$) represents the sign of the angular momentum for the bulk (induced) component.
The vorticity of the induced component, $\kappa_i$, is given by
$\kappa_i = \kappa_b + \chi_b - \chi_i$, where $\kappa_b$ is the vorticity of the bulk component. Here the $\chi_b - \chi_i$ accounts for the angular momentum difference between the bulk and induced components, and is reduced to $2\chi_b$. For the $(1/2, +1/4)$-HQV, the bulk component $\gamma_{-2}$ has the vorticity $\kappa_b = \kappa - n (-2) = 1$, and $\chi_b = - 1$. Thereby $\kappa_i = \kappa_b + 2\chi_b = -1$ successfully explains the vorticity of the induced component shown in panel~(e). More striking feature is the three-fold symmetry in the amplitude of the induced component $|\gamma_0|$. Reflecting this discrete symmetry, the phase evolution of the induced component surrounding the vortex is non-linear. As seen from Eq.~\eqref{eq:hqv-spinsector}, the main structure of this HQV is in the spin-down sector.

\sub{Isolated HQV $(\kappa,n) = (1/2, -1/4)$}
The other HQV, characterized by $(1/2, -1/4)$, at $\bm{R}_2 = (0,0)$, is obtained by $\bm{R}_1\to (+\infty, 0)$, i.e., $\theta \to \pi$ in Eq.~\eqref{eq:bc-twohqvs}, in which the relative phase of $\gamma_{\pm2}$ is opposite because of the phase factor $\exp[i(1/2+ M/4)\pi]$ from the $(1/2, +1/4)$ HQV. 
\green{This HQV is $(1/2, C_4^{-1})$ in the conjugacy class (V) in Eq.~\eqref{eq:conj-nematic}, and its}
 disgyration is schematically drawn in Fig.~\ref{fig:1/4}(f).
In contrast to the previous case, 
a conventional vortex structure appears in the $M = +2$ sector shown in panels (g) and (h), 
while an almost uniform structure without phase winding is in the $M = -2$ sector. 
In panel (j) the phase winding of the induced component is $+ 3$, which can be understood as in the previous case:
$\kappa_i = \kappa_b + 2 \chi_b = 3$ for $\kappa_b = \kappa + M/4 = 1$ and $\chi_b = +1$. The phase evolution surrounding the vortex is again non-linear reflecting the discrete symmetry of the amplitude $|\gamma_0|$, which is five-fold as shown in panel (h). The major vortex structure of this HQV is in the spin-up sector.

\sub{Molecule of HQVs (stability)}
The two types of internal structures in the $M=0$ component induced for the HQVs with $n = \pm1/4$ are modulated by the connection of these two HQVs. This modulation causes an interaction between the two HQVs.
The interaction energy of the two HQVs are defined by $
\Delta \mathcal{J}_{\mathrm{sn}}(d_{\mathrm{v}}) = 
\mathcal{J}_{\mathrm{sn}}(\bm{R}_{1},\bm{R}_{2}) - 
\mathcal{J}_{\mathrm{sn}}^{+}(\bm{R}_{1}) - 
\mathcal{J}_{\mathrm{sn}}^{-}(\bm{R}_{2})$ based on the Luttinger--Ward energy functional $\mathcal{J}_{\mathrm{sn}}$~\cite{Vorontsov2003}. Here 
the two HQVs are at $\bm{R}_{1}=(d_\mathrm{v}/2,0)$ and $\bm{R}_{2}=(-d_\mathrm{v}/2,0)$, 
schematically shown in Fig.~\ref{fig:hqv_all}(a), 
whose energy is denoted by $\mathcal{J}_{\mathrm{sn}}(\bm{R}_{1},\bm{R}_{2})$. 
The energies of the isolated HQVs $(\kappa,n)=(1/2, +1/4)$ at $\bm{R}_{1}$ and $(1/2,-1/4)$ at $\bm{R}_{2}$ 
are denoted by $\mathcal{J}_{\mathrm{sn}}^{+}(\bm{R}_{1})$ and $\mathcal{J}_{\mathrm{sn}}^{-}(\bm{R}_{2})$, respectively.

Figure~\ref{fig:hqv_all}(b) shows the interaction energies of the two HQVs as a function of intervortex distance $d_{\mathrm{v}}$. 
When $d_{\mathrm{v}} = 0$, the $d$ vortex (the singly quantized vortex with double core structure) is realized. The gap structure is show in Fig.~\ref{fig:d-vortex}. 
Importantly, the positive energy of the $d$ vortex means that two isolated HQVs $(d_\mathrm{v} \to \infty)$ are more stable than the $d$ vortex.
The actual interaction energy takes the minimum at finite $d_{\mathrm{v}}$, indicating the instability of the $d$ vortex into a bound state of the two HQVs like a molecule.  \green{As discussed in Sec.~\ref{sec:vortex-examples}, a singly quantized vortex $(\kappa,n) = (1,0)$ in the conjugacy class (I) in Eq.~\eqref{eq:conj-nematic} splits into two HQVs $(1/2,+1/4)$ and $(1/2, -1/4)$ in the class (V). }

The stabilization mechanism in this molecule state is 
a deformation in the non-axisymmetric induced component $\gamma_0$, as shown in Figs.~\ref{fig:hqv_all}(c-1) and \ref{fig:hqv_all}(c-2), realized in the strongly spin-orbit coupled Cooper pairs~\cite{Masaki2022}. This mechanism is purely intrinsic and possible even in the weak coupling limit, different from those in other systems such as the
SF ${}^{3}$He~\cite{PhysRevLett.55.1184} and unconventional SCs~\cite{Chung:2007zzc}: 
In the SF ${}^{3}$He-A phase, Volovik and Salomaa phenomenologically introduced some corrections in spin mass 
to stabilize the HQV~\cite{PhysRevLett.55.1184}; This correction might be regarded as a kind of strong coupling effect through the Fermi liquid correction, but another strong coupling effect is known to destabilize the HQV~\cite{Kawakami2009,Kawakami2010,Kawakami2011,mizushimaJPSJ16}. 
Vakaryuk and Leggett unveiled that the HQVs in equal spin pairing states, such as $^3$He-A and $D_4$-BN, are accompanied by a nonzero spin polarization even in the absence of external Zeeman coupling~\cite{vak09}. The coupling of such spin polarization to external magnetic fields may affect the stability of HQVs.
Another example of the stabilization mechanism is an extrinsic one in the polar and polar distorted phases of the SF ${}^3$He because of strongly anisotropic impurities, as discussed in Sec.~\ref{sec:plat}.

\subsection{Non-Abelian anyon in non-Abelian vortex}
\label{sec:majorana-in-hqv}

\begin{figure}[b!]
\begin{center}
\includegraphics[width=\hsize]{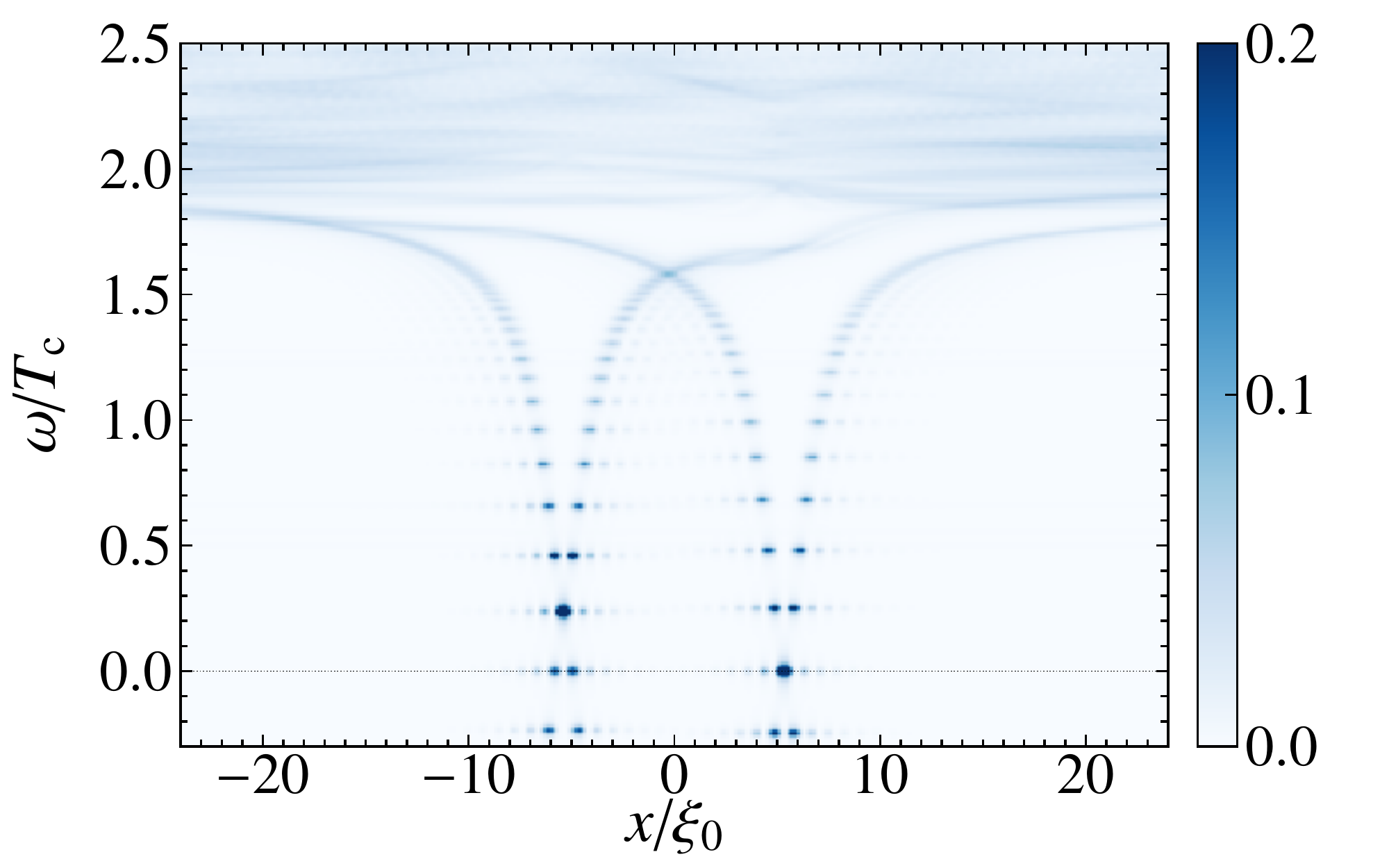}
\caption{
	Local density of states $\nu_{k_{z} = 0}(\bm{R};\omega)$ at $k_{z} = 0$ and $y = 0$ for a pair of HQVs located at $(x,y)=(\pm d_{\mathrm{v}}/2,0)$ with $d_{\mathrm{v}} \simeq10.7\xi_0$. The spatial distributions of the two zero energy states are different. Figures are adapted from Ref.~\cite{Masaki2022} \copyright 2022 American Physical Society.
}
\label{fig:ldos}
\end{center}
\end{figure}

In this section, first the zero energy bound states in above-discussed HQVs are demonstrated. Second we summarize the discrete symmetries in the presence of vortices, and finally discuss the topological protection of the zero energy states and their non-Abelian nature.

\sub{Existence of zero energy states} To investigate fermionic bound states at discrete energy levels, it is necessary to solve the BdG equation~\eqref{eq:bdg}. By assuming the spatial uniformity along the $z$-direction, the quantum number is labeled as $\nu = (\alpha,k_z)$, and only the $k_z = 0$ sector is focused on to seek for zero energy bound states. The direct numerical calculation gives us a spectral function including the discrete bound states around the HQV-cores, as shown in Fig.~\ref{fig:ldos}(a). There are two zero energy states localized in both of the cores. Here, the molecule state of the two HQVs with $d_{\mathrm{v}} \simeq 10.7 \xi_0$ is used for a gap function in the BdG equation, and the spectral function, also  known as the local density of state, is defined as 
$\nu_{k_{z} =0}(\bm{R};\omega) = \sum_{\alpha,\sigma}|u_{\alpha,k_{z}=0,\sigma}(\bm{R})|^{2}\delta(\omega- \epsilon_{\alpha,k_{z}=0})$ along $\bm{R} = (x, 0)$,
where $u_{\alpha,k_z,\sigma}(\bm{R})$ is the particle component with spin $\sigma$ of the eigenfunction $\vec{u}_{\alpha,k_z}(\bm{R})$. In the calculation, the parameter $k_\F \xi_0$, characterizing the discreteness of the bound states,  is set to $5$. The Majorana condition, $u_{\alpha,k_z=0, \sigma}(\bm{R}) = [v_{\alpha,k_z=0,\sigma}(\bm{R})]^*$, for the wave functions of the zero energy state can be confirmed~\cite{Masaki2022}. 

\sub{Discrete symmetry of vortices}
We utilize the semiclassical approximation to clarify the symmetry and topology of vortices, following Refs.~\cite{tsutsumiPRB15,mizushimaJPSJ16,shiozakiPRB14}. In the semiclassical approximation, the spatial modulation due to a vortex line is treated as adiabatic changes as a function of the real-space coordinate surrounding the defect with the angle $\theta$. Then, the BdG Hamiltonian in the base space, $(\bm{k},\theta)$, is obtained from Eq.~\eqref{eq:bdg} as 
\begin{align}
\check{\mathcal{H}}({\bm k},\theta) 
= \begin{pmatrix}
\hat{h}_N({\bm k}) & \hat{\Delta}({\bm k},\theta) \\
\hat{\Delta}^{\dag}({\bm k},\theta) & -\hat{h}_{N}^*({\bm k})
\end{pmatrix},
\label{eq:bdg2}
\end{align}
where the $^{3}P_{2}$ pair potential within a semiclassical approximation is given by 
\begin{align}
\hat{\Delta}({\bm k},\theta) = i \hat{\sigma}_{\mu} A_{\mu i}(\theta) k_i/k_{\F} \hat{\sigma}_{y}.
\end{align}

Let us now summarize the discrete symmetries of the BdG Hamiltonian in Eq.~\eqref{eq:bdg2}. In the presence of vortex, the BdG Hamiltonian breaks the time-reversal symmetry $\hat{\mathcal{T}}=-i\hat{\sigma}_{y}K$, but holds the particle-hole symmetry
\begin{align}
\check{\mathcal{C}} \check{\mathcal{H}}({\bm k},\theta) \check{\mathcal{C}}^{-1} = -\check{\mathcal{H}}(-{\bm k},\theta),
\label{eq:phs}
\end{align}
where $\check{\mathcal{C}} = \check{\tau}_{x} K$ and $K$ is the complex conjugation operator. 
In addition to the particle-hole symmetry, three discrete symmetries, the $\{P_{1},P_{2},P_{3}\}$, are pointed out to be relevant to vortices of the SF $^3$He-B phase~\cite{sal85,salomaaRMP}. 
Their representations are given by
\begin{align}
\hat{P}_{1}=PU_{\pi}\hat{1}, ~ \hat{P}_{3} = \hat{\mathcal{T}}\hat{C}_{2,x},
\end{align}
and $\hat{P}_{2}=\hat{P}_{1}\hat{P}_{3}$. Here, $P$, $U_\pi = e^{i (\kappa + 1 )\pi}$, $\hat{C}_{2.x}=e^{-i\hat{J}_{x}\pi}$ are the spatial inversion, the discrete phase rotation, and the $\pi$-rotation in the spin-orbit space about the $x$-axis. The physical meanings of $P_1$, $P_2$, and $P_3$ symmetries are, the inversion, magnetic reflection, and magnetic $\pi$-rotation symmetries respectively. From the definition, $P_1P_2P_3 = 1$. 
Under these symmetry operations, a momentum $\bm{k}$ and a spin $\bm{\sigma}$ are transformed as
\begin{align}
P_1:\bm{k} 
&\to 
(-k_x, -k_y, -k_z),~\bm{\sigma} \to
(\sigma_x, \sigma_y, \sigma_z),\\
P_2:\bm{k}
&\to 
(k_x, -k_y, -k_z),~\bm{\sigma} \to 
(-\sigma_x, \sigma_y, \sigma_z),\\
P_3: \bm{k}
&\to 
(-k_x, k_y, k_z),~
\bm{\sigma} \to 
(-\sigma_x, \sigma_y, \sigma_z).
\end{align}
Under these transformation,
the order parameters are transformed as
\begin{align}
P_1: \gamma_M(\rho,\theta)&\to -\gamma_M(\rho,\theta + \pi),\\
P_2: \gamma_M(\rho,\theta)&\to (-1)^{\kappa + M}[\gamma_M(\rho,\pi-\theta)]^*,\\
P_3: \gamma_M(\rho,\theta)&\to (-1)^M[\gamma_M(\rho,-\theta)]^*.
\end{align}
For axisymmetric vortices, we confirm the non-zero components of $\gamma_M(\rho)$ summarized in Table.~\ref{tab:axisym-vortex} by using 
the relation $\gamma_M(\rho,\theta) = \gamma_M(\rho)e^{i(\kappa - M)\theta}$.
For example, 
$P_2:\gamma_M(\rho)e^{i(\kappa - M)\theta}\to
\gamma_M^*(\rho)e^{i(\kappa - M)\theta}$, and
when the $P_2$ symmetry is preserved, $\gamma_M(\rho) = \gamma_M^*(\rho)$ can be obtained. In the case of the $d$ vortex, the $P_3$ symmetry is broken because of the non-zero $\gamma_{\pm1}$ components, and the $P_2$ symmetry is the only preserved symmetry among these three. In the case of isolated HQVs, 
even though $\gamma_{\pm1} = 0$,
because of the three-fold, and five-fold symmetry in the gap amplitude of $\gamma_0$, the $P_1$ and $P_2$ symmetries are broken, but the $P_3$ symmetry is preserved.

Another discrete symmetry is the mirror reflection symmetry about the plane perpendicular to the vortex line, that is, the $xy$-plane, which is already discussed in Sec.~\ref{sec:SPnAayon}. The mirror refection symmetry $M_{xy}$
transforms the momentum and the spin as
\begin{align}
M_{xy}: \bm{k} \to (k_x,k_y,-k_z), \bm{\sigma}\to(-\sigma_x,-\sigma_y,\sigma_z),
\end{align}
and thus the order parameter is transformed as
\begin{align}
M_{xy}:\gamma_M(\rho,\theta)\to(-1)^{M+1}\gamma_M(\rho,\theta).
\end{align}
Therefore, in the presence of $M_{xy}$, even $M$ and odd $M$ cannot be mixed. 
In the case of the $D_4$-BN state with the point nodes along the $z$-direction, $\gamma_{M=\pm2} \neq 0$, and the only possible mirror operation
that commutes with the BdG Hamiltonian at the mirror invariant momentum $\bm{k}_M=(k_x,k_y,0)$ is $\eta = -$ in Eq.~\eqref{eq:commut} when $\gamma_{\pm1} = 0$. The above-discussed vortices of a \tPt SF which preserve the $M_{xy}$ symmetry are the axisymmetric $o$ and $u$ vortices, and the HQVs.

\sub{Topology of the vortex bound states}
On the basis of the above-discussed discrete symmetries, the zero energy state of each HQV is shown to be protected by two topological number: one based on the chiral symmetry, a combined symmetry between the particle-hole symmetry and the $P_3$ symmetry, and the other based on the mirror reflection symmetry.  

When the $P_3$ symmetry is preserved, the semiclassical BdG Hamiltonian is transformed under the combined symmetry operation $\Gamma = \mathcal{C}P_3$ with $\Gamma^2 = +1$ as
\begin{align}
\check{\Gamma}\check{\mathcal{H}}(\bm{k},\theta)\check{\Gamma}^{-1} = - \check{\mathcal{H}}(k_x,-k_y, -k_z, -\theta).
\end{align}
Particularly, for $k_y = k_z = 0$ and $\theta = 0$ or $\pi$, the BdG Hamiltonian anticommutes with $\Gamma$, which can be regarded as chiral symmetry~\cite{sat11,miz12,tsutsumiJPSJ13,shiozakiPRB14,tsutsumiPRB15,Masaki:2019rsz,mizushimaJPSJ16}. As in the bulk nematic state, as long as the chiral symmetry is preserved, one can define the one-dimensional winding number for $\bm{k}_x = (k_x,0,0)$ and $\theta = 0$ or $\pi$ as 
\begin{align}
w_{\mathrm{1d}}(\theta) = -\frac{1}{4\pi i} \int dk_{x} \mathrm{tr}\left[
\check{\Gamma}\check{\mathcal{H}}^{-1}({\bm k}_x,\theta)\partial _{k_{x}} \check{\mathcal{H}}({\bm k}_x,\theta)
\right].
\end{align}
The vortices which preserve the $P_3$ symmetry are $o$, and $w$ vortices and the non-Abelian HQVs. Among them, the winding numbers $(w_{\mathrm{1d}}(0), w_{\mathrm{1d}}(\pi))$ of  $o$ and $w$ vortices are $(2,-2)$, and thus the topological invariant $w_{\mathrm{1d}} = (w_{\mathrm{1d}}(0) - w_{\mathrm{1d}}(\pi))/2 = 2$ ensures the existence of two zero energy states. In the case of non-Abelian HQVs, $(w_{\mathrm{1d}}(0),w_{\mathrm{1d}}(\pi))$ is $(2,0)$ for $(\kappa,n) = (1/2,+1/4)$
and $(0,-2)$ for $(1/2,-1/4)$\footnote{
\green{The choice of the relative phase between $\gamma_{\pm2}$ as $\pi$ by $\theta_1 \to \pi$ leads to $(w_{\mathrm{1d}}(0),w_{\mathrm{1d}}(\pi))=(0,-2)$ rather than $(2,0)$ for the HQV of $(\kappa,n) = (1/2, -1/4)$.}
}. In both cases, the topological invariant $w_{\mathrm{1d}}$ becomes $1$, which ensures the existence of one zero energy
state in each HQV core.

Next we explain the topological invariant based on the mirror reflection symmetry $M_{xy}$. As discussed above, the BdG Hamiltonian $\check{\mathcal{H}}(\bm{k}_M,\theta)$ for the mirror invariant momentum $\bm{k}_M$ commutes with the mirror operator $\check{\mathcal{M}}^{\eta}$ for $\eta = -$ in Eq.~\eqref{eq:commut}. Thus, $\check{\mathcal{H}}(\bm{k}_{\mathrm{M}},\theta)$ can be block diagonalized with respect to the mirror eigenstates with eigenvalues $\lambda = \pm i$ as 
\begin{align}
\check{\mathcal{H}}({\bm k}_{\mathrm{M}},\theta) = \bigoplus_{\lambda}\tilde{\mathcal{H}}_{\lambda}({\bm k}_{\mathrm{M}},\theta).
\end{align}
Significantly, even for the BdG Hamiltonian
in each mirror subsector, $\tilde{\mathcal{H}}_{\lambda}$, the reduced particle-hole symmetry $\tilde{\mathcal{C}}$ is preserved, though it is not for the other mirror symmetry operator $\check{\mathcal{M}}^{\eta = +}$. Therefore,
$\tilde{H}_{\lambda}$ belongs to the class D as well as spinless chiral SCs~\cite{Schnyder:2008tya}, and the $\mathbb{Z}_2$ invariant, $\nu_{\lambda}$ can be constructed in each mirror subsector~\cite{teo,qiPRB08}. Non-trivial (odd) value of $\nu_{\lambda}$ ensures that the vortex has a single Majorana zero mode which behaves as a non-Abelian (Ising) anyon~\cite{ueno,sato14,tsutsumiJPSJ13}.

The $\mathbb{Z}_2$ invariant $\nu_{\lambda}$ is defined on the base space $(S^2 \times S)$ composed of the two dimensional mirror invariant momentum space $\bm{k}_{\mathrm{M}}$ and the angle in the real space surrounding the vortex $\theta$, 
and constructed by the dimensional reduction of the second Chern number
defined in the four dimensional base space $(S^3\times S)$ composed of $\bm{k}$ and $\theta$. The $\mathbb{Z}_2$ invariant in each mirror subsector is given by the integral of the Chern-Simons form:
\begin{align}
\nu_{\lambda} &= \left(\frac{i}{\pi} \right)^{2} \int _{S^{2}\times S^{1}} \tilde{Q}_{3,\lambda}~ \mod 2,\\
\tilde{Q}_{3,\lambda} &= \mathrm{tr}[\tilde{\mathcal{A}}_{\lambda}\wedge d~\!\!\tilde{\mathcal{A}}_{\lambda} + \frac{2}{3}\tilde{\mathcal{A}}_{\lambda}\wedge\tilde{\mathcal{A}}_{\lambda}\wedge\tilde{\mathcal{A}}_{\lambda}].
\end{align}
A non-Abelian Berry connection  $\tilde{\mathcal{A}}_{\lambda}$ is given by
\begin{align}
i [\tilde{\mathcal{A}}_{\lambda}]_{nm} = \sum_{\alpha}\braket{\tilde{u}_{\lambda,n}({\bm k}_{\mathrm{M}},\theta)| \partial _{\alpha} \tilde{u}_{\lambda,m}({\bm k}_{\mathrm{M}},\theta)} d~\!\! \alpha, 
\end{align}
where $\alpha$ denotes $(k_x,k_y,\theta)$, and 
$\ket{\tilde{u}_{\lambda,m}({\bm k}_{\mathrm{M}},\theta)}$ is the $m$th eigenstate of $\tilde{\mathcal{H}}_{\lambda}(\bm{k},\theta)$. 
Within the semiclassical approximation in the $D_4$-BN state, the $\mathbb{Z}_2$ invariant is calculated as
\begin{align}
\nu_{\lambda} = \ell_\lambda \kappa_\lambda ~\mod 2,
\end{align}
where $\ell_{\lambda}$ and $\kappa_{\lambda}$ are the first Chern number and the vorticity in the mirror subsector $\lambda$.
The former characterizes the bulk topology and
$\ell_{\lambda = \pm i} = \pm 1$. The latter is given by $\kappa_{\lambda} = \kappa - nM$ for the bulk component $M$ in the $\lambda$ subsector. In the case of the isolated HQVs,
$(\nu_{+i}, \nu_{-i})= (0, -1)$ [$(+1,0)$] for $(\kappa,n) = (1/2, + 1/4)$ [$(1/2, - 1/4)$]. 
The two HQVs host Majorana fermions in different mirror sectors, and thus the $\mathbb{Z}_2$ invariant of the molecule state is given by $(\nu_{+i}, \nu_{-i})= (+1, -1)$. The $\mathbb{Z}_2$ invariant of the $o$ vortex has the same one, and the pairwise Majorana fermions are localized at the same core~\cite{Masaki:2019rsz}.

The zero energy state bound in each HQV core is protected by the two topological invariants based on the mirror reflection symmetry and the chiral symmetry, and it is identified as non-Abelian (Ising) anyon of the Majorana fermion. Therefore, the HQV in the $D_4$-BN state has two fold non-Abelian nature: one from the non-Abelian Ising anyon, and the other from non-Abelian first homotopy group.

\section{Summary}\label{sec:summary}
We have presented various types of 
non-Abelian anyons 
in topological SCs/SFs 
and other systems.
Non-Abelian anyons are attracting a great attention because of possible applications to topological quantum computations, 
where quantum computations are realized by 
braiding of non-Abelian anyons.
The simplest non-Abelian anyons are 
Ising anyons realized by Majorana fermions hosted by vortices 
(or edges) of topological superconductors, 
$\nu =5/2$ quantum Hall states,
spin liquids, 
and high density quark matter. 
These are, however, insufficient for universal quantum computations. 
The other anyons which can be used for universal quantum computations
is given by Fibonacci anyons 
which exist in $\nu =12/5$ quantum Hall states. 
There are also Yang-Lee anyons which are non-unitary counterparts of Fibonacci anyons.
Another class of non-Abelian anyons 
of the bosonic origin can be given by 
non-Abelian vortex anyons 
realized by non-Abelian vortices supported by a non-Abelian first homotopy group. 
These vortex anyons exist in BN liquid crystals, 
cholesteric liquid crystals, 
spin-2 spinor BECs, 
and high density quark matter. 
There is a unique system simultaneously 
admitting 
two kinds of non-Abelian anyons, which is  
the Majorana fermions (Ising anyons) 
and non-Abelian vortex anyons. 
That is \tPt SFs,
spin-triplet, $p$-wave paring of neutrons, 
expected to exist 
in neutron star interiors  
as the largest 
topological quantum matter in universe.

\section*{Acknowledgements}
This work was supported by a Grant-in-Aid for Scientific Research on Innovative Areas ``Quantum Liquid Crystals'' (Grant No. JP22H04480) from JSPS of Japan and JSPS KAKENHI (Grants No. JP18H01217, No. JP19K14662, No. JP20K03860, No. JP20H01857, No. JP21H01039, and No. JP22H01221), and 
the WPI program "Sustainability with Knotted Chiral Meta Matter (SKCM$^2$)" at Hiroshima University.

\bibliographystyle{agsm}
\bibliography{reference}

\end{document}